\documentclass[twocolumn]{aastex631}
\usepackage{listings}
\usepackage{multirow}
\usepackage{amsmath}
\begin{document}

\title{Surface Brightness Fluctuations in Two SPT clusters: a Pilot Study}
\author[0000-0001-5725-0359]{Charles E. Romero}
\altaffiliation{E-mail: \href{mailto:charles.romero@cfa.harvard.edu}{charles.romero@cfa.harvard.edu} }
\affiliation{Center for Astrophysics $\vert$ Harvard \& Smithsonian, 60 Garden Street, Cambridge, MA 02138, USA}

\author[0000-0003-2754-9258]{Massimo Gaspari}
\affiliation{Department of Astrophysical Sciences, Princeton University, 4 Ivy Lane, Princeton, NJ 08544-1001, USA}

\author[0000-0002-4962-0740]{Gerrit Schellenberger}
\affiliation{Center for Astrophysics $\vert$ Harvard \& Smithsonian, 60 Garden Street, Cambridge, MA 02138, USA}

\author[0000-0002-5108-6823]{Bradford A. Benson}
\affiliation{Fermi National Accelerator Laboratory, MS209, P.O. Box 500, Batavia, IL 60510, USA}
\affiliation{Department of Astronomy and Astrophysics, University of Chicago, 5640 South Ellis Avenue, Chicago IL 60637, USA}
\affiliation{Kavli Institute for Cosmological Physics, University of Chicago, 5640 South Ellis Avenue, Chicago, IL 60637, USA}

\author[0000-0001-7665-5079]{Lindsey E. Bleem}
\affiliation{High Energy Physics Division, Argonne National Laboratory, 9700 South Cass Avenue, Lemont, IL, 60439, USA}
\affiliation{Kavli Institute for Cosmological Physics, University of Chicago, 5640 South Ellis Avenue, Chicago, IL 60637, USA}

\author[0000-0002-7619-5399]{Esra Bulbul}
\affiliation{Max Planck Institute for Extraterrestrial Physics, Giessenbachstrasse 1, 85748 Garching, Germany}

\author[0000-0002-8248-4488]{Matthias Klein}
\affiliation{University Observatory, Faculty of Physics, Ludwig-Maximilians-Universit\"{a}t, Scheinerstr. 1, 81679 Munich, Germany}

\author[0000-0002-0765-0511]{Ralph Kraft}
\affiliation{Center for Astrophysics $\vert$ Harvard \& Smithsonian, 60 Garden Street, Cambridge, MA 02138, USA}

\author[0000-0003-0297-4493]{Paul Nulsen}
%\altaffiliation{ICRAR, University of Western Australia, 35 Stirling Hwy, Crawley, WA 6009, Australia}
\affiliation{Center for Astrophysics $\vert$ Harvard \& Smithsonian, 60 Garden Street, Cambridge, MA 02138, USA}
\affiliation{ICRAR, University of Western Australia, 35 Stirling Hwy, Crawley, WA 6009, Australia}

\author[0000-0003-2226-9169]{Christian L. Reichardt}
\affiliation{School of Physics, University of Melbourne, Parkville, VIC 3010, Australia}
%\affiliation{ARC Centre of Excellence for All Sky Astrophysics in 3 Dimensions (ASTRO 3D), Australia}

\author{Laura Salvati}
\affiliation{Universit\'{e} Paris-Saclay, CNRS, Institut d'Astrophysique Spatial, F-91405, Orsay, France}

\author[0000-0003-3521-3631]{Taweewat Somboonpanyakul}
\affiliation{Kavli Institute for Particle Astrophysics and Cosmology, Stanford University, 452 Lomital Mall, Stanford, CA 94305, USA}
\affiliation{Department of Physics, Faculty of Science, Chulalongkorn University
254 Phayathai Road, Pathumwan, Bangkok Thailand. 10330}

\author[0000-0002-3886-1258]{Yuanyuan Su}
\affiliation{Department of Physics and Astronomy, University of Kentucky, 505 Rose Street, Lexington, KY 40506, USA}

%\affiliation{Various}

\begin{abstract}

    %Sunyaev-Zel'dovich (SZ) observations have been
    Studies of surface brightness fluctuations in the intracluster medium (ICM) present an indirect probe of turbulent properties such as the turbulent velocities, injection scales, and the slope of the power spectrum of fluctuations towards smaller scales. With the advancement of Sunyaev-Zel'dovich (SZ) studies and surveys relative to X-ray observations, we seek to investigate surface brightness fluctuations in a sample of SPT-SZ clusters which also have archival \textit{XMM-Newton} data. Here we present a pilot study of two typical clusters in that sample: SPT-CLJ0232-4421 and SPT-CLJ0638-5358.
    We infer injection scales larger than 500 kpc in both clusters and Mach numbers $\approx 0.5$ in SPT-CLJ0232-4421 and Mach numbers $\approx 0.6 - 1.6$ in SPT-CLJ0638-5358, which has a known shock. We find hydrostatic bias values for $M_{500}$ less than 0.2 for SPT-CLJ0232-4421 and less than 0.1 for SPT-CLJ0638-5358. These results show the importance to assess its quantitative values via a detailed multiwavelength approach and suggest that the drivers of turbulence may occur at quite larger scales.
    %of $0.15 \pm 0.12$ and $0.01 \pm 0.48$ 
    %\MG{I don't understand such a low value for the disturbed system; also we quote another value in the Conclusions, please double check/improve} for SPT-CLJ0232-4421 and SPT-CLJ0638-5358, respectively.
\end{abstract}

\section{Introduction} \label{sec:intro}

    Surface brightness fluctuations of the intracluster medium (ICM) provide a window to probe the turbulent properties of the ICM. The surface brightness fluctuations of the ICM in the X-ray and millimeter  \citep[via the Sunyaev-Zel'dovich, SZ, effect][]{sunyaev1972} %(via the Sunyaev-Zel'dovich\citep[SZ;][]{sunyaev1972} effect) 
    regimes can be deprojected to thermodynamic quantities because the ICM is optically thin. The SZ signal, parameterized in terms of Compton $y$, is proportional to the electron pressure, $P_e$ along the line of sight:
    \begin{equation}
        y = \frac{\sigma_T}{m_e c^2} \int P_e dz,
        \label{eqn:Compton_y}
    \end{equation}
    where $\sigma_T$ is the Thomson cross section, $m_e$ is the electron mass, $c$ is the speed of light, and $z$ is taken to be the axis along the line of sight. The X-ray surface brightness is proportional to the emission integral, EI:
    \begin{equation}
        \text{EI} \equiv \int n_p n_e dz,
        \label{eqn:ei}
    \end{equation}
    where $n_p$ is the proton density and $n_e$ is the electron density. This value is similar to the classically defined emission measure, EM\footnote{EM is the integral of the expression in Equation~\ref{eqn:ei} over a volume, $dV$, rather than $dz$}; one can consider the EI to be the EM per unit area. The proportionality is modulated by the cooling function\footnote{Formally the cooling function is bolometric; i.e. it encapsulates the emission over all frequencies.} defined for a given energy band $b$, $\Lambda_b(Z,T_g)$, which is a function of metallicity $Z$ and gas temperature $T_g$. Noting that $n_p$ will have some proportionality to $n_e$ depending on $Z$, we see that EM $\propto n_e^2$. When measuring the X-ray surface brightness of hot clusters (such as those in our sample) in a soft-energy band (e.g.,0.5 to 2 keV), the cooling function (and thus the surface brightness) is relatively temperature insensitive \citep[e.g.;][]{sarazin1988}. 
    
    Due to these proportionalities, the surface brightness of the SZ and X-ray (in soft bands) can be deprojected to determine pressure \citep[e.g.;][]{khatri2016} and density \citep[e.g.;][]{churazov2012} fluctuations, respectively. This deprojection can be done in Fourier space \citep[][]{peacock1999} such that a power spectrum of the surface brightness fluctuations can be transformed into the power spectrum of the respective thermodynamic quantity governing the surface brightness signal. The power spectrum of thermodynamic quantities is particularly informative about the physics of the ICM plasma as it will depend on the injection scale, turbulent motions, and transport properties such as viscosity.
    %thermal conduction. 

    In practice, probing the viscosity (turbulent cascade) is very difficult due to the required dynamic range, both in angular scales and amplitudes. That is, these measurements generally require high resolution and high sensitivity \citep[e.g.;][]{churazov2012}. In contrast, estimating the turbulent velocity is less stringent, since in a stratified atmosphere we expect a linear relation between thermodynamic fluctuations and turbulent velocities, $v$, \citep[i.e.~Mach numbers, $v / c_s$, that quantify gas motions relative to the speed of sound, $c_s$;][]{Gaspari2013_PS,Gaspari2014_PS,Zhuravleva2014,zhuravleva2023,simonte2022}, where this relation is dominated by the peak of the amplitude spectra.
    %with the Mach scaling being dominated by the peak of the amplitude spectra. 
    Thus, assuming an injection scale, Mach numbers can be discerned over a sample of clusters via Fourier analysis of surface brightness fluctuations \citep[e.g.,][]{eckert2017}. Similarly, \citet{hofmann2016} binned regions by photon count and obtained deprojected thermodynamic profiles and took the scatter in those profiles to reflect the fluctuations and then infer Mach numbers.

    %%%%%%%%%%%%%%%%%%%%%%%%%%%%%%%%%%%%%%%%%%%%%%%%%%%%%%%%%%%%%%%%%%%%%%%%%%%%%%%%%%%%%%%%%%%%%%%%%%%%%%%%%%%%%%%%%%%%%%%%%%%%%%%%%%%

    As a complementary application of power spectra of ICM fluctuations, we can investigate the hydrostatic mass bias within clusters. From simulations and observations, many studies find average hydrostatic mass biases between 10\% to 30\% \citep[e.g.,][]{mahdavi2013,martino2014,applegate2014,hoekstra2015,mantz2016,hurier2018,siegel2018}, while a hydrostatic mass bias of over 40\% would be needed to resolve the tension between \textit{Planck} cosmology and \textit{Planck} cluster number counts (\citealt{Planck2016_XXIV}; see also \citealt{Pratt2019} for a recent review of cluster mass estimates and their cosmological impact). The majority of recent observational constraints on hydrostatic mass bias values compare some hydrostatic mass estimate (from X-ray or SZ data) and a total mass estimate from weak lensing, though one can also estimate the mass bias from other methods; e.g.,calculating total mass via caustic methods \citep[e.g.,][]{maughan2016}, or relating an observed gas fraction to an assume gas fraction \citep[e.g.,][]{allen2008,eckert2019,wicker2023}.

    In order to infer a hydrostatic mass bias from thermodynamic fluctuations, we seek to relate the fluctuations to the non-thermal pressure support. Given that we expect the non-thermal pressure support to be predominantly from turbulent motions, we can relate the Mach numbers, which we infer from the spectra, to the non-thermal pressure support as
        \begin{equation}
        \frac{P_{\text{NT}}}{P_{\text{th}}} = (\gamma / 3) \mathcal{M}_{\text{3D}}^2,
        \label{eqn:NTpp_Mach}
    \end{equation}
    where $P_{\text{th}}$ is the thermal pressure, $\gamma$ is the adiabatic index, and $\mathcal{M}_{\text{3D}}$ is the Mach number of the turbulent velocity in 3D \citep[e.g.,][]{lau2009,khatri2016}. When calculating the canonical hydrostatic mass one needs the slope of the thermal pressure (with respect to radius); similarly, one needs the slope of the non-thermal pressure to infer a hydrostatic mass bias.
    %As with the thermal pressure support and hydrostatic mass, we need the slope of the non-thermal pressure to infer the hydrostatic mass bias. 
    In practice we calculate the logarithmic slope of the turbulent Mach number (see Section~\ref{sec:PS_results}) in the calculation of a hydrostatic mass bias. As such, we must determine Mach numbers which can be associated with at least two distinct radii.

    In this study, we build on our previous theoretical and observational investigations (e.g., \citealt{Gaspari2013_PS,Gaspari2014_PS,khatri2016,romero2023}) and seek to leverage key scaling relations between the relative thermodynamic fluctuations and the turbulent kinematics in the ICM via the Fourier power spectra (e.g., Mach numbers, injection scales, and spectral slopes) when applied to a cluster sample. In particular, we seek to apply these scaling relations to both SZ and X-ray data where both datasets have overlap in angular (physical) scales accessed. 

    We discuss our full and pilot sample selection in Section~\ref{sec:sample_selection}. Together with \citet{Gaspari2014_PS,khatri2016,romero2023}, this work serves as our last pilot SZ/X-ray spectral study to complete the foundational methodology to be applied to the full sample. As investigations of surface brightness fluctuations for the ICM require the relative fluctuations, i.e. $\delta I/\bar{I}$ or the residual surface brightness divided by some model, we discuss how we obtain $\bar{I}$ in Sections~\ref{sec:Xray_analysis} and ~\ref{sec:SPT_img_analysis} for \textit{XMM-Newton} and SPT, respectively. Section~\ref{sec:PS_analysis} introduces the $\delta I/\bar{I}$ images and discusses the methodology to obtain 2D amplitude spectra and subsequent deprojection to obtain 3D amplitude spectra. Results and inferences are presented in Section~\ref{sec:PS_results} and we reflect on the impact of methodological choices on our results in Section~\ref{sec:Choices}. We provide concluding remarks in Section~\ref{sec:conclusions}.

    We adopt a concordance cosmology: $H_0 = 70$~km~s$^{-1}$~Mpc$^{-1}$, $\Omega_M = 0.3$, $\Omega_{\Lambda} = 0.7$. Unless otherwise stated all uncertainties are reported as the standard deviation (for distributions taken to be symmetric) or the distance from the median to the 16th and 84th percentiles (when allowing for asymmetric distributions).

\section{Sample Selection}
\label{sec:sample_selection}

    \begin{figure}[!h]
        \begin{center}
            \includegraphics[width=0.45\textwidth]{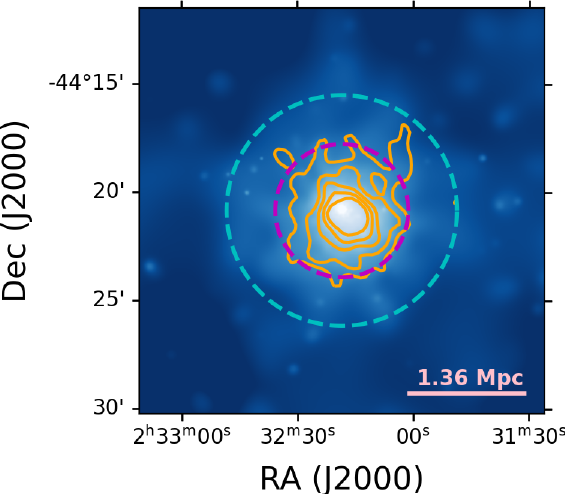}
            \includegraphics[width=0.45\textwidth]{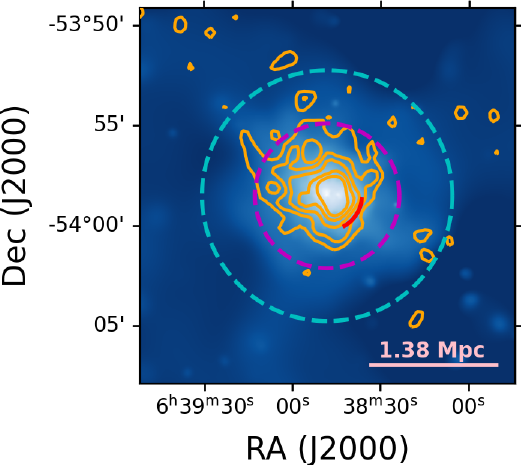}
        \end{center}
        \caption{X-ray (\textit{XMM}) images of SPT-CLJ0232-4421 (left) and SPT-CLJ0638-5358 (right) with SZ (SPT) contours in orange. The X-ray images are adaptively smoothed; the SZ contours begin at $3\sigma$ with $2\sigma$ intervals. The physical scale of $R_{500}$ is captioned above a (pink) bar showing the angular scale of $R_{500}$. The dashed cyan circles enclose $R_{500}$ and the dashed magenta circles separate the inner and outer regions (rings; see Section~\ref{sec:PS_analysis}). The red curve in the image of SPT-CLJ0638-5358 indicates the shock location as found in \citet{botteon2018}.}
        \label{fig:Xray-w-SZoverlay}
    \end{figure}
    
    We seek to analyze a sample of clusters across a wide redshift range and which have been observed both via the SZ effect and in the X-rays. The SPT-SZ Survey \citep{bleem2015} identified 516 galaxy clusters in 2500 square degrees which span redshifts $0 < z < 1.8$ and with masses of $M_{500} \geq 3 \times 10^{14}$ M$_{\odot}$. We opt to select SPT clusters which have been observed with \textit{XMM-Newton} as photon counts are likely to be a limiting factor. Such a selection has already been studied with respect to thermodynamic profiles as determined from X-ray data in \citet{bulbul2019} and thus provides an excellent sample which complements our study of fluctuations. Our sample selection is thus that used in \citet{bulbul2019}, though additional archival \textit{XMM-Newton} will augment the sample size from the 59 used in \citet{bulbul2019}.

    \begin{table*}
    \centering
        \begin{tabular}{c c | c c c c c c}
        \hline
            Cluster & Alternative names & $z^a$ & M$_{500}^{a}$ & $R_{500}^{a}$ & $\theta_{500}$ & $k T_{ce}^{a}$ & $k T_{ce}^{b}$ \\
                    &                & & ($10^{14} M_{\odot}$) & (Mpc) & (arcmin) & (keV) & (keV) \\
                 \hline
            \multirow{2}{*}{SPT-CLJ0232-4421} & RXCJ0232.2-4420 &  0.28 & 9.45 & 1.36 & 5.32 & $7.19_{-0.50}^{+0.46}$ & $10.06 \pm 2.31$ \\
                                          & SPT-CLJ0232-4420 & & & & \\
            SPT-CLJ0638-5358 & Abell S0592 & 0.23 & 9.42 & 1.38 & 6.26 & $8.44_{-0.48}^{+0.86}$ & $9.45 \pm 0.95$  

        \end{tabular}
        \caption{Fundamental properties of the clusters in our pilot sample. $^{a}$values are from \citet{bulbul2019}, which uses \textit{XMM-Newton} data and $^{b}$ values are from \citet{mantz2010b}, which uses \textit{Chandra} data.}
        \label{tab:cluster_properties}
    \end{table*}
    
     Of this sample, we choose two clusters which are well resolved by SPT to pilot our analysis of surface brightness fluctuations. We visually inspected a subsample of clusters for which $R_{500}$ is several times the full-width at half maximum (FWHM) of the SPT beam (FWHM = $1.^{\prime}25$). We wish to have a dynamically relaxed and a disturbed cluster, which we crudely estimate from the SPT Compton $y$ images. Based on this visual inspection, we choose SPT-CLJ0232-4421 as a dynamically relaxed cluster and SPT-CLJ0638-5358 as our dynamically disturbed cluster. Figure~\ref{fig:Xray-w-SZoverlay} presents adaptively smoothed X-ray images with SPT contours overlaid. We list $M_{500}$ and corresponding $R_{500}$ as reported in the SPT-SZ catalog \citep{bleem2015} in Table~\ref{tab:cluster_properties}; the angular scale of $R_{500}$ is computed with our assumed cosmology.
     
    %\citet{bulbul2019} presents a table of many derived X-ray quantities, but we do not find many comments on individual clusters, specifically our pilot clusters. 
    In SPT-CLJ0232-4421 \citet{parekh2021} find substructure about 1 arcminute southwest of the cluster core using \textit{XMM-Newton} and \textit{Chandra} data and posit that while the cool-core remains relatively undisturbed (and which \citet{hudson2010} classify as a weak cool-core) this substructure could be related to the detected radio halo \citep{kale2019}. More recently MeerKAT has revealed candidate relics \citep{kale2022}: an eastern relic which lies within $R_{500}$ and a southern relic which lies at 1.9 Mpc ($\sim R_{200}$) from the cluster center. However, \citet{kale2022} suspect that their origin is not from a cluster merger.

    SPT-CLJ0638-5358, also listed as RXC J0638.7-5358 and Abell-S 592, had been noted as having a high ICM temperature, a mass $M_{500} \approx 10^{15}$ M$_{\odot}$, and a non-circular shape in data from the Atacama Cosmology Telescope \citep[ACT; e.g.,][]{hincks2010,mantz2010b,menanteau2010}. More recently \citet{botteon2018} have found a shock to the south-west of the cluster core (indicated with a red curve in Figure~\ref{fig:Xray-w-SZoverlay}) using \textit{Chandra} data. They derive a Mach number of $\mathcal{M} = 1.72_{-0.12}^{+0.15}$ from surface brightness (i.e. density jump) and they derive a consistent Mach value ($\mathcal{M} = 1.6_{-0.42}^{+0.54}$) from the temperature jump. Despite searching for a counterpart shock to the north east, they do not find such a counterpart. \citet{botteon2018} also identify two cool cores (visible in their temperature map of SPT-CLJ0638-5358) and note the low entropies in those cool-cores despite the clear dynamical disturbance present in the cluster.

\section{\textit{XMM-Newton} data analysis}
    \label{sec:Xray_analysis}

There are two \textit{XMM-Newton} observations (Obs.IDs) of SPT-CLJ0232-4421: 0042340301 and 0827350201. There is a single \textit{XMM-Newton} observation of SPT-CLJ0638-5358 with Obs.ID of 0650860101. Our data processing and analysis is nearly identical to that used in \citet{romero2023} where we use heasoft v6.28 and SAS 19.0 and the Extended Source Analysis Software (ESAS) data reduction package \citep{snowden2008} to produce event files and eventually images for the three EPIC detectors: MOS1, MOS2, and pn.

\begin{table}[h]
    \centering
    \begin{tabular}{c|c|cc}
         \multirow{2}{*}{Cluster:} & SPT-CLJ & \multicolumn{2}{c}{SPT-CLJ} \\
                                   & 0638-5358 & \multicolumn{2}{c}{0232-4421} \\
         Obs ID & 0650860101 & 0042340301 & 0827350201 \\
         \hline
         %0108670401 & 2000-12-05 & 10.8 ks & TBD & TBD \\
         Date & 2010 May 22 & 2002 July 11 & 2021 January 1 \\
         Exposure  & 46.8 (ks) & 13.4 (ks) & 27.8 (ks) \\
         \hline
         \multirow{2}{*}{Clean Exp} & MOS1: 24.6 & MOS1: 11.6 & MOS1: 24.6 \\
        \multirow{2}{*}{(ks)}& MOS2: 31.7 & MOS2: 12.1 & MOS2: 25.7 \\
          & pn: 7.6 & pn: 6.8 & pn: 18.0 \\
          \hline
         pn mode &  FF & EFF & FF \\
         PI & J. Hughes & \multirow{2}{*}{H. B\"{o}hringer} & M. Arnaud \\
            &           &                  & \& S. Ettori 
    \end{tabular}
    \caption{Overview of imaging \textit{XMM-Newton} observations of SPT-CLJ0232-4421 and SPT-CLJ0638-5358. FF refers to the "full frame" mode.}

    \label{tbl:xmm_obs}
\end{table}

\subsection{Image creation}
\label{sec:img}

We choose to extract images in the [0.4-1.25] keV and [2.0-5.0] keV bands thus avoiding the energies where fluorescent lines are present in EPIC cameras. Images and vignetted exposures are extracted for each detector over the entire detector area whilst masking point sources (see Section~\ref{sec:ptsrc_exc} for point source identification) via the task \lstinline{mos-spectra} or \lstinline{pn-spectra}. 
%Though we do not find a need for unvignetted exposures in the profile fitting (Section~\ref{sec:xray_profile}), we create these images with the task \lstinline{eexpmap withvignetting=no}.

%\subsection{Constrained background components}
%\label{sec:cbc}
%
%The relevant particle backgrounds are calculated for the desired energy band via the tasks \lstinline{mos_back} and \lstinline{pn_back}.  For the pn detector, we extract a separate spectrum (via \lstinline{pn-spectra}) over the cluster region, which we take to be a radius of 5 arcminutes about the cluster center. While we treat the residual soft proton spectrum as a single power law, we must fit several other components to the spectrum: a thermal plasma component (\lstinline{apec}) for each of the local (Solar) hot bubble, Galactic emission, and the ICM in Zwicky 3146. In addition to this, we also consider Gaussian components for fluorescent lines. A soft proton background is then made with the task \lstinline{proton} and added to the particle background with the task \lstinline{farith}. For the pn detector, we also consider the out-of-time (OOT) contribution. Depending on the full frame mode, we multiply our resultant pn image with randomized columns by 0.063 or 0.023 for full frame and extended frame modes, respectively, to have an OOT component which we incorporate into the pn background. These background images will be subtracted from the respective images when extracting profiles.

\subsection{Point Source exclusion}
\label{sec:ptsrc_exc}

We generate a list of point sources using \lstinline{cheese} and initially mask the sources with circles of radius 25\arcsec. If the signal-to-noise ratio (SNR) of the point source is above 50 we modify the masking radius to 35\arcsec. Finally, we perform a manual inspection to identify any remaining point sources or any necessary modifications to the masking circles. In the case of SPT-CLJ0232-4421 a very bright source at (RA, Dec) of (38.158650, -44.363767), in degrees, requires masking out to a radius of 80\arcsec.

\subsection{Profile fitting}
\label{sec:xray_profile}

We use the Python package \lstinline{pyproffit} \citep{eckert2017} to determine centers, ellipticity, and extract profiles for each of our images. As we have low energy([0.4-1.25] keV) and high energy ([2.0-5.0] keV) images; per camera and per ObsID, we have 12 images in total for SPT-CLJ0232-4421 and 6 images in total for SPT-CLJ0638-5358. We adopt the average centroid (across all respective images) as our X-ray center, where \lstinline{pyproffit} accounts for masked regions: point sources and chip gaps, where we consider the gaps to include low-exposure pixels \citep{romero2023}.

    \begin{figure}[!h]
        \begin{center}
        \includegraphics[width=0.46\textwidth]{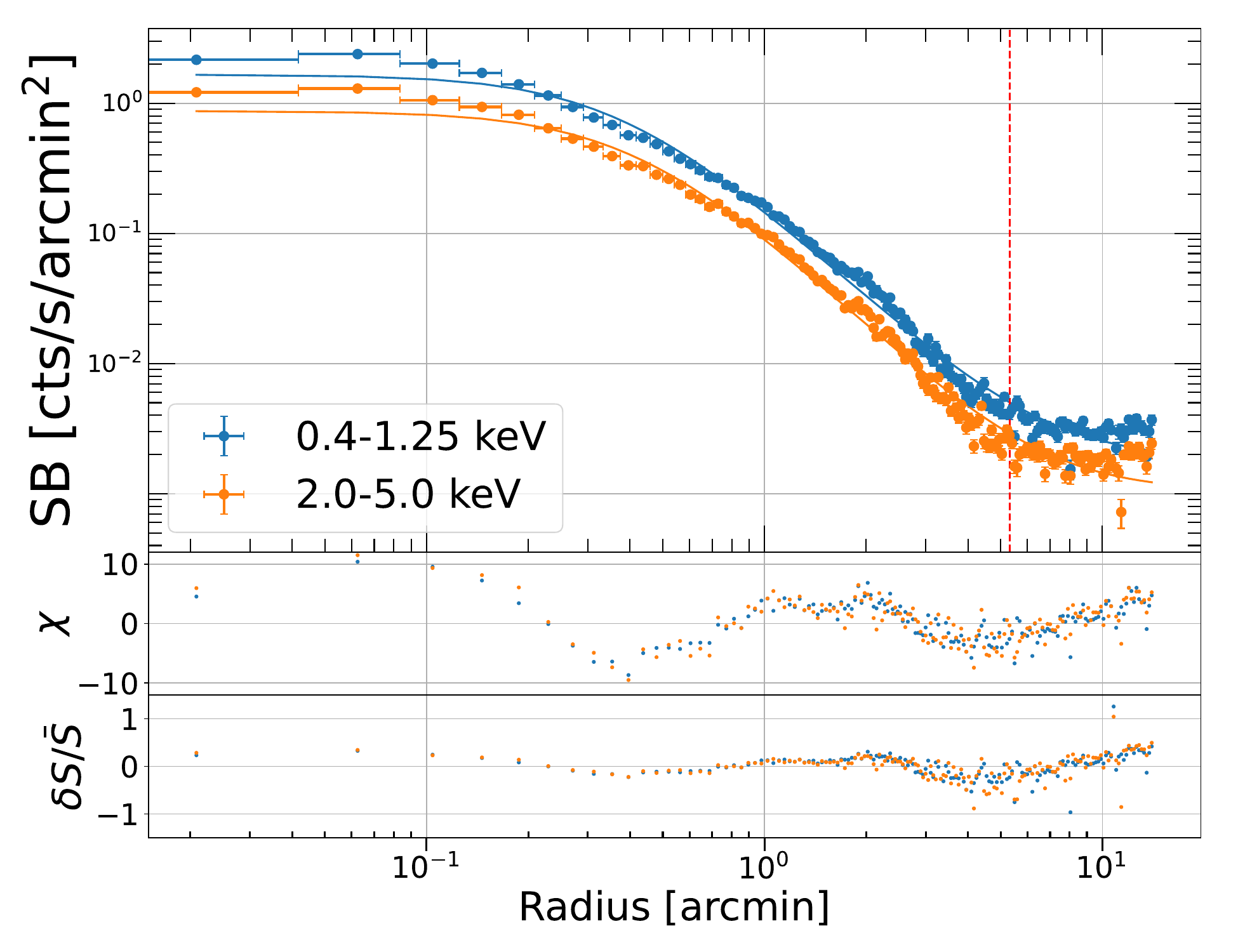}
        \includegraphics[width=0.46\textwidth]{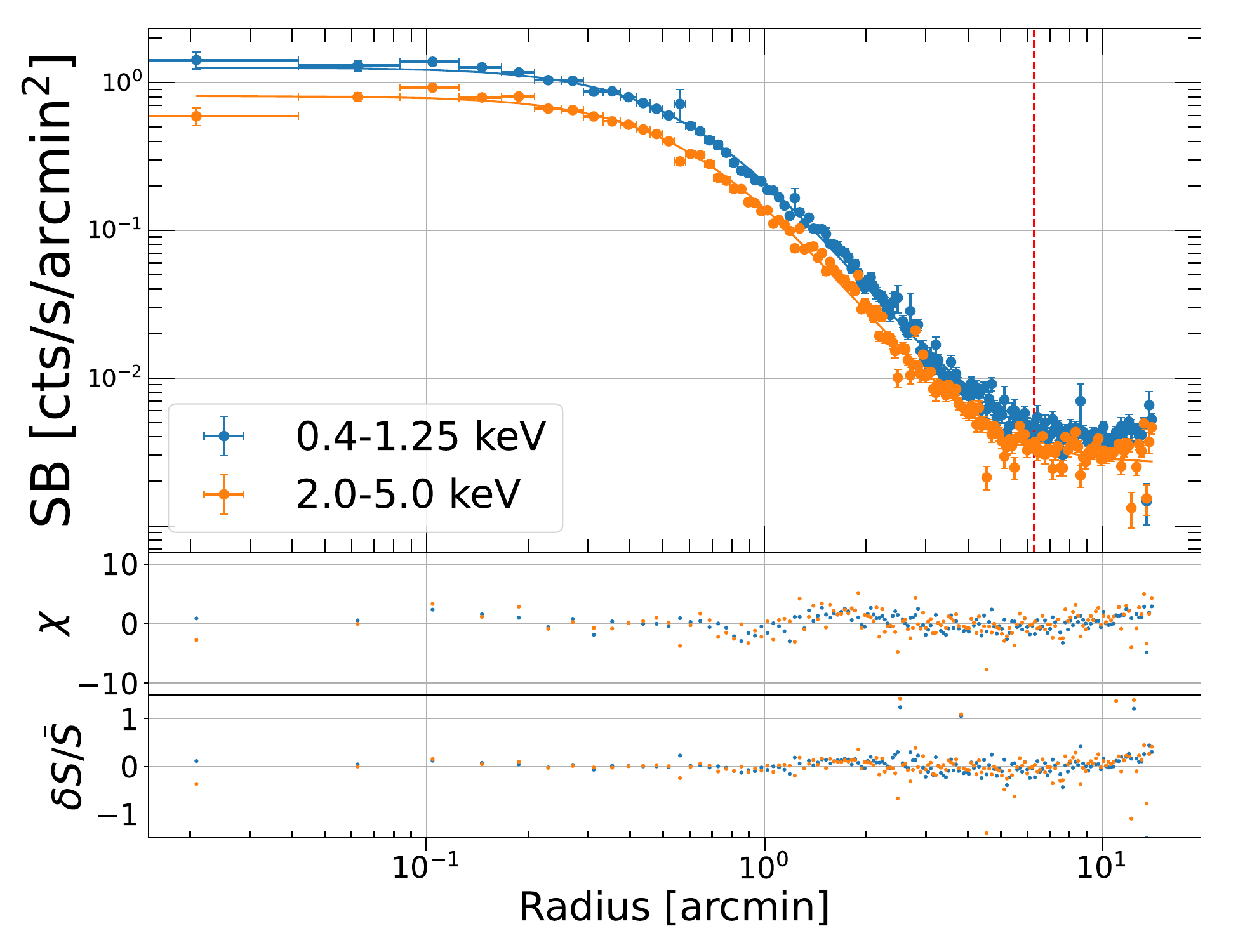}
        \end{center}
        \caption{Surface brightness profiles for SPT-CLJ0232-4421 (left) and SPT-CLJ0638-5358 (right) and the respective single beta model fits and residuals. For brevity, the profiles presented here are the combination of profiles from each CCD and each ObsID. In practice the fits are done to each individual CCD and ObsID. Residuals are shown both in terms of $\chi$ and perhaps more relevant to the fluctuations, as $\delta S/\bar{S}$. The vertical red line marks $R_{500}$.}
        \label{fig:xmm_sb_profiles}
    \end{figure}

The best fit profile parameters and uncertainties are determined via \lstinline{emcee} \citep{foreman2013} separately for each detector, each energy band, and each observation.  We adopt a circular $\beta$-model with the form:
\begin{equation}
    S(r) = S_0 [(1+(r/r_{c})^2]^{-3 \beta + 0.5} + B,
    \label{eqn:XR_beta}
\end{equation}
where $r$ is the radius, $r_{c}$ is the ``core'' (scaling) radius, $S_0$ is the surface brightness normalization, and $B$ is the background. We find that a uniform background component is sufficient and that to constrain the background we should fit (from $r=0$) out to at least 10 arcminutes. The $\beta$-model leaves evident residuals in Figure~\ref{fig:xmm_sb_profiles}, where we have combined the profiles of each CCD across ObsIDs for brevity; the parameter results of these abbreviated profiles are also listed in Table~\ref{tab:beta_fits}. To reiterate, we fit the $\beta$-models to each CCD and each ObsID separately. In all cases we find that aside from the innermost radii the $\beta$-model does quite well. More specifically, when considering the context of developing a method to be applied to a sample of clusters with varying depth, we wish to adopt a model which should do well across the sample. Finally, the actual impact of the residuals in the center is likely small given the area it subtends relative to the regions used for power spectra extraction (see Sections~\ref{sec:PS_analysis} and \ref{sec:Choices}). We also explore the impact of ellipticity in Section~\ref{sec:ProfileChoices}.

\section{SPT Image Analysis}
    \label{sec:SPT_img_analysis}

The SPT-SZ \citep[][]{bleem2015} products are available in spherical coordinates (HEALpix format) or a flat-sky (Sanson-Flamsteed projection)\footnote{\url{https://lambda.gsfc.nasa.gov/product/spt/spt_sz_comp_maps_get.html}}. As the angular size of $R_{500}$ for our clusters is less than 10 arcminutes, we expect that the flat sky projection is adequate and allows for the application of the same analysis tools on X-ray and SZ data. There are also several flavors of Compton-y maps available: minimum variance, CMB-Nulled, and CMB-CIB-Nulled \citep{bleem2022}. When confined to the vicinity of a cluster, the CMB and CIB are subdominant on the scales probed and so we opt for the minimum variance maps which provide the best resolution and signal significance.

    \subsection{Characterizing SPT Noise}
    \label{sec:SPTnoise}

The SPT Compton $y$ maps are provided in fields (over 10 degrees on a side), 19 contiguous sub-patches \citep[e.g.][]{bleem2015}. The noise in each field is expected to be homogeneous. To characterize the noise (on scales of $20^{\prime}$ and smaller), we identify 100 regions within a field that are devoid of point sources or other evident signal. This is achieved by avoiding regions which are masked\footnote{Masks are provided at \url{https://lambda.gsfc.nasa.gov/data/suborbital/SPT/bleem_2021/masks.zip}.}

\begin{figure}[!h]
    \begin{center}
        \includegraphics[width=0.5\textwidth]{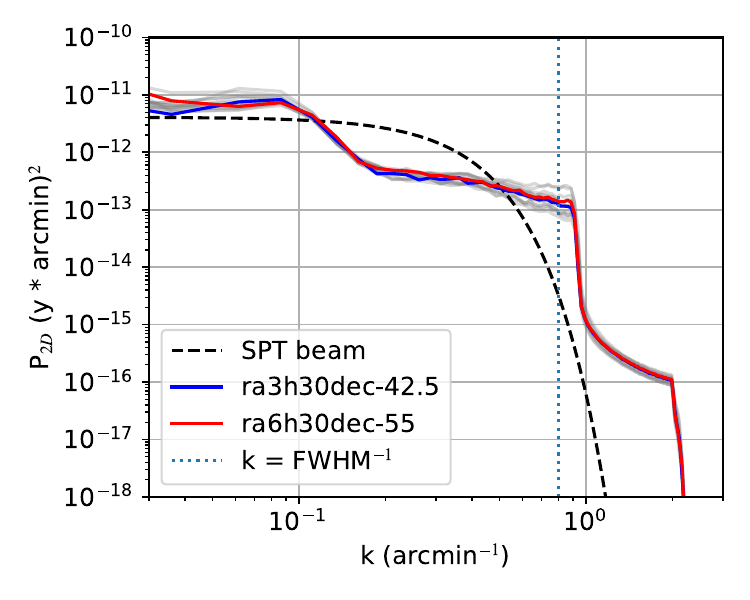}
    \end{center}
    \caption{The power spectra of the SPT-SZ fields for the minimum variance Compton-y maps. Comments on the departure(s) from a flat spectrum are given in the text. The fields \lstinline{ra3h30dec-42.5} and \lstinline{ra6h30dec-55} contain SPT-CLJ0232-4421 and SPT-CLJ0638-5358, respectively. The dotted blue line is at 0.8 arcmin$^{-1}$ corresponding to the reported FWHM of $1.^{\prime}25.$ The black dashed curve shows an arbitrarily normalized power spectrum for a Gaussian with FWHM of $1.^{\prime}25.$}
    \label{fig:Covariance_both}
\end{figure}

The noise in the maps is assumed to be stationary; as such, the power spectrum of the 100 blank regions identified above serve to characterize the noise. To further ensure that we avoid signal in our noise characterization, we calculate the power spectrum on the blank regions of difference map (half 1 - half 2). In particular, we average over the 100 regions to calculate the average power spectrum shown in Figure~\ref{fig:Covariance_both}.
We identify three features in the noise power spectrum that warrant comment. (1) At $k \approx 10^{-1}$ arcmin$^{-1}$ the steepening of the slope is due to the transition of weights between SPT and \textit{Planck}, (with \textit{Planck}'s weight increasing towards lower values of $k$; \citealt{bleem2022}), where the weight is especially dominated by the 217 GHz band in \textit{Planck} data. (2) Just below $k = 1$ arcmin$^{-1}$
the power drops quickly (towards higher $k$), and (3) a similar drop occurs again around $k \approx 2$ arcmin$^{-1}$. These are features of the projection from HEALpix to flat-sky (Sanson-Flamsteed projection) that are inconsequential in our analyses.

    \subsection{Beta profile fitting}
    \label{sec:betaFittingSPT}

\begin{figure}[!h]
    \begin{center}
        \includegraphics[width=0.45\textwidth]{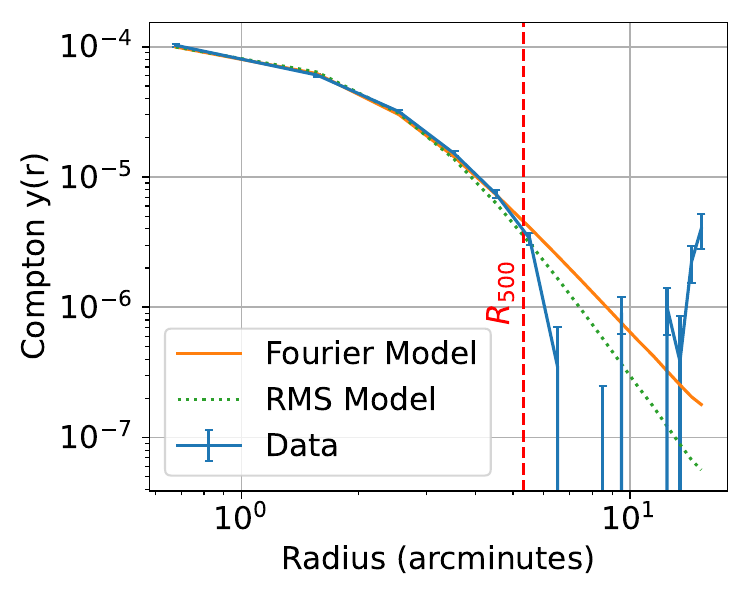}
        \includegraphics[width=0.45\textwidth]{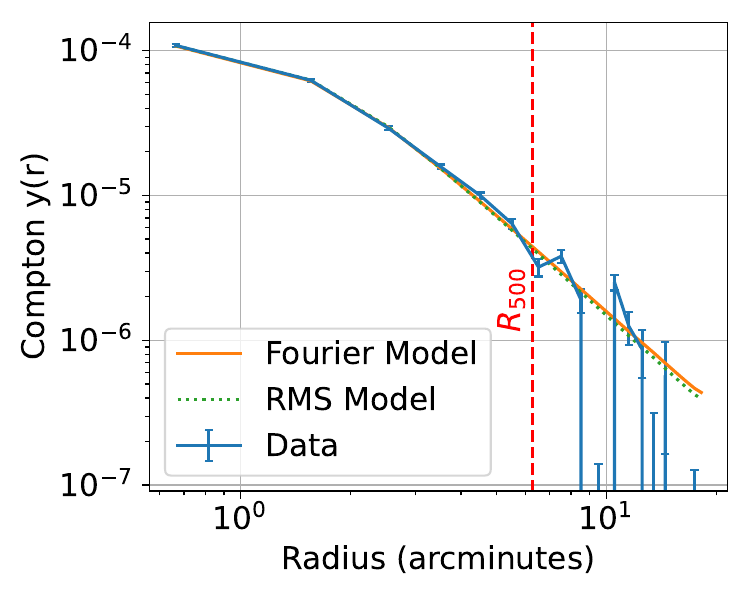}
    \end{center}
    \caption{Profiles (data and models) for SPT-CLJ0232-4421 (left) and SPT-CLJ0638-5358 (right). The model profile includes the effect of beam-convolution. In the case of SPT-CLJ0232-4421, the positive Compton $y$ values well beyond $R_{500}$ are due to primary CMB fluctuations as this feature is not present in the CMB-nulled map.}
    \label{fig:SZ_profiles}
\end{figure}

To fit the cluster profile we extract cutouts out to $2 R_{500}$, i.e. the sides are $4 R_{500}$ (rounding up for selecting pixels). Although no point sources are found within $R_{500}$ for the two clusters investigated, we do see a point source beyond $R_{500}$ in the cutout of SPT-CLJ0232-4421. We thus pass the cutout images through the \lstinline{Daophot} utility within astropy. A point source map is then constructed from the locations and peaks as found with the \lstinline{Daophot} utility. We assume a circular Gaussian with FWHM of $1.^{\prime}25$ for each point source and subtract any sources found from the maps.

\begin{table*}[]
    \centering
    \begin{tabular}{c |c c c | c c c}
          & \multicolumn{3}{c}{SPT-CLJ 0232-4432} & \multicolumn{3}{c}{SPT-CLJ 0638-5531} \\
           & RA (deg) & Dec (deg) & $\Delta (^{\prime\prime})$ & RA (deg) & Dec (deg) & $\Delta (^{\prime\prime})$ \\
         \hline
         XMM$_c$ & 38.072751 & -44.348505 & --   & 99.699212 & -53.976286 & --  \\
         XMM$_p$ & 38.078971 & -44.346976 & 16.9 & 99.702212 & -53.974323 & 9.5 \\
         SPT     & 38.070109 & -44.354129 & 21.4 & 99.697766 & -53.974866 & 6.0 \\
         SZ      & 38.069225 & -44.349572 & 9.4  & 99.697749 & -53.976731 & 3.5 \\

         %%%%%%%%%%%%%%%%%%%%%%%%%%%%%%%%%%%%%%%%%%%%%%%%%%%%%%%%%%%%%
         %SPT & 38.070109 & -44.354129 & 19.6 & 99.697766 & -53.974866 & 6.5 \\
         %XMM & 38.076498 & -44.351169 & --   & 99.700679 & -53.975414 & --  \\
         %SZ  & 38.073131 & -44.352256 & 9.5  & 99.699209 & -53.975808 & 3.4 \\
    \end{tabular}
    \caption{Centers for the two clusters taken to be the centroid as found with \lstinline{pyproffit} (XMM$_c$), taken to be the peak as found with \lstinline{pyproffit} (XMM$_p$), as indicated in the SPT-SZ catalogue (SPT),  and as found when allowing the center to be a parameter in the beta model fit (SZ).}
    \label{tab:centroids}
\end{table*}

To fit a surface brightness profile to our point source subtracted maps, we employ \lstinline{emcee}\citep{foreman2013} and forward-model a circular $\beta$-model of the cluster. The forward modelling includes the convolution with the beam (circular Gaussian with FWHM of $1.^{\prime}25$) and transfer function (provided as a function of $\ell$\footnote{\url{https://lambda.gsfc.nasa.gov/data/suborbital/SPT/bleem_2021/sptsz_trough_filter_1d.txt}}). The impact of the point source (if not subtracted) on the profile fit is negligible and so we are not concerned with a more detailed treatment of point sources (e.g. such as simultaneous fitting with the cluster).

    \begin{figure}[!h]
        \begin{center}
            \includegraphics[width=0.45\textwidth]{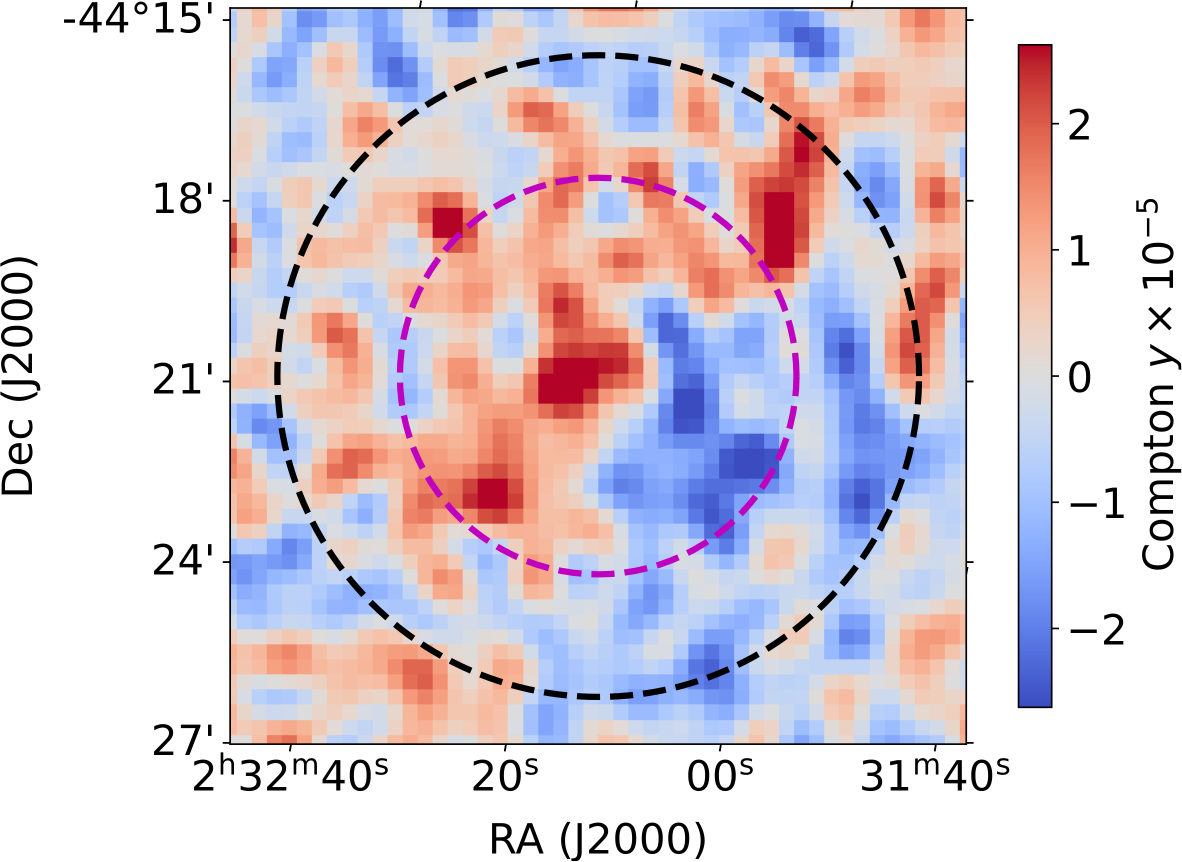}
            \includegraphics[width=0.45\textwidth]{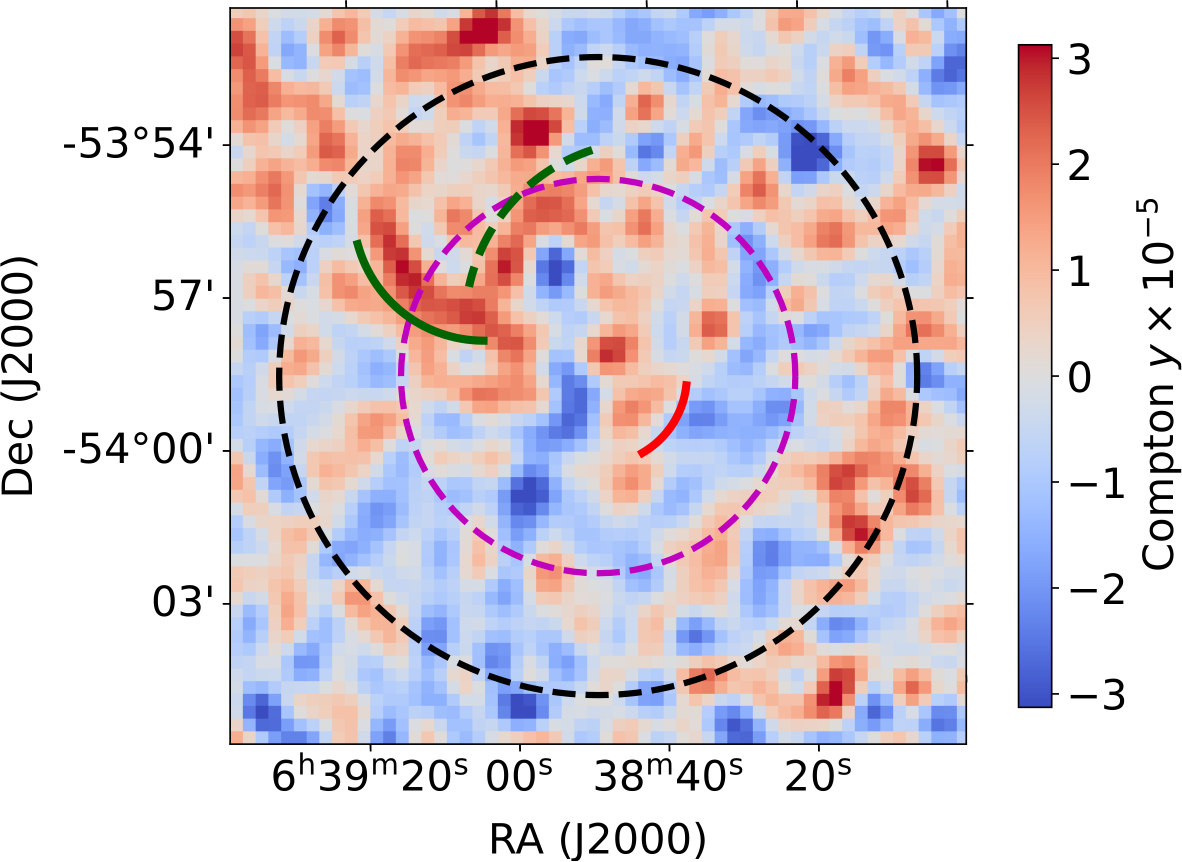}
        \end{center}
        \caption{\textbf{left:} The residual (Compton $y$) image of SPT-CLJ0232-4421. \textbf{right:} The residual (Compton $y$) image of SPT-CLJ0638-5358 with the known shock front location \citep{botteon2018} in red and plausible shock fronts from the SPT data in dark green. In both clusters, $R_{500}$ is shown as the black dashed circle. The division between Ring 1 and Ring 2 to be used in Section~\ref{sec:PS_analysis} is marked by the dashed magenta circle. The RMS within the fields for both clusters is roughly $1.8e{-5}$ in Compton $y$.}
    \label{fig:residualImages}
    \end{figure}

We perform the fitting in image space. We take our principal results to be those which assume the centroid from \textit{XMM-Newton} data and a Fourier (power spectrum) characterization of the noise.\footnote{Again, this choice comes from the expectation that this will provide good consistency across our full sample.} We check both assumptions with fits where both the centroid is allowed to vary (see Table~\ref{tab:centroids}) and fits where we assume the noise is uniformly white, which we denote by ``RMS" (see Figure~\ref{fig:SZ_profiles}). Fitted parameters of both (abbreviated) X-ray and SZ $\beta$-models are presented in Table~\ref{tab:beta_fits}. The profiles do not change much with the change of centroid nor with noise characterization, though we do note a tendency for the uniform white noise to prefer a steeper slope near $R_{500}$.

\begin{table*}[]
    \centering
    \begin{tabular}{c c c |c c c }
          Cluster & Dataset &  Subset & $I_0$ & $\theta_c$ & $\beta$ \\
         \hline
         \multirow{4}{*}{SPT-CLJ 0232-4432} & \multirow{2}{*}{SPT} & RMS     & ${1.34}_{-0.04}^{+{0.04}} \times 10^{-4} $ & ${2.49}_{-0.17}^{+0.19}$ & ${1.11}_{-0.10}^{+0.13}$ \\
         & & Fourier & ${1.39}_{-0.02}^{+0.02} \times 10^{-4} $ & ${1.97}_{-0.09}^{+0.09}$ & ${0.76}_{-0.05}^{+0.05}$ \\
         & \multirow{2}{*}{\textit{XMM}} & 0.4 - 1.25 keV & ${1.66}_{-0.03}^{+0.03}$ & ${0.38}_{-0.01}^{+0.01}$ & ${0.560}_{-0.003}^{+0.003}$ \\
         & & 2 - 5 keV & ${0.87}_{-0.01}^{+0.01}$ & ${0.42}_{-0.01}^{+0.01}$ & ${0.572}_{-0.003}^{+0.003}$ \\
         \hline
         \multirow{4}{*}{SPT-CLJ 0638-5358} & \multirow{2}{*}{SPT} & RMS     & ${1.69}_{-0.15}^{+0.16} \times 10^{-4} $ & ${1.29}_{-0.18}^{+0.21}$ & ${0.44}_{-0.07}^{+0.08}$  \\
         & & Fourier & ${1.73}_{-0.04}^{+0.04} \times 10^{-4} $ & ${1.21}_{-0.05}^{+0.05}$ & ${0.41}_{-0.02}^{+0.02}$ \\

         & \multirow{2}{*}{\textit{XMM}} & 0.4 - 1.25 keV & ${1.21}_{-0.02}^{+0.03}$ & ${0.68}_{-0.02}^{+0.02}$ & ${0.67}_{-0.01}^{+0.01}$ \\
         & & 2 - 5 keV & ${0.76}_{-0.01}^{+0.01}$ & ${0.73}_{-0.02}^{+0.02}$ & ${0.69}_{-0.01}^{+0.01}$  \\
         \hline

    \end{tabular}
    \caption{$\beta$-model fit parameters for SZ (SPT) and X-ray (\textit{XMM}) datasets, where $I_0$ is in units of Compton $y$ for the SZ fits and is in units of $\rm photon\ counts\ s^{-1}\ arcmin^{-2}$ for the X-ray fits. In the SZ case, the two subsets (RMS and Fourier) refer to the same data but two different characterizations of noise when fitting.}
    \label{tab:beta_fits}
\end{table*}

Residual images of both clusters are shown in Figure~\ref{fig:residualImages} and we see that both clusters exhibit substructure. Additionally, Figure~\ref{fig:residualImages} shows regions to be used in the spectral analysis (Section~\ref{sec:PS_analysis} and for SPT-CLJ0638-5358 we have also marked the location of a known shock \citep{botteon2018} in red and possible shocks as seen in the SPT residuals in dark green.

\section{Power Spectrum Analysis}
\label{sec:PS_analysis}

    \begin{figure*}[]
        \begin{center}
            \includegraphics[width=0.45\textwidth]{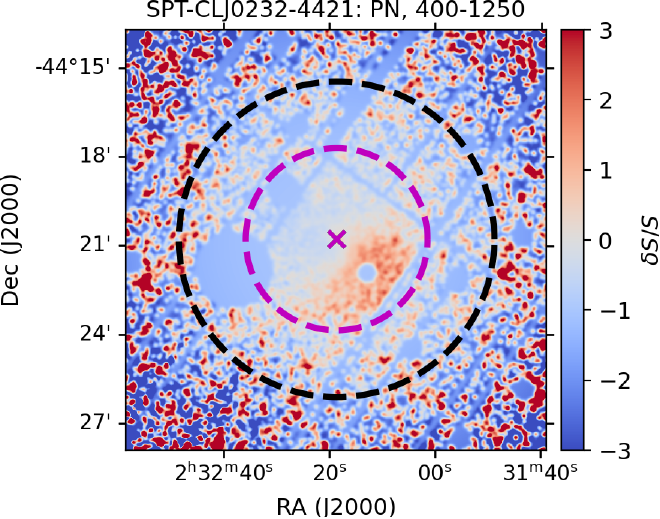}
            \includegraphics[width=0.45\textwidth]{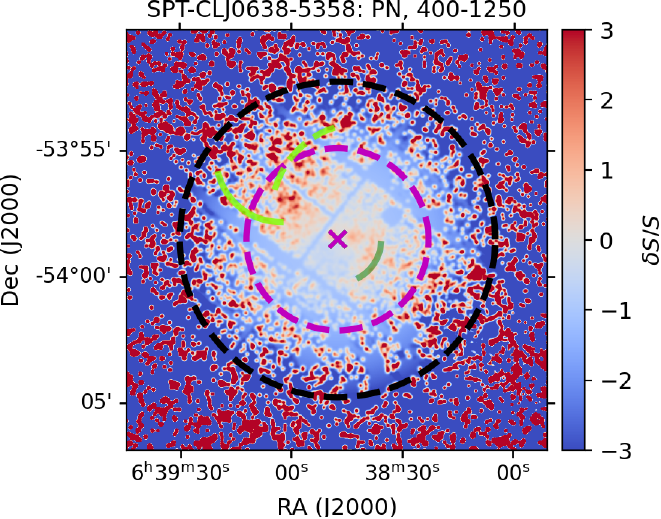}
            \includegraphics[width=0.45\textwidth]{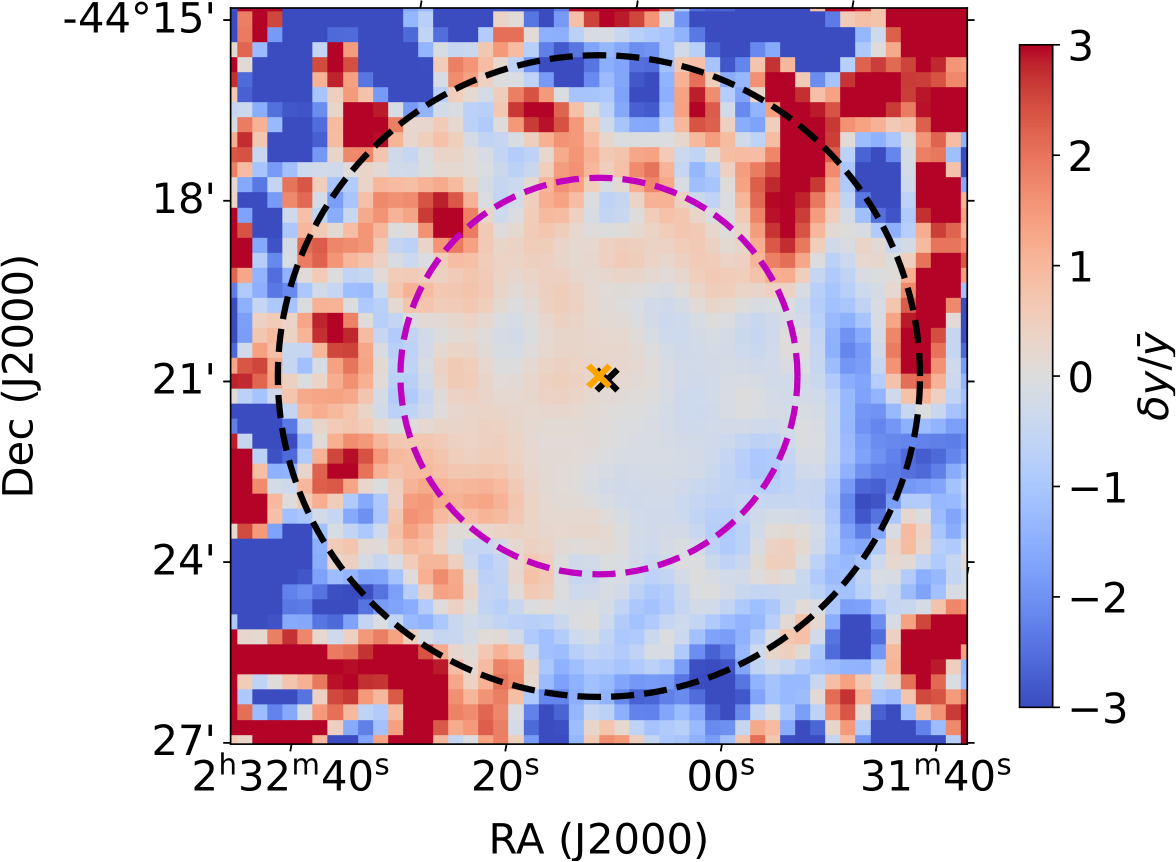}
            \includegraphics[width=0.45\textwidth]{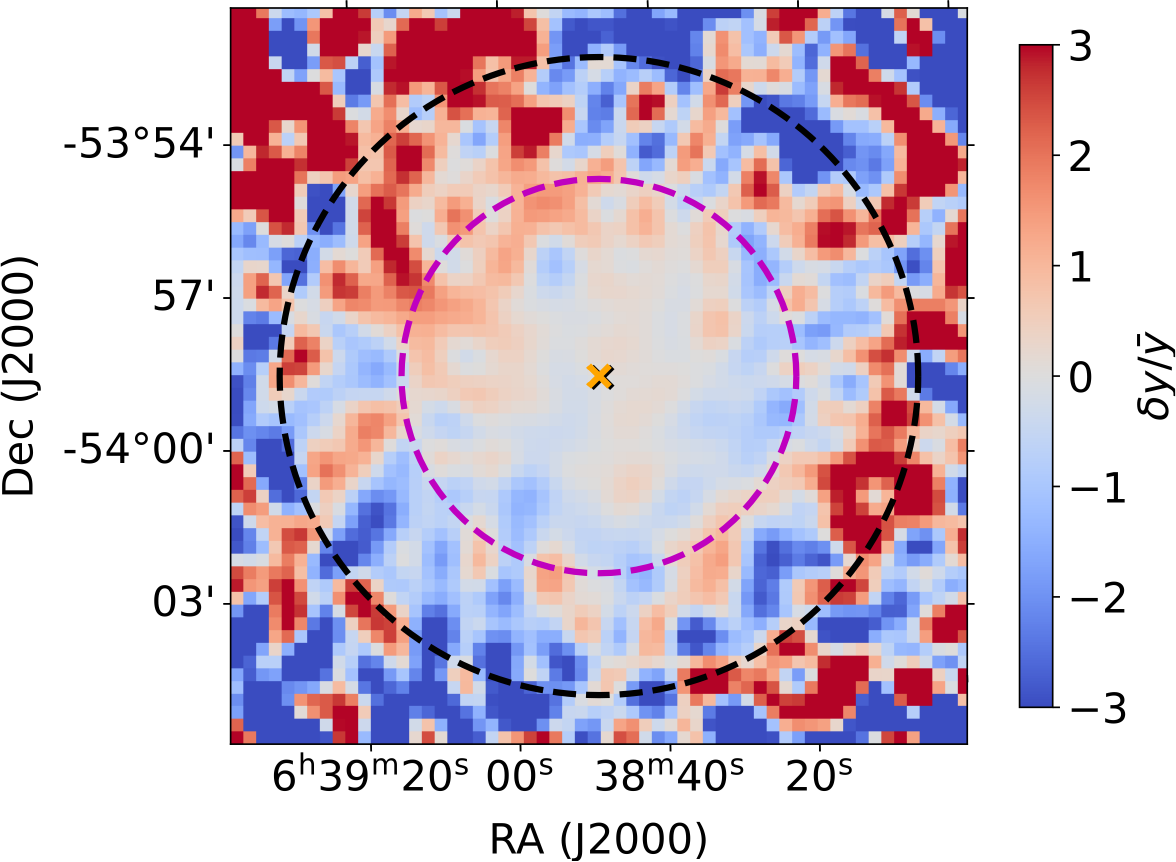}
        \end{center}
        \caption{\textbf{left:} fractional residuals for SPT-CLJ0232-4421; \textbf{right:} fractional residuals for SPT-CLJ0638-5358. The top panels are from \textit{XMM-Newton} data ($\delta S/S$), specifically images from the pn CCD for our low energy band (0.4-1.25 keV for a single ObsID). The bottom panels are from SPT data ($\delta y / y$). The known and candidate shock fronts in SPT-CLJ0638-5358 presented in Figure~\ref{fig:residualImages} are marked in green on the X-ray fractional residual.}
    \label{fig:FractionalResiduals}
    \end{figure*}
    
    As indicated in Section~\ref{sec:intro}, our first step to calculating spectra of surface brightness fluctuations is to obtain fractional residual images. Sections ~\ref{sec:Xray_analysis} and \ref{sec:SPT_img_analysis} have discussed how we obtain surface brightness images, fit and subtract a model to derive residual images. For each residual image, we need only divide by the respective fitted ICM surface brightness model (i.e. excluding a fitted background component if present); these fractional residuals are presented in Figure~\ref{fig:FractionalResiduals}

    To calculate power spectra we employ the $\Delta$-variance method presented in \citet{arevalo2012}, which is designed to provide low-resolution spectra for arbitrary data masking. As noted in \citet{arevalo2012}, \citet{Zhuravleva2015}, and \citet{romero2023}, there are biases associated with this method. To the extent that one knows the underlying spectra (or can at least make a reasonable estimates of them), these biases can be corrected.     
    We consider power spectra of the form $P(k) \propto e^{-k_c/k} k^{-\alpha}$, with a low cutoff wavenumber $k_c = 1/(5 R_{500})$ such that we effectively consider a single power-law within our sampled range. In particular, we correct for the bias when $\alpha = 3$ and we consider the spread of biases over the range ($2 < \alpha < 4$) as a source of systematic uncertainty that we fold into our reported uncertainties as in \citet{romero2023}. We privilege the bias for $\alpha=3$ because this will be the slope of the power spectrum at the peak of the 3D amplitude spectrum (by definition; see Equation~\ref{eqn:as3d}) and Section~\ref{sec:PS_results} clarifies our interest in the peak of the amplitude power spectrum. Appendix~\ref{sec:appendix_masking} investigates biases associated with large scales due to masking. Here again we apply corrections for the inferred biases for underlying power spectra of $\alpha=3$. Independent of the $\Delta$-variance method, our images are the convolution of each instrument's PSF, which we correct for by dividing by the normalized power spectrum of the appropriate PSF (where we consider CCD position and use \lstinline{ellbeta} PSF model to determine this for \textit{XMM} data).
    
    We wish to sample scales between (roughly) the FWHM of the PSF of each instrument and $R_{500}$ and we space apart the angular nodes where the power spectra are sampled by roughly a factor of two. These constraints lead us to sample the SPT data over four nodes and the \textit{XMM} data over seven nodes, of which we keep the four smallest nodes (largest scales) to match those used for SPT data. 
    As mentioned in Section~\ref{sec:intro}, if we hope to constrain the hydrostatic mass bias we will need an estimate of the radial slope of the RMS turbulent velocity (Mach number) and thus we desire at least two regions: an inner circle and outer annulus (or annuli). If we are to sample scales of $R_{500}$, then we expect that our inner circle should have a radius of at least $r = R_{500}/2$. Conversely, we don't want any region (annulus) to be thinner than the largest beam width (taken as the FWHM) of the instruments and so we set a minimum width of annuli to 1.\arcmin25. There are additional considerations with respect to the division of regions which are discussed and investigated in Section~\ref{sec:Choices} which still do not offer a clear choice. Left with some ambiguity, we opt to divide our rings at $R_{500}/\phi$, where we take $\phi = (1 + \sqrt{5})/2$, the golden ratio. We thus define Ring 1 to be the circle of radius 3.$^{\prime}3$ for SPT-CLJ0232-4421 and the circle of radius 3.$^{\prime}9$ for SPT-CLJ0638-5358 and Ring 2 to be the annulus from the edge of Ring 1 to each cluster's respective $R_{500}$ ($5.^{\prime}3$ and $6.^{\prime}3$).
    
    As elsewhere, we report our results in the form of amplitude spectra; given 2D ($P_{\rm 2D}$) and 3D ($P_{\rm 3D}$) power spectra, amplitude spectra are given as:
    \begin{align}
        A_{2D}(k) &= [(2 \pi) k^2 P_{\rm 2D}]^{1/2} \label{eqn:as2d}, \\
        A_{3D}(k) &= [(4 \pi) k^3 P_{\rm 3D}]^{1/2} \label{eqn:as3d},
    \end{align}
    where $k = 1/\lambda$ is the wavenumber which corresponds to (angular or physical) scales of length $\lambda$. 

    \subsection{Application to \textit{XMM} data}
    \label{sec:XR_ps}
    
        As with our profile fits, power spectra are measured on each fractional image (i.e. per CCD, per ObsID, per energy band) and mask the same pixels as masked in the profile fits. We do not measure the spectra on a single image (e.g., as presented in Figure~\ref{fig:FractionalResiduals}), but for each CCD, energy band, and ObsID, we take 100 surface brightness models from the respective MCMC chain (of the surface brightness profiles, long after burn-in). For each set of model parameters, a single Poisson noise realization is generated, from which we can create a noise realization of the fractional residual in addition to the (data) fractional residual for that model. The spectra of these two images are measured and their difference is our debiased spectrum. The debiased spectra also have the same treatment to account for faint point sources as in \citet{romero2023}. 

        \begin{figure}[]
            \begin{center}
                \includegraphics[width=0.45\textwidth]{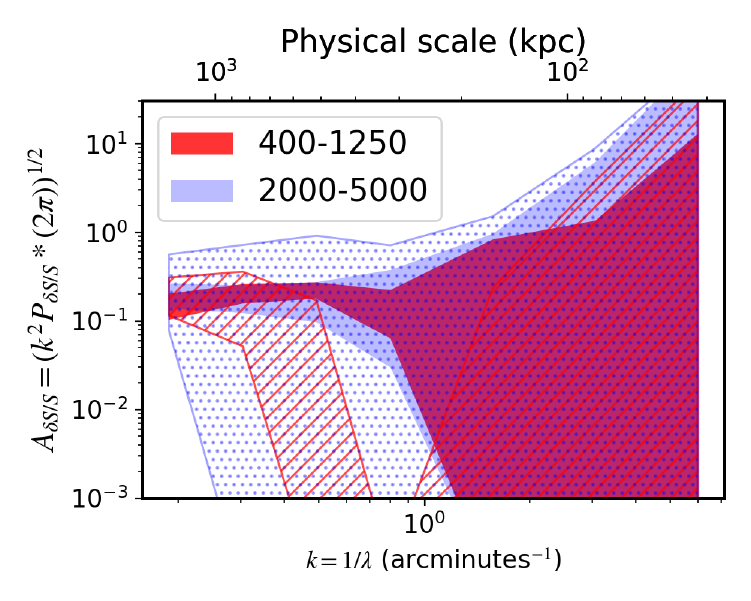}
                \includegraphics[width=0.45\textwidth]{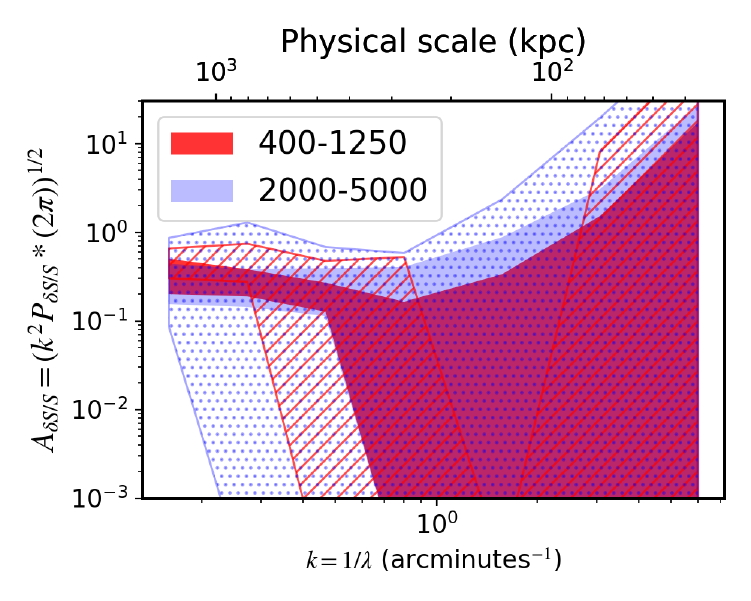}
            \end{center}
            \caption{\textbf{left:} The 2D amplitude spectra for SPT-CLJ0232-4421 color coded by energy band (legend reports in eV). The solid-color filled regions show the $1\sigma$ range for Ring 1 and the hatched and stippled regions show $1\sigma$ range for Ring 2; \textbf{right:} The 2D amplitude spectra for SPT-CLJ0638-5358 with the same scheme for energy bands and rings.
            %\MG{I am a bit concerned that we get (again) quasi flat amplitude spectra already in 2D; I thought we could improve this with SPT data... any insight?}
            %\MG{Another comment: the band collapsing below axis limit is a bit confusing for the external reader; can we find a better way to show upper limits instead of coloring such a large portion of the plot?}
            }
        \label{fig:A2D_XR}
        \end{figure}

        We take the expected values and associated uncertainties to be the mean and standard deviation, respectively, of the debiased spectra for each CCD and ObsID. We combine the spectra for each energy band across CCDs and ObsIDs as their weighted averages. Figure~\ref{fig:A2D_XR} shows mild (more so visually than statistically) differences in the amplitude spectra of each energy band for both clusters and both rings.

    \subsection{Application to SPT data}
    \label{sec:SZ_ps}

\begin{figure}[!h]
            \begin{center}
                \includegraphics[width=0.45\textwidth]{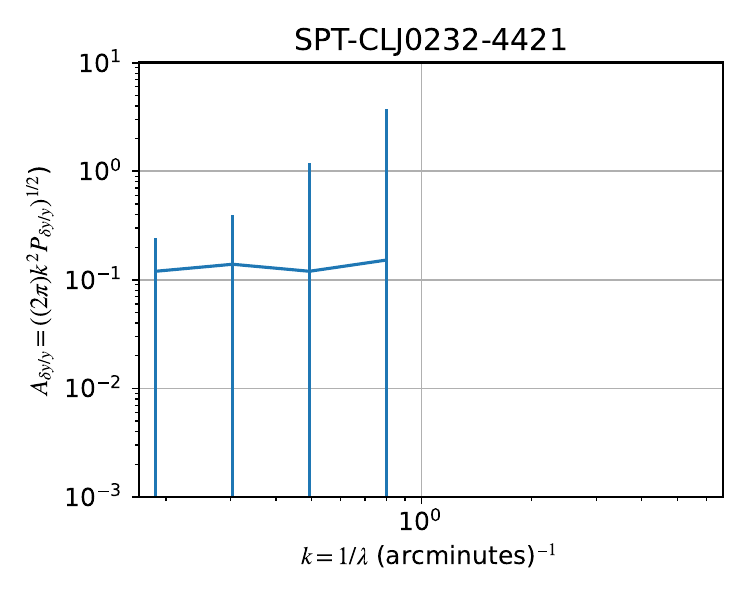}
                \includegraphics[width=0.45\textwidth]{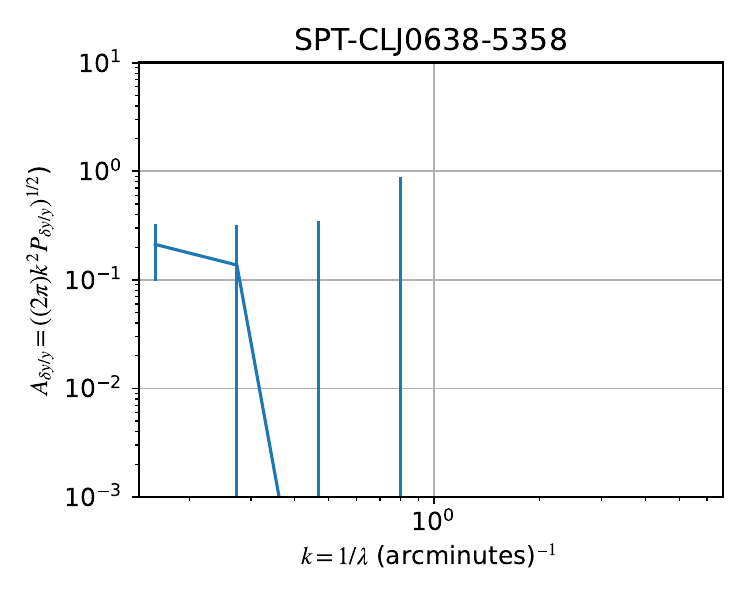}
            \end{center}
            \caption{Amplitude spectrum of $\delta y/y$ in Ring 1 for SPT-CLJ0232-4421 (left) and SPT-CLJ0638-5358 (right).}
        \label{fig:AK_dyy}
        \end{figure}

        We take the power spectrum of each SPT cluster to be that derived from the fractional residuals of its
        best-fit profile. In order to debias the spectrum, we make fractional residuals on each cluster's respective half maps and take the cross spectrum of the two half maps for each cluster. The uncertainty is taken to be the standard deviation of the cross spectra of the half-map blank regions (from the respective fields). In case of any leakage of instrumental noise, we subtract the mean across these noise cross spectra (the effect of this is negligible). Spectra are not reported for Ring 2 as none are statistically significant (Figure~\ref{fig:AK_dyy}).
 
    \subsection{Deprojection to 3D}
    \label{sec:Deproj}

        We wish to deproject our 2D fluctuation spectra to 3D fluctuations. To do this we use a common formalism \citep[e.g.,][]{peacock1999,churazov2012,khatri2016}:
        \begin{equation}
            P_{\text{2D}}(k_\theta) = \int P_{\text{3D}}(\sqrt{k_{\theta}^2 + k_{z}^2}) |\tilde{W}(k_z)|^2 dk_z,
            \label{eqn:deproj}
        \end{equation}
        where $z$ is the axis along the line of sight, $k_{\theta}^2 = k_{x}^2 + k_{y}^2$ is in the plane of the sky, and $|\tilde{W}(k_z)|^2$ is the 1D power spectrum of the window function, which normalizes the distribution of the relevant (unperturbed) 3D signal generation to the (unperturbed) 2D surface brightness. In the case of Compton $y$, the units of our SZ surface brightness maps, we already have Equation~\ref{eqn:Compton_y}. For our X-ray surface brightness, $S$, maps in units of $\rm photon\ counts\ s^{-1}\ arcmin^{-2}$, we can write:
        \begin{equation}
            S = \int \epsilon dz,
        \end{equation}
        where $\epsilon$ encapsulates the emission integral, EI, and band-averaged cooling function, $\Lambda_b$ presented in Section~\ref{sec:intro}. We denote the smooth (i.e. model) 3D distribution of $P_e$ and $\epsilon$ as $\bar{P_e}$ and $\bar{\epsilon}$, respectively, which when integrated along the line of sight, produce $\bar{y}$ and $\bar{S}$, the 2D (circular, unperturbed) surface brightness models.
        The SZ and X-ray window functions are respectively:
        \begin{align}
            W_{\text{SZ}}(\theta,z) &\equiv \frac{\sigma_{\text{T}}}{m_{\text{e}} c^2} \frac{\bar{P_e}(\theta,z)}{\bar{y}(\theta)} \text{ and} \label{eqn:Wsz} \\
            W_{\text{X}}(\theta,z) &\equiv \frac{\bar{\epsilon}(\theta,z)}{\bar{S}(\theta)}. \label{eqn:Wx}
        \end{align}

        \begin{figure}[!h]
            \begin{center}
                \includegraphics[width=0.47\textwidth]{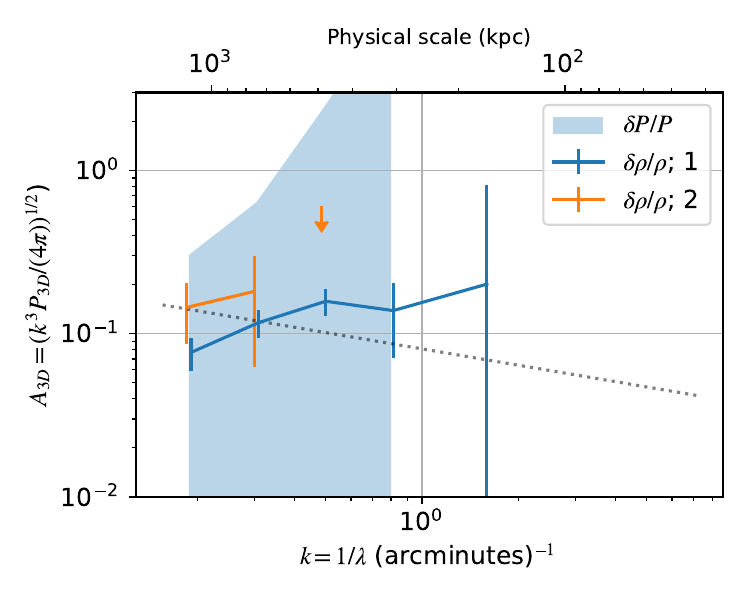}
                \includegraphics[width=0.47\textwidth]{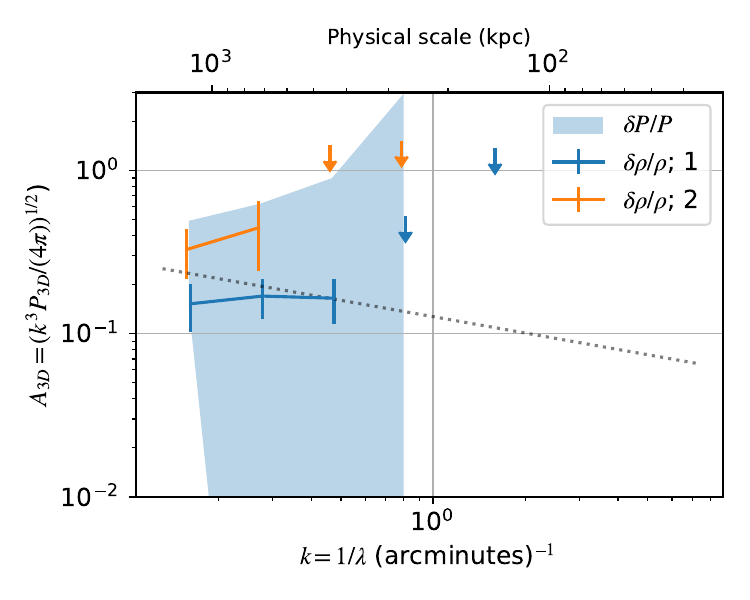}
            \end{center}
            \caption{Amplitude spectra of $\delta P/P$  and $\delta \rho / \rho$ for SPT-CLJ0232-4421 (left) and SPT-CLJ0638-5358 (right). The dotted lines have a spectral index, $\alpha = 11/3$, which equates to a slope of $-1/3$ in the amplitude spectra. Arrows indicate $3\sigma$ upper limits. The constraints on the amplitude spectra are not adequate to constrain the cascade (slope) of fluctuations at $k > k_{\text{peak}}$. The roughly flat spectra (plateau) in Ring 1 of both clusters is likely indicative of multiple injection scales (see Section~\ref{sec:PS_results}).
            }
        \label{fig:SZvsXraySpectra}
        \end{figure}

        Equation~\ref{eqn:deproj} can be approximated as
        \begin{equation}
            P_{\text{2D}}(k_\theta) \approx P_{\text{3D}}(k) \int |\tilde{W}(k_z)|^2 dk_z
            \label{eqn:deproj_approx}
        \end{equation}
        \citep[e.g.,][]{churazov2012,khatri2016}, where $k = \sqrt{k_{\theta}^2 + k_{z}^2}$. We adopt the notation used in \citet{khatri2016,romero2023} and define 
        \begin{equation}
            N(\theta) \equiv \int |\tilde{W}(k_z)|^2 dk_z.
            \label{eqn:window_approx}
        \end{equation}
        This approximation is quite robust \citep[e.g.,][]{khatri2016,romero2023} for most scales probed; we verify the validity of this approximation for our clusters in Appendix~\ref{sec:appendix_deproj}. Another concern with deprojection is the evolution of the window function within a region \citep[][]{Zhuravleva2015} which we consider in Appendix~\ref{sec:appendix_masking} where the result is that we weight a sampling of $N(\theta)$ by annular area to derive an effective value for a given ring, $N_{\text{eff}}$. 
        
        Once we have $N_{\text{eff}}$, from the best-fit profiles of each image (again, for \textit{XMM} data this is considered per CCD, per energy band, and per ObsID.), we calculate the weighted average of deprojected spectra from individual deprojected spectra as before with the 2D spectra. We further combine both energy bands considering the overall agreement. If we simply divide $P_{\rm 2D,X-ray}$ for a given ring by the respective $N_{\text{eff}}$, then we've only calculated the power, $P_{\rm 3D}$ for the underlying emmisivity fluctuations. Dividing by a further factor of four \citep[e.g.,][]{churazov2012} yields the density fluctuations.\footnote{This holds especially well for low photon energies in hot, massive clusters where the photon counts depend negligibly on gas temperature.}
        The 3D spectra ($P_{\rm 3D}$) for SPT are deprojected from their 2D counterparts shown in Figure~\ref{fig:AK_dyy} and directly yield pressure fluctuations. We present both SZ and X-ray 3D amplitude spectra in Figure~\ref{fig:SZvsXraySpectra} with dotted grey lines indicating a Kolmogorov spectrum ($\alpha = 11/3$ for the convention $P_{\text{3D}}(k) \propto k^{-\alpha}$), thus having a logarithmic slope of -1/3 when plotting amplitude spectra.

\section{Results and Inferences from Spectra}
\label{sec:PS_results}

    \begin{table*}[] 
   \centering  
   \begin{tabular}{c c c | c c c c c} 
    Cluster & Dataset & Ring & A$_{\text{3D}} (k_{\text{peak}})$ & $\sigma_{\text{3D}}$ & $\sigma_{\ln}$ & $k_{\text{peak}} $ & $\lambda_{\text{peak}}$ \\
    & & & & &  & (arcmin$^{-1}$) & (kpc) \\ 
   \hline 
   \multirow{2}{*}{SPT-CLJ0232-4421} & X-ray & Ring 1 & $0.16 \pm 0.03$ & $0.21 \pm 0.17$ & $0.20 \pm 0.11$ & 0.49 & 516 \\ 
   & X-ray & Ring 2 & $0.14 \pm 0.06$ & $0.11 \pm 0.06$ & $0.11 \pm 0.05$ & 0.30 & 837 \\ 
   \hline
   \multirow{3}{*}{SPT-CLJ0638-5358} & X-ray & Ring 1 & $0.17 \pm 0.05$ & $0.17 \pm 0.04$ & $0.17 \pm 0.04$ & 0.27 & 807 \\ 
    & SZ & Ring 1 & $0.19 \pm 0.10$ & $0.15 \pm 0.88$ & $0.15 \pm 0.31$ & 0.16 & 1380 \\
    & X-ray & Ring 2 & $0.44 \pm 0.20$ & $0.28 \pm 0.11$ & $0.27 \pm 0.11$ & 0.27 & 807 \\ 
    \hline
   \end{tabular} 
   \caption{Key values from amplitude or power spectra.}  
   \label{tbl:ps_products}
\end{table*} 
       %%%%%%%%%%%%%%%%%% Table of amplitude spectra properties

    We are able to confidently constrain the spectra for both clusters at the largest scales we probe, and we do capture the peaks of the amplitude spectra. We recover injection scales ($\lambda_{\rm peak}$) of several hundreds of kpc (see Table~\ref{tbl:ps_products}), which is largely within expectations \citep[cf.][]{Gaspari2014_PS,khatri2016,eckert2017}. The values of the amplitude spectra at scales smaller than the inferred injection scale are generally smaller, with the exception of amplitude spectrum of SPT-CLJ0232-4421 at $k = 1.6$ arcmin$^{-1}$. Unfortunately the uncertainties at scales smaller than our inferred injection scales are often large and thus we cannot exclude with high confidence the possibility of slightly smaller injection scales.
    %However, the uncertainties are large, and so we do not believe that this warrants further scrutiny. 
    Similarly, in  both clusters and both rings, our uncertainties at scales of 1 arcminute or smaller do not allow us to constrain the slope of a turbulent cascade.
    %Both clusters appear to have fluctuations which have dissipated significantly towards scales of 100 kpc. The drop off is unexpected 
    %\MG{Here it is worth to expand further; is it due to the telescope resolution/noise? The density fluctuations appear rather flat before the cutoff, so I would state here that this is more of an observational limitation, rather than a physical process (such as thermal conduction washing out fluctuations)}
    %relative to studies such as \citet{Gaspari2014_PS} where we might expect spectral indices $3 \lesssim \alpha \lesssim 4$ at scales moderately smaller than the injection scale. 
    Even so, we have drawn a guide line with arbitrary normalization and a spectral slope of $\alpha = 11/3$ (logarithmic slope of -1/3 for 3D amplitude spectra), i.e. a Kolmogorov spectrum, in Figure~\ref{fig:SZvsXraySpectra}. 
    %This appears to be primarily a matter of data quality, though we also note the potential impact of the known shock in SPT-CLJ0638-5358 (discussed more below); the uncertainties are such that the data are consistent with a Kolmogorov spectrum, with the possible exception of Ring 2 in SPT-CLJ0232-4421 whose spectrum suggests a clear lack of power on scales of $\sim 1^{\prime}$. However, this aligns with the circular mask of the point source; the mask having a radius of $80^{\prime\prime}$ (see Section~\ref{fig:xmm_sb_profiles}). 

    There are two principle avenues to derive turbulent velocities from power spectra: use either the peak of the amplitude spectrum \citep[e.g.,][]{Gaspari2013_PS,khatri2016,eckert2017} or some measure of the dispersion of fluctuations in resolution elements of numerical simulations \citep[e.g.,][]{zhuravleva2012,mohaptra2021a,simonte2022}. The (total) variance can be calculated from the integrated power spectrum:
    \begin{equation}
        \sigma_{\text{3D}}^2 = \int P_{\rm 3D}(k) 4 \pi k^2 dk.\label{eqn:variance_from_PS}
    \end{equation}
    We report the linear standard deviation ($\sigma_{\text{3D}}$) and equivalent (natural) logarithmic standard deviation ($\sigma_{\ln}$) in Table~\ref{tbl:ps_products} along with the peak of the amplitude spectra of density fluctuations. Though the linear and logarithmic standard deviations are nearly identical, we bring our distribution values into compliance with the width in \citet{zhuravleva2023}, which is equivalent of a full-width at half maximum of a log$_{10}$-normal distribution ($\delta \xi / \xi$ in their notation) and note that this width is equal to $1.02\,\sigma_{\ln}$. The integrated spectra are calculated up until $P_{\text{3D}} < 0$.

    \begin{table*}[] 
   \centering  
   \begin{tabular}{c c c | c c c} 
    Cluster & Dataset & Ring & $\mathcal{M}_{\text{3D,peak}}$ & $\mathcal{M}_{\text{3D,int}}$ & $\mathcal{M}_{\text{3D,comb}}$ \\ 
   \hline 
   \multirow{2}{*}{SPT-CLJ0232-4421} & X-ray & Ring 1 & $0.63 \pm 0.12$ & $0.43 \pm 0.23$ & $0.53 \pm 0.10 \pm 0.11$ \\ 
   & X-ray & Ring 2 & $0.45 \pm 0.18$ & $0.24 \pm 0.10$ & $0.34 \pm 0.11 \pm 0.09$ \\ 
   \hline
    \multirow{3}{*}{SPT-CLJ0638-5358} & X-ray & Ring 1 & $0.60 \pm 0.17$ &  $0.35 \pm 0.08$ & $0.48 \pm 0.12 \pm 0.07$ \\ 
     & SZ & Ring 1 & $0.59 \pm 0.32$ & $0.40 \pm 0.83$ & $0.50 \pm 0.09 \pm 0.30$ \\ 
    & X-ray & Ring 2 & $1.58 \pm 0.72$ & $0.57 \pm 0.23$ & $1.07 \pm 0.51 \pm 0.22$ \\ 
    %\multirow{3}{*}{SPT-CLJ0638-5358} & X-ray & Ring 1 & $0.68 \pm 0.19$ &  $0.35 \pm 0.26$ & $0.52 \pm 0.16 \pm 0.15$ \\ 
    % & SZ & Ring 1 & $0.46 \pm 0.25$ & $0.26 \pm 0.2.42$ & $0.36 \pm 0.10 \pm 0.25$ \\ 
    %& X-ray & Ring 2 & $1.78 \pm 0.81$ & $0.59 \pm 0.80$ & $1.18 \pm 0.60 \pm 0.57$ \\    \hline
   \end{tabular} 
   \caption{Derived Mach numbers.}  
   \label{tbl:mach_numbers}
\end{table*}

    Unfortunately much of the recovered pressure fluctuations are not statistically significant; however, the largest scale probed in SPT-CLJ0638-5358 is significant and so we report spectral properties of pressure fluctuations for SPT-CLJ0638-5358 in Table~\ref{tbl:ps_products}. In all cases we require that any selected peak be at least $2 \sigma$ significance, that is, we select the largest amplitude that meets this criterion.

    To derive Mach numbers from peak amplitude spectra we use as reference the linear relation first found via high-resolution hydrodynamical simulations by \citet{Gaspari2013_PS}:
    \begin{align}
    \begin{split}
        \mathcal{M}_{\text{3D,peak}} &\approx 4 \, A_{\rho}(k_{\text{peak}}) \left(\frac{l_{\text{inj}}}{\text 500 kpc} \right)^{\alpha_H} \\
        &\approx 2.4 \, A_{P}(k_{\text{peak}}) \left(\frac{l_{\text{inj}}}{\text 500 kpc} \right)^{\alpha_H},
        \label{eqn:mach_from_peak}
    \end{split}
    \end{align}
    where $l_{\text{inj}} = 1/k_{\text{peak}}$ (here in units of kpc), and $0.2 < \alpha_H < 0.3$ is the slight dependence on injection scale. We also test a relation from large-scale cosmological simulations (with 80 clusters; \citealt{zhuravleva2023}); noting that
     $\mathcal{M}_{\text{3D}} = \sqrt{3} \mathcal{M}_{\text{1D}}$, they find the following radially-averaged ($r < 2 R_{500}$) scaling with the total standard deviation:    
    \begin{equation}
       %\mathcal{M}_{\text{3D}} \approx 2.1 \, \delta \rho/\rho. \approx 2.4 \, \delta P/P.
        \mathcal{M}_{\text{3D,int}} \approx \sqrt{3} * \langle \eta_{\rho} \rangle * 1.02 * \sigma_{\ln,\rho} \approx \sqrt{3} * \langle \eta_{P} \rangle *1.02 * \sigma_{\ln,P},
        \label{eqn:mach_from_int}
    \end{equation}
    where $\eta_{\rho}$ and $\eta_P$ are functions of radius, cluster geometry (spherical or ellipsoidal), and dynamical state (relaxed, in-between, or unrelaxed). With respect to the choice of $\eta$, we thus utilize the value for the in-between state for SPT-CLJ0232-4421 and the unrelaxed state for SPT-CLJ0638-5358. For the spherical case, this gives the radially averaged values of $\langle \eta_{\rho} \rangle = 1.2$ in both the in-between state and unrelaxed state, and $\langle \eta_{P} \rangle = 1.4$ for the in-between state and $\langle \eta_{P} \rangle = 1.5$ for the unrelaxed state.
    Mach numbers corresponding to each estimation method are reported in Table~\ref{tbl:mach_numbers}; a combined value is also calculated as the average of the values derived from the two methods. The uncertainty on this combined value takes the spread (difference) between the two methods as a systematic uncertainty. Further quantities derived from $\mathcal{M}_{\text{3D,comb}}$ propagate the combined uncertainties (systematic and statistical uncertainties are summed).
    %\MG{After thinking more about this; the Gaspari+ relations are a bit higher because if you integrate the spectrum there is an additional $\sim$1.7 factor (see G14) -- maybe a good idea is to unify the G13/14-Z23 density relations such as $\mathcal{M}_{\text{1D}} = \delta\rho/\rho\ \pm$ scatter arising from the difference between the two relations. This way we can provide a single Mach number and plot (with bands representing the systematic modeling error), to facilitate any external reader too. For the pressure scaling we can use the average between adiabatic and isobaric case, with a related scatter enveloping the pure isobaric and pure adiabatic regime.}
    
    %\input{HydrostaticBiasTable}
    \begin{table*}[] 
   \centering  
   \begin{tabular}{c c | c c c} 

      & & $-b_{\mathcal{M},\text{peak}}$ & $-b_{\mathcal{M},\text{int}}$ & $-b_{\mathcal{M},\text{comb}}$ \\ 
   \hline 
   \multirow{2}{*}{SPT-CLJ0232-4421} & Ring 1 & $0.34 \pm 0.20$ & $0.26 \pm 0.26$ & $0.30 \pm 0.27$ \\ 
    & Ring 2 & $0.13 \pm 0.10$ & $0.05 \pm 0.04$ & $0.08 \pm 0.09$ \\ 
    \hline
    \multirow{2}{*}{SPT-CLJ0638-5358} & Ring 1 & $-0.42 \pm 0.65$ & $-0.01 \pm 0.09$ & $-0.16 \pm 0.38$ \\ 
    & Ring 2 & $0.02 \pm 0.73$ & $0.08 \pm 0.10$ & $0.10 \pm 0.44$ \\ 
    
   % & & $-b_{\mathcal{M}}$ & $d \ln P / d \ln r$ & $d \ln \mathcal{M}_{\text{3D}}/ d \ln r$ \\ 
   %\hline 
   %\multirow{2}{*}{SPT-CLJ0232-4421} & Ring 1 & $0.05 \pm 0.19$ & $-0.75 \pm 0.01$ & \multirow{2}{*}{$0.08 \pm 1.22$} \\ 
   % & Ring 2 &  $0.06 \pm 0.10$ &  $-3.42 \pm 0.09$ &  \\ 
   % \hline
   % \multirow{2}{*}{SPT-CLJ0638-5358} & Ring 1 & $-0.17 \pm 0.37$ & $-1.22 \pm 0.01$ & \multirow{2}{*}{$1.19 \pm 1.00$} \\
   %  & Ring 2 & $0.13 \pm 0.37$ & $-3.04 \pm 0.11$ &  \\
   \end{tabular} 
   \caption{Derived hydrostatic biases via two Mach scaling relations, as well as a third estimation from the combination of the Mach estimation.}  
   \label{tbl:bms}
\end{table*} 

    By including non-thermal pressure support due to turbulence in the standard hydrostatic equilibrium equation and the relation of non-thermal pressure, thermal pressure, and Mach numbers in Equation~\ref{eqn:NTpp_Mach} one can derive the hydrostatic bias in terms of a Mach number, logarithmic Mach slope, and logarithmic pressure slope. We briefly reiterate this here and use the notation and convention\footnote{NB that the sign convention of $b_{\mathcal{M}}$ is opposite that of the conventional hydrostatic bias, $b$.} from \citet{khatri2016} for the hydrostatic bias, $b_{\mathcal{M}} \equiv M_x / M_{\text{tot}} - 1$, derived in this manner: 
    \begin{equation}
        b_{\mathcal{M}} = \frac{-\gamma \mathcal{M}_{\text{3D}}^2}{3} \frac{ d \ln P_{\text{NT}}}{d \ln P_{\text{th}}} \left( 1 +  \frac{\gamma \mathcal{M}_{\text{3D}}^2}{3}\frac{ d \ln P_{\text{NT}}}{d \ln P_{\text{th}}} \right)^{-1},
        \label{eqn:mach_bias}
    \end{equation}
    where $\gamma$ is the adiabatic index, taken to be 5/3 for the ICM. We can reformulate $ (d \ln P_{\text{NT}}) / (d \ln P_{\text{th}})$ in terms of variables more directly calculated from our SZ surface brightness fits and derived Mach numbers (Table~\ref{tbl:mach_numbers}):
    \begin{equation}
        \frac{ d \ln P_{\text{NT}}}{d \ln P_{\text{th}}} = \frac{ d \ln P_{\text{NT}} / d \ln r}{d \ln P_{\text{th}} / d \ln r} = 1 + 2 \frac{ d \ln \mathcal{M}_{\text{3D}} / d \ln r}{d \ln P_{\text{th}} / d \ln r}.
        \label{eqn:Pnt_mach}
    \end{equation}
    Taking the $\mathcal{M}_{\text{3D,peak}}$ values, we derive hydrostatic biases which we report in Table~\ref{tbl:bms}; the logarithmic slopes of Mach numbers and pressure profiles are reported in Appendix~\ref{sec:appendix_Supplemental}. As we only have two rings, we can only estimate a single power-law for $\mathcal{M}_{\text{3D}}(r)$ across all radii of concern. 
    For the logarithmic pressure slope we calculate the average slope within a ring based on our deprojected SZ $\beta$-model.

    \subsection{Interpretation}
    \label{sec:interpretation}

    We selected our clusters in part based on a visual inspection of the SZ and X-ray images to assess dynamical states; we note that while there is correlation of the cluster morphology with the dynamics of the ICM \citep[e.g.,][]{battaglia2012a,zhuravleva2023}, cluster morphology is not strictly indicative of the dynamics of the ICM. We note that turbulence itself does not require a clear visual disturbance (such as an elongation or surface brightness edge, as would emerge for a shock or cold front). Even the turbulence driven inside-out via AGN feeding or feedback can occur in a quasi-spherical manner over several cycles, superposed to the large-scale cosmological chaotic motions \citep[][]{gaspari13,gaspari20,lau2017,wittor2020,wittor23}. 
    Moreover, shocks or other injection mechanisms may exist but be masked via several observational effects, e.g., due to projection along the line of sight.
    
    The Mach numbers for both clusters span $0.5 < \mathcal{M}_{\rm 3D} < 1.6$, with both clusters having $\mathcal{M}_{\rm 3D} \approx 0.6$ in Ring 1. It is perhaps surprising that the Mach numbers are similar in Ring 1 for the two clusters given that the known shock in SPT-CLJ0638-5358 \citep{botteon2018} occurs within two arcminutes of the cluster center, thus in Ring 1. However, neither the length of the detected X-ray edge associated with the shock front nor its radial extent is large. That is, the effect of the shock, which is already sub-dominant (see below and Appendix~\ref{sec:appendix_slices}), is dampened by a filling factor of its area relative to the area of the region (Ring 1). Even so, taking the X-ray Mach values, $\mathcal{M}_{\rm 3D} = 0.64 \pm 0.03$ is not much larger than the derived value for SPT-CLJ0232-4421 in Ring 1 of $\mathcal{M}_{\rm 3D} = 0.51 \pm 0.06$. Again, we come to the somewhat ambiguous state of SPT-CLJ0232-4421; many X-ray metrics indicate that it is relaxed \citep[e.g.,][]{lovisari2017}, but the substructure found in \citet{parekh2021} is evident in our data too and undoubtedly contributes to our derived Mach number. In Ring 2, both clusters show substructure (prominent residuals) in the SPT data, but distinct features are not readily visible in the X-ray (fractional) residuals. That said, fluctuations appear to have larger amplitude to the southwest and northeast of the cluster and an investigation of fluctuations separating directions in the cluster (see Appendix~\ref{sec:appendix_slices}) echos this. 

    Our Mach numbers are generally larger than those found in other studies [e.g., \citealt{Gaspari2013_PS} ($\mathcal{M}_{\text{3D}} \simeq 0.45$), \citealt{Zhuravleva2015} ($\mathcal{M}_{\text{3D}} \approx 0.2$), \citealt{Hitomi2016} ($\mathcal{M}_{\text{3D}} \approx 0.3$), \citealt{hofmann2016} ($\mathcal{M}_{\text{3D}} \approx 0.3$), \citealt{eckert2017} ($\mathcal{M}_{\text{3D}} \approx 0.3$), \citealt{zhang2023} ($\mathcal{M}_{\text{3D}} \approx 0.2$), \citealt{dupourque2023} ($\mathcal{M}_{\text{3D}} \approx 0.2$)]. As noted in \citet{hofmann2016}, Mach numbers may be expected to grow with cluster-centric radius and with the scales being probed. That is, at larger cluster-centric radii we expect larger non-thermal pressure support \citep[e.g.,][]{battaglia2012a} and thus larger Mach numbers. Similarly, the Mach number inferred at scales below the injection scale will be less than the Mach number inferred at the injection scale, which tends towards scales of a few hundred kpc (as most of the studies quoted above; e.g., \citealt{hofmann2016,Zhuravleva2015}). Our analysis probes scales out to $0.62 R_{500} \simeq 0.85$ Mpc for the inner ring and scales up to $R_{500}$, where we recover injection scales $l_{\text{inj}} \geq 500$ kpc; thus, on both accounts we might expect larger Mach numbers than found in many other studies. Even if one considers that due to poor spectral constraints, the injection scales may be smaller than inferred, one is still left with substantial fluctuations at our inferred injection scales. While not many studies probe our large scale, a similar multiwavelength analysis by \citet{khatri2016} find $\mathcal{M}_{\text{3D}} \simeq 0.8$ over the Mpc scale.
    %However, \citet{eckert2017,zhang2023,dupourque2023} probe scales of a few hundred kpc (and beyond Mpc for \citet{khatri2016}) and still find $\mathcal{M}_{\text{3D}} \lesssim 0.4$ for the former three studies. %\MG{I removed this: the issue is that these three study still have a limited spatial range; the only actual Mpc-scale study other than our current study is KG16}  

    %As discussed in Section~\ref{sec:MachAffairs}, %\MG{moved here the discussion below, keep it more focused, as it's not the goal of the paper to compare infinite number of scalings}
    Another aspect when comparing Mach numbers across the literature is that there is a significant scatter in the Mach scaling relation \citep[][]{Gaspari2013_PS,Gaspari2014_PS,Zhuravleva2014,zhuravleva2023,mohaptra2021a,simonte2022}.
    Albeit nearly all of the literature results agree on a linear conversion between ICM fluctuations and turbulent velocities, the different numerical and reduction techniques can introduce differences in normalization. 
    %For most of the comparisons, we still find larger fluctuations, where again we are generally probing larger scales than most other studies. A notable difference from the above comparison is that the value of the fluctuations found in \citet{dupourque2023} are actually quite similar to the fluctuations we find, where indeed this seems to be attributable to a shallower Mach scaling relation: that of \citep{simonte2022}. %{MG: removed this -- see my email on simonte22 relation, which has some key issues in both the analysis and numerics adopted}
    However, the inferred Mach numbers as determined via the peak and via the integrated spectra are generally in agreement, though the inferred values from integrated spectra tend to be slightly less than those from the peak (Table~\ref{tbl:mach_numbers}). A similar trend was seen in Zwicky 3146 \citep{romero2023}, although the pressure spectrum in their Ring 1 yield a larger Mach number from the integrated relation than the peak relation. Thus it appears that scaling relations derived from high-resolution hydrodynamical simulations \citep{Gaspari2013_PS,Gaspari2014_PS} are in agreement with those derived from large-scale cosmological simulations \citep{Zhuravleva2014,zhuravleva2023}. 
    An outlier to this converged trend is the scaling from \citet{simonte2022} cosmological runs, which find a shallower sample relation (by roughly $2\times$). This can be understood by the different ad-hoc filtering of `clumps'/sub-structures (not driven by turbulence) which must be adopted in large-scale runs with necessarily coarse resolution.
    E.g., in \citet{Zhuravleva2014}, the clump filtering is minimal with a 3.5\,RMS cut of the right tail in the density distribution, implying that 99.98\% of the distribution is retained. In \citet{simonte2022}, the filtering is instead aggressive, with a cut above $\sim$1.6\,RMS. They report that the related density fluctuation threshold cut is $\sigma_{\rho,{\rm 3D}}^2 = 0.22$, implying that the potentially included Mach numbers remain below 0.5, hence the retrieved low scaling normalization and small pressure support in their study, also compared with our findings.
    %between $\mathcal{M}_{\text{3D}}$ and $\sigma_{\rho,\rm {3D}}$
    %\footnote{That is, $\sigma_{\rho} = \sigma_{\text{3D}}$ as tabulated for the X-ray datasets in Table~\ref{tbl:ps_products}} 
    %than what is found in \citet{Zhuravleva2014} or \citet{Gaspari2013_PS}.

    We know that our recovered spectrum of fluctuations cannot be interpreted as solely due to turbulence, especially in the case of SPT-CLJ0638-5358, which has a known shock \citep{botteon2018} and for which we see suggestive hints of edges in the SZ (SPT) data (see Section~\ref{sec:SPT_img_analysis}). Nonetheless, the observed X-ray surface brightness fluctuations from the known shock are sub-dominant (see Appendix~\ref{sec:appendix_slices}) though the known shock may contribute to flattening the amplitude spectrum at smaller scales than $l_{\text{inj}} \approx 800$ kpc, but otherwise does not significantly impact the results we present (i.e. Mach numbers or hydrostatic mass biases). Were we to probe the turbulent cascade (with deeper data), it would be necessary to mask the shock to properly constrain fluctuations due to turbulent motions.
    %\citep[e.g.,][ recalling $\mathcal{M}_{\text{3D}} = \sqrt{3}\mathcal{M}_{\text{1D}}$]{Gaspari2013_PS,Zhuravleva2015,hofmann2016,eckert2017}.

    %\MG{Please expand here saying that the visual relaxation state does not necessarily imply a quiescently dynamical cluster (in terms of weather ICM); this is a key point to emphasize in the Conclusions. I'm fairly confident we will find such discrepancies all over the place in the full SPT sample, which is very good since it's a novel result.}\\

    %We may also compare our derived Mach numbers to those from other studies of fluctuations. 
    %\MG{Also, please expand this section comparing with more results, e.g., \citealt{hofmann2016} Mach numbers.}

    %%%%%%%%%%%%%%%%%%%%%%%%%%%%%%%%%%%%%%%%%%%%%%%%%%%%%%%%%%%%%%%%%%%%%%%%%%%%%%%%%%

    We find that the hydrostatic bias for $M_{500}$ (those for Ring 2) derived for SPT-CLJ0232-4421 is well within the expected range \citep[e.g.,][]{hurier2018}, while the hydrostatic bias (for Ring 2, i.e. $M_{500}$) for SPT-CLJ0638-5358 is nearly 0, which is plausible but rare for clusters. It's generally expected that a hydrostatic mass will underestimate the mass because it fails to account for non-thermal pressure. However, this perspective may overlook the conditions in which a hydrostatic mass may \textit{overestimate} the mass (which are likely to be conditions in which pressure equilibrium itself is not entirely accurate). In particular, studies such as \citet{wik2008,krause2012} show that the both central Compton $y$ and integrated Compton $Y$ parameter (proxies for mass) will increase rapidly with the generation of merger shocks. In the case of boosted SZ signal, we can expect that a hydrostatic mass derived either from a scaling relation (e.g., $Y$-$M$) or through an explicit calculation of thermal pressure support will have boosted $Y$ and $P_e$ values and would overestimate the true mass. By extension, inferences from X-ray observations should be similarly affected: if $P_e$ is being boosted, under roughly adiabatic conditions, then both $n_e$ and $T_e$ will be boosted.

    %In recent simulation works that show the distribution of hydrostatic mass biases we indeed see values close to zero and even some values below zero \citep[e.g][]{biffi16,ruppin2019,jennings2023}. Moreover, \citet{ruppin2019} classify clusters into relaxed and disturbed and find that both have a median hydrostatic mass bias $b = 0.29$ while the standard deviation of the hydrostatic mass bias is narrow for the relaxed sample and broad for the disturbed sample. In other words, when explicitly studying hydrostatic mass bias, simulations appear to corroborate the case for sufficient departures from equilibrium resulting in low (and even negative) hydrostatic mass biases. 

    Numerical simulations which have estimated the hydrostatic mass bias have found average bias values of $0.1 \lesssim b \lesssim 0.3$ \citep[e.g.,][]{nagai2007,lau2009,rasia2012,biffi16,ruppin2019,gianfagna2023,jennings2023}. However, they all find that the hydrostatic mass bias will have a distribution (and will also depend on the radius or density contrast at which the mass is being measured), and nearly all of these studies find that for an individual cluster, the bias for $M_{500}$ can be close to zero, and in some cases even negative. Given that they all find (positive) non-thermal pressure support, the small biases cannot be due to a lack of non-thermal pressure support. Moreover, \citet{ruppin2019} classify clusters into relaxed and disturbed and find that both have a median hydrostatic mass bias $b = 0.29$ while the standard deviation of the hydrostatic mass bias is narrow for the relaxed sample and broad for the disturbed sample. In other words, simulations appear to corroborate the case for sufficient departures from equilibrium resulting in low (and even negative) hydrostatic mass biases. 
    
    %such that $Y$ is greater than the merged cluster's true mass would suggest when in equilibrium. Thus, we can imagine that SPT-CLJ0638-5358 may have a boost in Compton $Y$ due to its merger state which led to a higher mass estimate, which happens to yield a low hydrostatic mass bias. On the other hand, the derivation of the hydrostatic mass bias calculation here assumes the non-thermal pressure support acts as a turbulent pressure support that is roughly uniform at a given radius. Whatever turbulence shocks may induce, they are initially localized and so we should be cautious in accepting the values of $b_{\mathcal{M}}$ for both rings in SPT-CLJ0638-5358.

    \subsection{Other Hydrostatic mass estimates for our clusters}

    We present here a variety of alternative calculations of the hydrostatic mass bias leaning heavily on mass estimates in the literature which have been compiled in Appendix~\ref{appendix:mass_estimates}. Our additional methods of calculating a hydrostatic bias are: (1) employing the gas fraction, $f_{\text{gas}}$, (2a) comparing hydrostatic masses (from X-ray data) to total masses as calculated by weak lensing, (2b) comparing hydrostatic masses (from SZ data) to calibrated total masses; i.e. some scaling of the SZ data with either a lensing calibration \citep[e.g.,][abbreviated as H21]{hilton2021} or via abundance matching \citep[e.g.,][abbreviated as B15, H16, Bu19, and B19, respectively]{bleem2015,deHaan2016,bulbul2019,bocquet2019}.

    \begin{figure}[!h]
        \begin{center}
            \includegraphics[width=0.45\textwidth]{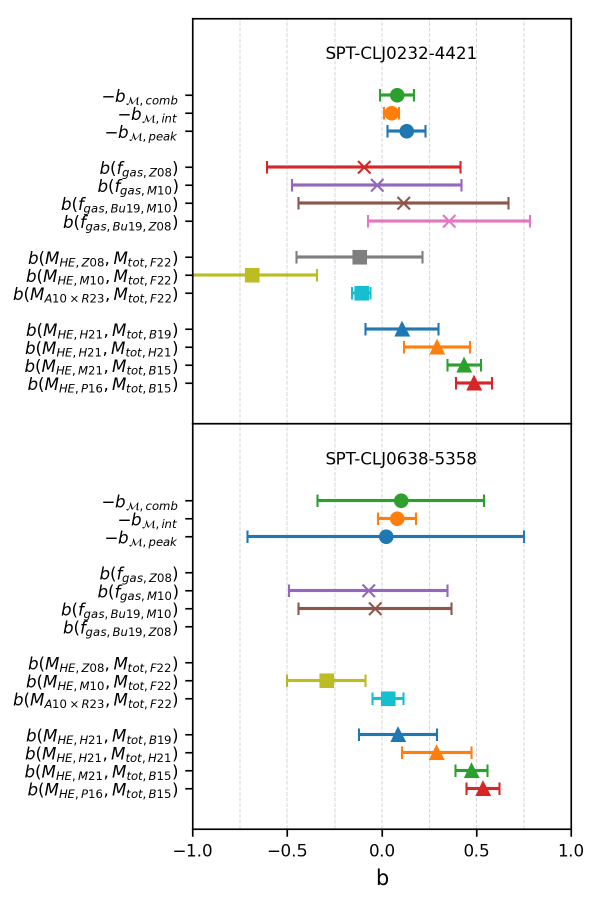}
        \end{center}
        \caption{Hydrostatic Mass Biases calculated via various methods for SPT-CLJ0232-4421 (top) and SPT-CLJ0638-5358 (bottom). We remind the reader that the sign convention for hydrostatic mass biases calculated via our fluctuation analyses ($b_{\mathcal{M}}$ is opposite that of the more common $b = 1 - M_{\rm HE}/M_{\rm tot}$.) Biases are grouped by method; the top group are those derived from this work and estimation from spectra of surface brightness fluctuations. The second grouping is derived from the gas fraction. The third grouping shows biases when comparing HE mass estimates to total (WL) mass estimates from \citet{fox2022}. The final grouping shows HE mass estimates relative to cosmological abundance-matched total masses. Citation abbreviations (indicated in the text) reflect corresponding references for HE and total (or abundance-matched) masses in the third and fourth groupings, or $M_{\rm gas}$ and $M_{\rm HE}$, respectively for biases from $f_{\rm gas}$. The mass estimate subscripted with A10$\times$R23 is a HE mass estimate described below and in Appendix~\ref{appendix:mass_estimates}.}
    \label{fig:HydMassBiases}
    \end{figure}
    
    In the case of deriving a hydrostatic mass bias from $f_{\text{gas}}$, we employ the relation presented in \citet{allen2008} and \citet{wicker2023} which give a theoretical expectation for the gas fraction:
    \begin{equation}
        f_{\text{gas,Theory}} = K \frac{\Upsilon(M,z)}{1-b} A(z) \left( \frac{\Omega_b}{\Omega_m} \right) \left( \frac{D_{\rm A}^{\rm ref}(z)}{D_{\rm A}(z)} \right)^{3/2} - f_{*},
    \end{equation}
    where $K$ is an instrumental calibration factor, $\Upsilon(M,z)$ is the baryon depletion factor, $b$ is the hydrostatic mass bias ($b = 1 - M_{\rm HE}/M_{\rm tot}$), $A(z)$ is an angular correction, $\Omega_b / \Omega_m$ is the universal baryon fraction, $D_{\rm A}$ is the angular diameter distance% (in a test cosmology versus a reference cosmology)
    , and $f_{*}$ is the stellar mass fraction. As we are not comparing different cosmologies (and the works from which we derive our masses assume the same cosmology), the terms $A(z)$ and $D_{\rm A}^{\rm ref}(z)/D_{\rm A}(z)$ can be taken as unity. We follow \citet{wicker2023} and take $K = 1 \pm 0.1$, $\Omega_b / \Omega_m = 0.156 \pm 0.03$ \citep{planck2020_VI}, $\Upsilon(M,z) = \Upsilon_0 = 0.85 \pm 0.03$ \citep{planelles2013}, and $f_{*} = 0.015 \pm 0.005$ \citep{eckert2019}. Mass biases derived from $f_{\rm gas}$ in Figure~\ref{fig:HydMassBiases} make use of $M_{\rm gas}$ and $M_{\rm HE}$ either from a single paper (if both gas mass and HE mass are reported in that paper), or a gas mass in one paper, cited first, and a HE mass from the second paper)

    We choose this method to derive a hydrostatic mass bias from $f_{\text{gas}}$, as opposed to the method employed in \citet{eckert2019}, because it requires no estimate of non-thermal pressure and its derivative. Here again, assumptions would need to be made (assume a non-thermal pressure profile as given in \citet{nelson2014a}) or we would need to incorporate values from our fluctuation analysis, but then the resulting hydrostatic mass bias would not be independent of our fluctuation analysis.

    The remaining method(s) of hydrostatic mass bias is simply using its definition ($b = 1 - M_{\rm HE}/M_{\rm tot}$) and varying what we take as a hydrostatic mass estimate and what we take as a total mass estimate. In one set, we take predominantly X-ray derived masses and compare to mass estimates from weak lensing \citep{fox2022}. In this category, we include an estimate of $M_{500}$ by fixing the pressure profile shape to the Universal Pressure Profile (UPP) shape found in \citet{arnaud2010} and fitting the profile to the SPT data, allowing only $M_{500}$ as a free parameter, which artificially constrains the uncertainty (see Appendix~\ref{appendix:mass_estimates}).  Finally, we also compare mass estimates from SZ surveys which are derived, in basis, from the assumption of hydrostatic equilibrium to mass estimates from SZ surveys where the mass is estimated from either weak lensing calibration or abundance matching. To avoid exploring all possible permutations, Figure~\ref{fig:HydMassBiases} shows only the lowest and highest biases one can calculate from comparable hydrostatic and total mass estimates for an assumed cosmology. This is discussed further in Appendix~\ref{appendix:mass_estimates}. Remaining citation abbreviations are Z08 and M10 which refer to \citet{zhang2008} and \citet{mantz2010b}, respectively.

    Figure~\ref{fig:HydMassBiases} then compares hydrostatic mass biases as calculated by our fluctuation analyses ($-b_{\mathcal{M}}$) to the other methods laid out in this section. We see the expected clustering of hydrostatic mass biases from SZ+HE-to SZ-abundance-matched between $0.25 < b < 0.55$. That is, mass biases when the total mass is matched to be consistent with a cosmology tend towards higher bias values ($b$), such as that found in \citet{Planck2016_XXIV} ($1 - b = 0.58 \pm 0.04$; work abbreviated as P16). By comparison, the other methods find bias values much closer to zero (for these two clusters), with a scattering of positive and negative values. Thus, beyond the statistical agreement of our fluctuation-derived hydrostatic mass bias values (in part due to large uncertainties, especially for SPT-CLJ0638-5358), the values themselves are easily within the range of values one can derive from mass estimates for individual clusters in the literature. Thus, a more significant comparison will arise when comparing the distribution of hydrostatic mass biases derived from fluctuations over the full SPT and \textit{XMM} sample to the distribution from other methods of estimating the hydrostatic mass bias.

    %OK, so far I've found that (1) the masses in the Planck catalog a "XMM-like" and (2) the SPT masses (being abundance-matched) are inherently "total".  Laura Salvatti's work effectively recalibrated the Planck catalog based on the SPT masses? It seems that Melin's PSZSPT catalog did the reverse; they scaled SPT masses to fit in the Planck's framework. Meanwhile, Matt Hilton's paper approaches the mass estimation independently. The calibrated masses are calibrated based on a richness-weak lensing scaling relation. Ostensibly they have richnesses for each cluster (certainly the two clusters of interest here). 

\section{Impact of choices on results}
\label{sec:Choices}

    Every step in analyzing surface brightness fluctuations (and those of their corresponding deprojected thermodynamic quantities) presents choices. We have addressed major algorithmic choices in corresponding analysis sections. Here we address choices (alternatively, assumptions) about the underlying morphology of the ICM and distribution of its fluctuations. In this section we consider the impact of (1) the assumed geometry for the surface brightness profiles and (2) which regions we use in calculating the spectra of our fluctuations.

    \subsection{Choice of profile fitting parameters}
    \label{sec:ProfileChoices}

        \begin{figure}[]
            \begin{center}
                \includegraphics[width=0.45\textwidth]{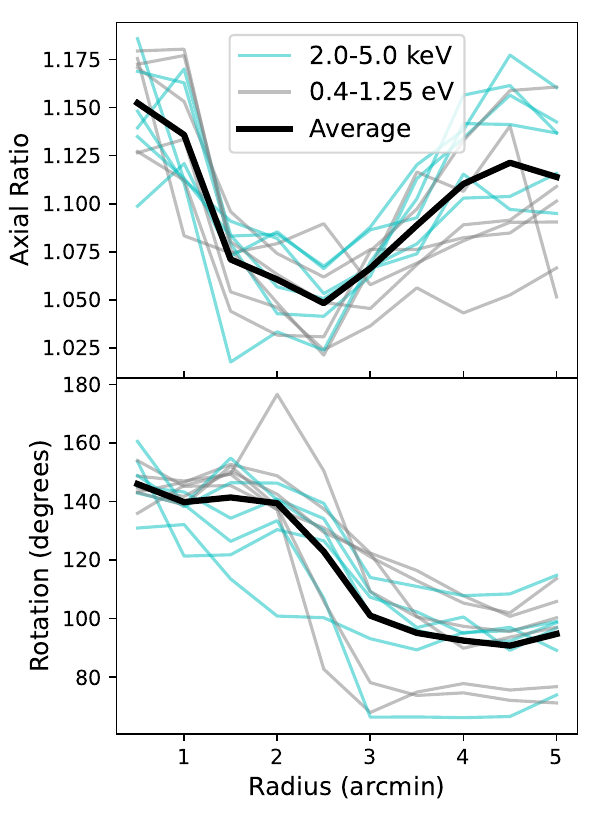}
                \includegraphics[width=0.45\textwidth]{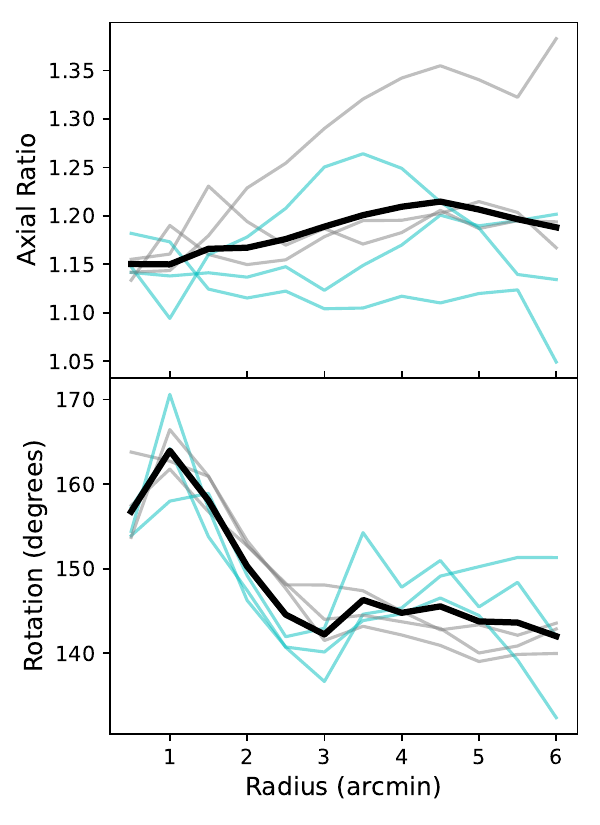}
            \end{center}
            \caption{Axial ratio and orientation versus radius for SPT-CLJ0232-4421 (left) and SPT-CLJ0638-5358 (right) as a function of fitting radius about the cluster centroid. The lines in all subplots are as indicated in the legend of the upper left plot. Neither the axis ratio nor rotation angle is constant.}
            \label{fig:EllipticityParams}
        \end{figure}
        
        For our surface brightness profiles we adopted circular $\beta$-models for both the SZ and X-ray images where we fixed the centers to be the X-ray centroid. Each of these choices ([1] choosing to fit a $\beta$-model to each image, [2] adopting circular symmetry, and [3] using the X-ray centroid) have been justified earlier in this article. Given that the $\beta$-model appears quite sufficient for both datasets, we do not explore the range of other parameterizations available in the literature. However, we do explore the impact of our center choice and elliptical geometry on the resultant X-ray power spectra.

        \begin{figure}[]
            \begin{center}
                \includegraphics[width=0.45\textwidth]{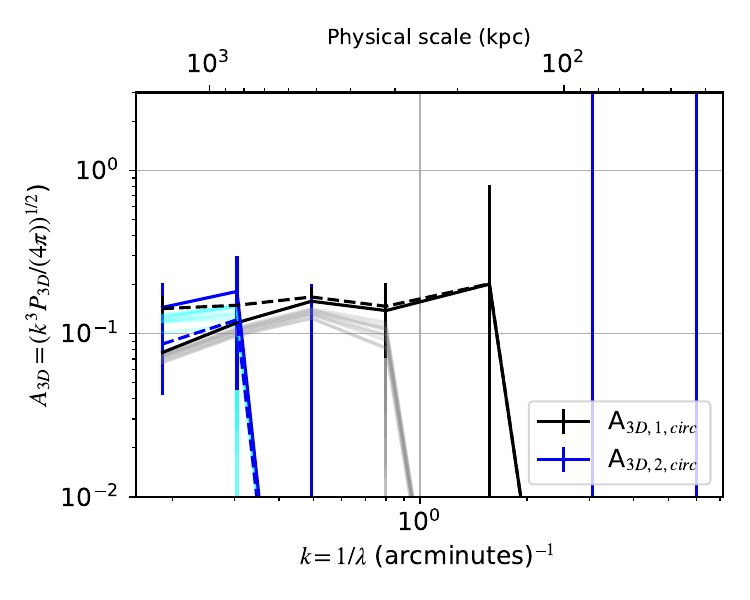}
                \includegraphics[width=0.45\textwidth]{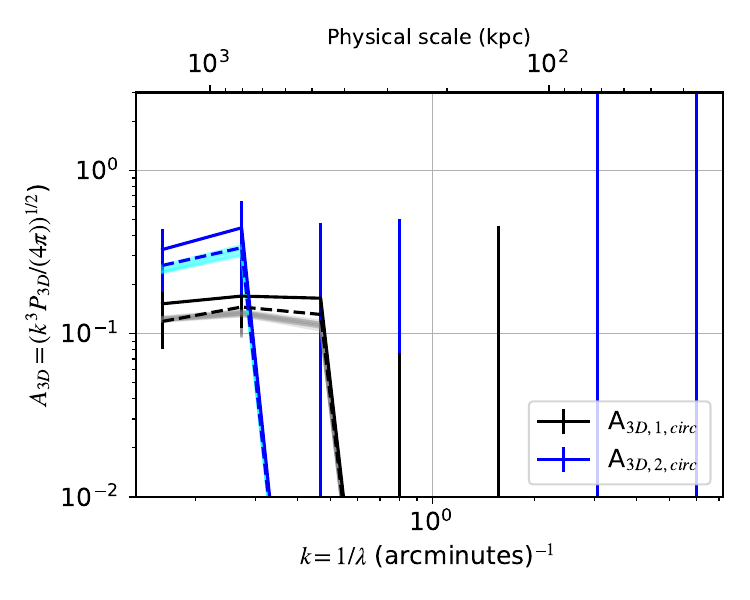}
            \end{center}
            \caption{Recovered 3D magnitude spectra for SPT-CLJ0232-4421 (left) and SPT-CLJ0638-5358. The solid black and blue curves are our reference (circular model centered on the centroid) model, the dashed curves correspond to circular fits about the X-ray peak, and the solid grey and cyan curves are for the various elliptical fits (about the centroid). The black and grey curves correspond to the inner Ring (i.e. circle) and the blue and cyan curves correspond to the outer Ring (annulus).}
            \label{fig:MS_vs_ellipticity}
        \end{figure}        

        \begin{figure}[]
            \begin{center}
                \includegraphics[width=0.45\textwidth]{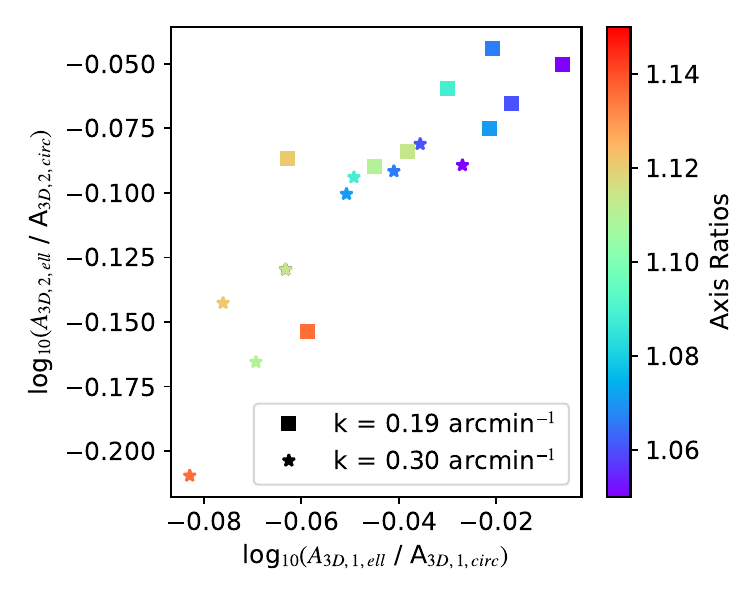}
                \includegraphics[width=0.45\textwidth]{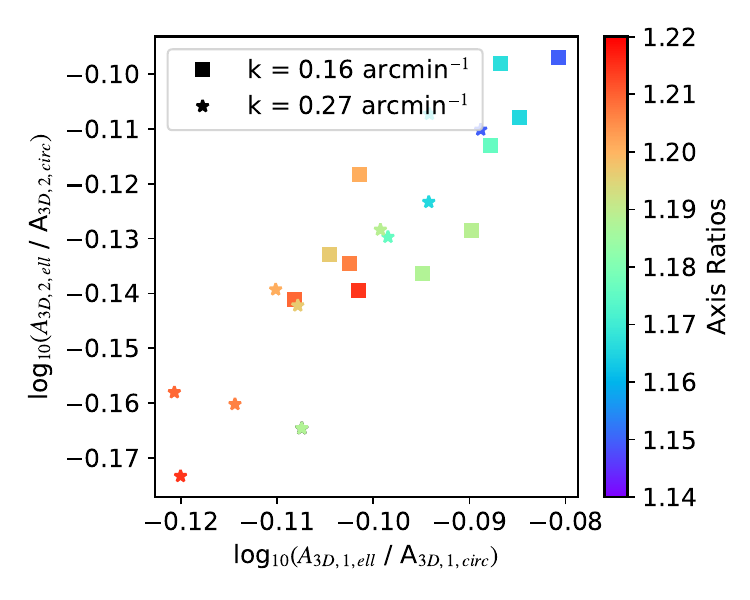}
            \end{center}
            \caption{A comparison of the recovered magnitude spectra in Ring 2 (ordinate) relative to the recovered magnitude spectra in Ring 1 (abscissa) for the two lowest wavenumbers (squares and stars; see legends) color-coded by axial ratio.}
            \label{fig:DeltaMS_vs_ell}
        \end{figure}

        For the center choice we consider the X-ray peak as an alternative to the centroid. Regarding an elliptical model, we have multiple choices as there is no singular choice of ellipse as illustrated in Figure~\ref{fig:EllipticityParams}. The axial ratios and rotation angles are calculated via \lstinline{pyproffit}\footnote{This, again, accounts for point source masking and the exposure map. Rotation angles are the degrees counterclockwise from west (such that 90$^{\circ}$ is north). The fitting of the elliptical parameters uses all (none-masked) pixels in the defined circle.} within a circle of radius $r$ about the centroid for each image. We vary $r$ in steps of 0.5 arcmin out to $R_{500}$.

        \begin{table*}[] 
   \centering  
   \begin{tabular}{c | c c c c c | c c c c c} 
                  & \multicolumn{4}{c}{SPT-CLJ0232-4421} & \multicolumn{4}{c}{SPT-CLJ0638-5358} \\
    $R_{\rm ell}$ & $a/b$ & $\theta_R$ & Ring & $-b_{\mathcal{M,\text{peak}}}$ & $-b_{\mathcal{M,\text{int}}}$ & $a/b$ & $\theta_R$ & Ring & $-b_{\mathcal{M},\text{peak}}$ & $-b_{\mathcal{M,\text{int}}}$ \\ 
   \hline 

\multirow{2}{*}{$\phi = \sqrt{3}$} & \multirow{2}{*}{1} & \multirow{2}{*}{0} & Ring 1 & $0.21 \pm 0.39$ & $0.25 \pm 0.20$ & 
\multirow{2}{*}{1} & \multirow{2}{*}{0} & Ring 1 & $-0.20 \pm 0.23$ & $-0.02 \pm 0.04$ \\ 
 & & & Ring 2 & $0.25 \pm 0.17$ & $0.06 \pm 0.04$ & & & Ring 2 & $0.03 \pm 0.38$ & $0.05 \pm 0.06$ \\ 

\multirow{2}{*}{$\phi = (1 + \sqrt{5})/2$} & \multirow{2}{*}{1} & \multirow{2}{*}{0} & Ring 1 & $0.34 \pm 0.20$ & $0.26 \pm 0.26$ & 
\multirow{2}{*}{1} & \multirow{2}{*}{0} & Ring 1 & $-0.42 \pm 0.65$ & $-0.01 \pm 0.09$ \\ 
 & & & Ring 2 & $0.13 \pm 0.10$ & $0.05 \pm 0.04$ &  & & Ring 2 & $0.02 \pm 0.73$ & $0.08 \pm 0.10$ \\ 

\multirow{2}{*}{$\phi = \sqrt{2}$} & \multirow{2}{*}{1} & \multirow{2}{*}{0} & Ring 1 & -- & $0.15 \pm 0.13$ &  \multirow{2}{*}{1} & \multirow{2}{*}{0} & Ring 1 & -- & $0.10 \pm 0.17$ \\ 
 & & & Ring 2 & --  & $0.04 \pm 0.04$ & & & Ring 2 & -- & $0.03 \pm 0.09$ \\ 

\multirow{2}{*}{Peak} & \multirow{2}{*}{1} & \multirow{2}{*}{0} & Ring 1 & -- & $0.42 \pm 0.25$ & 
\multirow{2}{*}{1} & \multirow{2}{*}{0} & Ring 1 & $-0.21 \pm 0.34$ & $-0.00 \pm 0.06$ \\ 
 & & & Ring 2 & -- & $0.03 \pm 0.02$ & & & Ring 2 & $0.10 \pm 0.38$ & $0.06 \pm 0.07$ \\ 

\hline 

\multirow{2}{*}{1.0} & \multirow{2}{*}{1.14} & \multirow{2}{*}{140} & Ring 1 & $0.28 \pm 0.19$ &  $0.07 \pm 0.06$ & 
\multirow{2}{*}{1.15} & \multirow{2}{*}{164} & Ring 1 & $-0.22 \pm 0.34$ & $-0.01 \pm 0.07$ \\ 
 & & & Ring 2 & $0.07 \pm 0.08$ & $0.01 \pm 0.02$ &  & & Ring 2 & $0.05 \pm 0.45$ & $0.06 \pm 0.08$ \\ 

\multirow{2}{*}{1.5} & \multirow{2}{*}{1.07} & \multirow{2}{*}{141} & Ring 1 & $0.28 \pm 0.19$ & $0.07 \pm 0.06$ & 
\multirow{2}{*}{1.17} & \multirow{2}{*}{158} & Ring 1 & $-0.20 \pm 0.30$ & $-0.01 \pm 0.07$ \\ 
 & & & Ring 2 & $0.09 \pm 0.08$ & $0.02 \pm 0.02$ & & & Ring 2 & $0.06 \pm 0.40$ & $0.06 \pm 0.08$ \\ 

\multirow{2}{*}{2.0} & \multirow{2}{*}{1.06} & \multirow{2}{*}{139} & Ring 1 & $0.30 \pm 0.19$  & $0.07 \pm 0.06$ & 
\multirow{2}{*}{1.17} & \multirow{2}{*}{150} & Ring 1 & $-0.22 \pm 0.30$ & $-0.01 \pm 0.06$ \\
 & & & Ring 2 & $0.10 \pm 0.08$ & $0.02 \pm 0.02$ & & & Ring 2 & $0.04 \pm 0.43$ & $0.06 \pm 0.07$ \\ 

\multirow{2}{*}{2.5} & \multirow{2}{*}{1.05} & \multirow{2}{*}{123} & Ring 1 & $0.30 \pm 0.20$ & $0.08 \pm 0.07$ & 
\multirow{2}{*}{1.18} & \multirow{2}{*}{145} & Ring 1 & $-0.20 \pm 0.29$ & $-0.01 \pm 0.06$ \\ 
 & & & Ring 2 & $0.11 \pm 0.09$ & $0.02 \pm 0.02$ & & & Ring 2 & $0.06 \pm 0.39$ & $0.06 \pm 0.07$ \\ 

%%%%%%%%%%%%%%%%%%%%%%%%%%%%%%%
\multirow{2}{*}{3.0} & \multirow{2}{*}{1.07} & \multirow{2}{*}{101} & Ring 1 & $0.27 \pm 0.18$ & $0.07 \pm 0.06$ & 
\multirow{2}{*}{1.19} & \multirow{2}{*}{142} & Ring 1 & $-0.20 \pm 0.29$ & $-0.01 \pm 0.07$ \\ 
 & & & Ring 2 & $0.11 \pm 0.08$ & $0.02 \pm 0.02$ & & & Ring 2 & $0.06 \pm 0.39$ & $0.06 \pm 0.08$ \\ 

\multirow{2}{*}{3.5} & \multirow{2}{*}{1.09} & \multirow{2}{*}{95} & Ring 1 & $0.27 \pm 0.19$ &  $0.06 \pm 0.06$ & 
\multirow{2}{*}{1.20} & \multirow{2}{*}{146} & Ring 1 & $-0.19 \pm 0.28$ & $-0.01 \pm 0.06$ \\  
 & & & Ring 2 & $0.10 \pm 0.08$ &  $0.02 \pm 0.02$ & & & Ring 2 & $0.06 \pm 0.38$ & $0.06 \pm 0.07$ \\ 

\multirow{2}{*}{4.0} & \multirow{2}{*}{1.11} & \multirow{2}{*}{92} & Ring 1 & $0.27 \pm 0.19$ & $0.07 \pm 0.06$ & 
\multirow{2}{*}{1.21} & \multirow{2}{*}{145} & Ring 1 & $-0.17 \pm 0.27$ & $-0.01 \pm 0.07$  \\
 & & & Ring 2 & $0.09 \pm 0.08$ & $0.02 \pm 0.02$ & & & Ring 2 & $0.06 \pm 0.37$ & $0.06 \pm 0.08$ \\ 

\multirow{2}{*}{4.5} & \multirow{2}{*}{1.12} & \multirow{2}{*}{91} & Ring 1 & $0.23 \pm 0.18$ & $0.05 \pm 0.05$ & 
\multirow{2}{*}{1.21} & \multirow{2}{*}{146} & Ring 1 & $-0.16 \pm 0.27$ & $-0.00 \pm 0.07$ \\
 & & & Ring 2 & $0.09 \pm 0.08$ & $0.02 \pm 0.02$ & & & Ring 2 & $0.08 \pm 0.36$ & $0.06 \pm 0.09$ \\ 

\multirow{2}{*}{5.0} & \multirow{2}{*}{1.11} & \multirow{2}{*}{95} & Ring 1 & $0.26 \pm 0.18$ & $0.06 \pm 0.06$ & 
\multirow{2}{*}{1.21} & \multirow{2}{*}{144} & Ring 1 & $-0.18 \pm 0.29$ & $-0.01 \pm 0.08$ \\ 
 & & & Ring 2 & $0.09 \pm 0.08$ & $0.02 \pm 0.02$ & & & Ring 2 & $0.06 \pm 0.38$ & $0.06 \pm 0.09$ \\ 

\multirow{2}{*}{5.5} & \multirow{2}{*}{--} & \multirow{2}{*}{--} & \multirow{2}{*}{--} & \multirow{2}{*}{--} & 
\multirow{2}{*}{--} & \multirow{2}{*}{1.20} & \multirow{2}{*}{144} & Ring 1 & $-0.18 \pm 0.29$ & $-0.01 \pm 0.07$ \\ 
& & & & & & & & Ring 2 & $0.06 \pm 0.38$ & $0.06 \pm 0.08$ \\ 

\multirow{2}{*}{6.0} & \multirow{2}{*}{--} & \multirow{2}{*}{--} & \multirow{2}{*}{--} & \multirow{2}{*}{--} & 
\multirow{2}{*}{--} & \multirow{2}{*}{1.19} & \multirow{2}{*}{142} & Ring 1 & $-0.16 \pm 0.30$ & $-0.00 \pm 0.08$ \\ 
& & & & & & & & Ring 2 & $0.09 \pm 0.38$ & $0.06 \pm 0.09$ \\ 

\hline
   \end{tabular} 
   \caption{Hydrostatic mass biases as derived for different ellipticities. $R_{\rm ell}$ is the radius at which ellipticity parameters are determined (as in Figure~\ref{fig:EllipticityParams}). $a/b$ is then the axis ratio and $\theta_R$ is the rotation angle (in degrees). The first four rows show results from circular cases and either different ring extents or different center; entries for $\phi = (1 + \sqrt{5})/2$ are the reprint of the hydrostatic biases in Table~\ref{tbl:bms}. All numbers for elliptical fits use $\phi = (1 + \sqrt{5})/2$ to define the ring extents. Empty entries for SPT-CLJ0638-5358 arise from a lack of nodes (in Ring 2) about $2\sigma$; empty entries for SPT-CLJ0232-4421 reflect that $R_{500} = 5.^{\prime}3$ and thus elliptical parameters are not pursued beyond $R_{500}$.}
   \label{tbl:bms_ell_both}
\end{table*} 

        For each cluster, we average the elliptical parameters at each radius from 1 arcminute out to $R_{500}$. We then follow the same procedures as in Section~\ref{sec:Xray_analysis}: we fit profiles extracted from each image (where we adopt the average parameters of the ellipse(s) across EPIC cameras with the center taken to be the circular centroid used in Section~\ref{sec:Xray_analysis}), create fractional residual images, and calculate power spectra within the same regions (rings) as before. For these elliptical models, we assume the cluster is a prolate ellipsoid with its major axis in the plane of the sky. We further assume that the deprojection can be calculated with a weighted average of $N$ (defined in Equation~\ref{eq:window_approx}) based on the semi-major axis (rather than circular radius; see Appendix~\ref{sec:appendix_deproj} for further details.)

        We find that the choice of center and ellipticity will impact the recovered 3D amplitude spectra as evidenced in Figure~\ref{fig:MS_vs_ellipticity}. Accounting for ellipticity does reduce the recovered power spectra as expected, and Figure~\ref{fig:DeltaMS_vs_ell} shows a trend that the larger the axial ratio used, the more reduced the amplitudes. The reduction in amplitudes is not uniform for the two clusters; that is, we cannot write a single equation relating the reduction in amplitude (at a given wavenumber) to the axial ratio that holds for both clusters. Furthermore, we note that, for a given cluster, the shapes of the amplitude spectra are quite similar among the circular or elliptical profiles fitted. That is, this inferred injection scale does not appear to be sensitive to the geometry used. However, these conclusions are drawn from a sample of merely two clusters which are not grossly elongated. This may extend to other clusters which are not particularly elongated, though we may anticipate that for larger axial ratios the impact is more severe and we should expect it to modulate the shape of the amplitude spectra, especially at lower $k$.

        We briefly digress into a discussion about the impact of elongation along the line of sight on the projection of power spectra, in particular in the case that an axis is aligned with the line of sight, then we can simply scale $W$ (from Equations~\ref{eqn:Wsz} and \ref{eqn:Wx}; we omit subscripts as the same scaling applies to both window functions) and $z$. In particular, let $z^{\prime} = z * c$, i.e. the cluster is elongated by a factor $c$ along the line of sight. In this case, $W^{\prime}(\theta,z^{\prime}) = W(\theta,z)/c$. Their respective Fourier transforms yield the relation $\tilde{W}^{\prime}(k_z/c) = \tilde{W}(k_z)$.
        %$W^{\prime}(\theta,z^{\prime}) = W(\theta,z)/c$
        %Their respective Fourier transforms yield the relation $\tilde{W}^{\prime}(c k_z) = \tilde{W}(k_z)$. 
        Integrating the square of $|\tilde{W}^{\prime}(k_z)|$ over $d k_z$ (Equation~\ref{eqn:window_approx}), it is evident that $N^{\prime}$ (the approximate scaling between $P_{\rm2D}$ and $P_{\rm3D}$ in the case of elongation along the line of sight) is reduced by a factor of $c$ relative to $N$. The relative bias between the full projection along the line of sight (Equation~\ref{eqn:deproj}) and the approximate equation (Equation~\ref{eqn:deproj_approx}) as shown in Appendix~\ref{sec:deproj_validity} remains unchanged with elongation. In short, the primary effect of elongation along the line of sight would be to boost the recovered $P_{\rm3D}$ by a factor of $c$. If we consider the axial ratios (Table~\ref{tbl:bms_ell_both}) in the plane of the sky serve as a naive upper-bound for $c$, we find that the recover $A_{\rm3D}$ will at most be boosted by $\sim10$\%, and thus we suspect that elongation along the line of sight is not a significant concern for these two clusters.
        
        For the circular case, the impact of center choice (peak vs. centroid) is mixed for Ring 1, but yields reduced amplitudes in Ring 2. Note that choosing the center to be the peak will also shift (translate) the center of the rings so that the spectra are taken from slightly different regions of the sky. All spectra from elliptical fits are derived from the same regions as the circular centroid case of $\phi = (1 + \sqrt{5})/2$.

        One of the main goal of this work is also to derive hydrostatic mass biases, so we proceed from our 3D spectra as in Section~\ref{sec:PS_results} and derive hydrostatic biases for the elliptical cases. To do this, we have adopted the same equivalences as in section~\ref{sec:PS_results} between the Mach number and the peak of the amplitude spectra and the integrated power spectrum (i.e. variance). Moreover, we use the same logarithmic pressure slopes as in the circular case (See section~\ref{sec:PS_results} and Table~\ref{tbl:log_slopes}). We maintain the $2\sigma$ threshold when selecting a peak, while $\langle \eta_{\rho} \rangle$ takes values of 1.0 and 1.3 for ellipsoidal geometry for in-between and unrelaxed clusters, respectively. In principle, the formulation for total pressure balance with gravitational potential should be reworked in each ellipsoidal coordinates. Though, we also lack a convention for assigning $R_{500}$ for an ellipsoidal cluster (i.e. do we still take some sphere, or do we pick one of the axes as our length metric?). Notwithstanding these caveats, our calculations suggest that even with reduced amplitude spectra (and hence reduced Mach numbers), there is less of a clear trend in hydrostatic mass biases. For SPT-CLJ0232, the elliptical results all suggest a lower hydrostatic bias; this is not true for SPT-CLJ0638-5358.

        \begin{figure*}[]
            \begin{center}
                \includegraphics[width=0.95\textwidth]{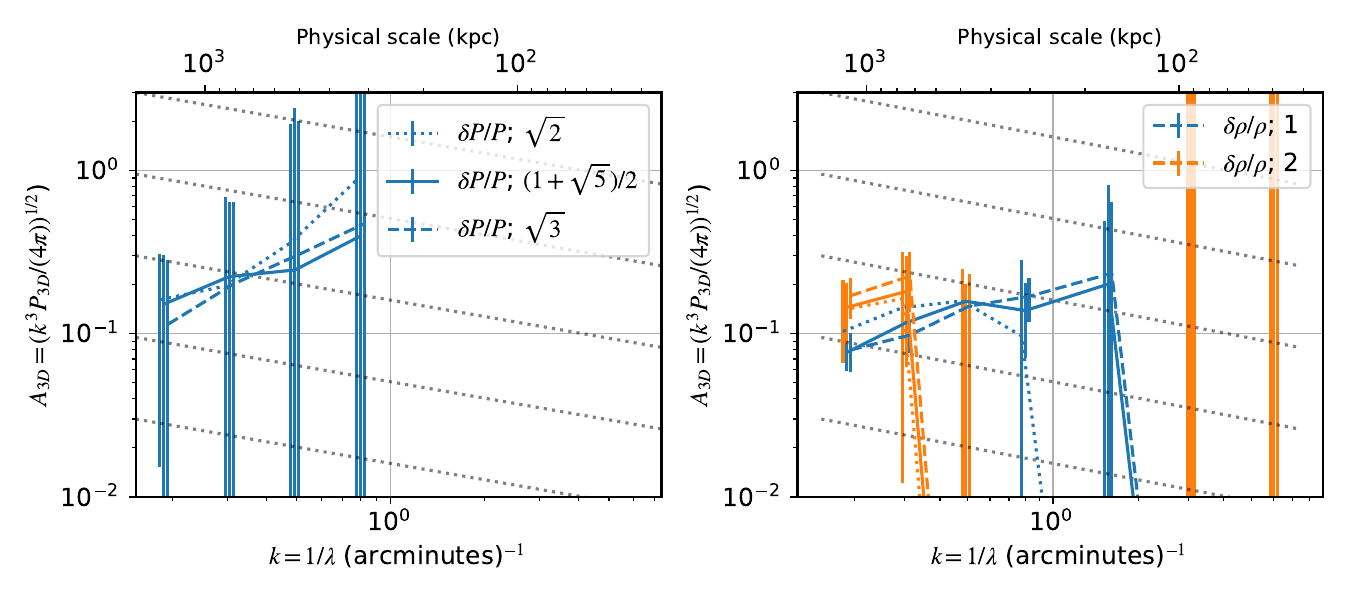}
                \includegraphics[width=0.95\textwidth]{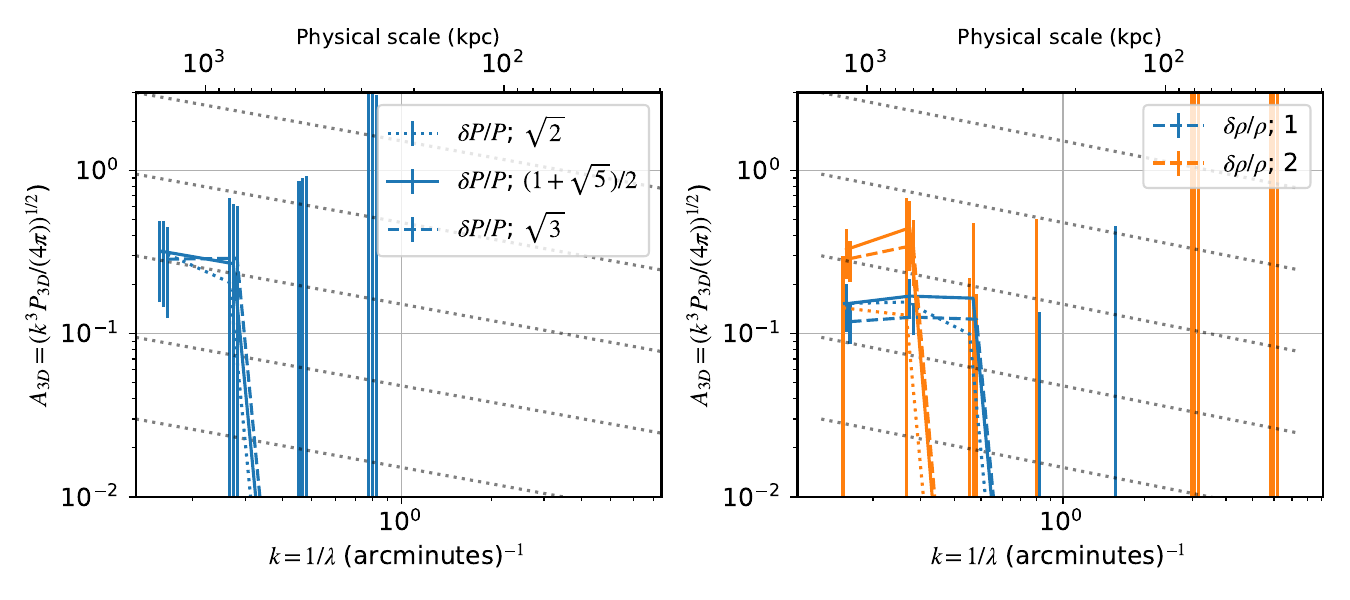}
            \end{center}
            \caption{A comparison of the recovered magnitude spectra as a function of the inner circle radius, noted in the legend of the left panel as the divisor of $R_{500}$. The same convention relating divisors and linestyles is used in the right panel. \textbf{Top:} Amplitude spectra for SPT-CLJ0232-4421; \textbf{left}: amplitude spectra of pressure fluctuations; \textbf{right:} amplitude spectra of density fluctuations. \textbf{Bottom:} Amplitude spectra for SPT-CLJ0638-5358 \textbf{left}: amplitude spectra of pressure fluctuations; \textbf{right:} amplitude spectra of density fluctuations.}
            \label{fig:MS_vs_AR}
        \end{figure*}        

        It is hopefully evident that when considering elliptical models, the parameter space is immediately inflated: geometrically one must choose which ellipticity is appropriate and what to assume about the third ellipsoidal axis and inclination. Similar to previous discussions \citep[e.g.,][]{Zhuravleva2015,romero2023}, it's unclear what would be correct, though some choice is taken in \citet{zhang2023,dupourque2023}, where we note \citet{zhang2023} models a circular $\beta$-model for the core and uses an elliptical $\beta$-model for the broader emission. There are further considerations in terms of deprojection, and deriving a hydrostatic mass bias for an ellipsoidal potential and how to define $R_{500}$ in this framework. Certainly, for our sample we conclude that adopting a spherical cluster model centered on the X-ray surface brightness centroid is the most appropriate choice. 

    \subsection{Choice of regions for spectra}
    \label{Sec:SpectralChoices}

        There are many factors to consider in choosing the regions within which to calculate the power spectra. Some obvious limitations come about from the depth of observations (i.e. the noise in the images) and the angular resolution of the instrument(s) used. Similarly, if one wants to constrain fluctuations on large scales, then the region ought to be large enough to well sample those scales.

        In our case, we wish to measure fluctuations out to $R_{500}$, both in the radial sense and in the Fourier sense. Our desire to sample scales close to $R_{500}$ motivates an inner circle of at least $r = R_{500}/2$. Conversely, the resolution of the instruments can also set a minimum width of annuli (1.\arcmin25 in our case). Even with these bounds in place, for the two clusters here, there are a range of options for two annuli (i.e. inner circle and outer annulus) that fit these criteria, though these criteria effectively rule out using three (or more) annuli.

        A simple geometric argument would be to divide the area equally between the inner circle and outer annulus. While this meets the above criteria, we again note that noise should be considered. That is, the noise for spectral calculations is not the map noise but the map noise divided by the surface brightness model. Thus, regions at larger radii will have larger intrinsic noise.

        From here we see the choice of regions as a vague endeavor. To limit the parameter space searched, we consider three options. We define $R_1$ as the radius of the inner circle (equally the inner radius of our outer annulus) with $R_1 = R_{500}/ \phi$; as such we try three values of $\phi$: [$\sqrt{2}$,$(1 + \sqrt{5})/2$,$\sqrt{3}$]. Figure~\ref{fig:MS_vs_AR} shows that indeed the region choice is mostly inconsequential. However, for SPT-CLJ0638-5358, we do see that there is a significant difference in the amplitude spectra of Ring 2, where $\phi = \sqrt{2}$ results in a lower spectrum than the other two values of $\phi$. This shows that there is a residual feature between $\phi = \sqrt{2}$ and $\phi = (1 + \sqrt{5})/2$ that is relevant. In particular, we infer that this feature is the plausible shock feature(s) to the northeast of the cluster center (highlighted in Figure~\ref{fig:residualImages}). The impact of $\phi$ on the resultant hydrostatic mass bias is difficult to infer with some values missing due to data quality (see Table~\ref{tbl:bms_ell_both}). The values presented in Table~\ref{tbl:bms_ell_both} suggest that this choice is not particularly important, especially for these two clusters.

\section{Conclusions}
\label{sec:conclusions}

    In this work we have outlined the methodology to study surface brightness fluctuations of both SZ and X-ray data that will be applied to a sample of SPT-selected galaxy clusters which have archival \textit{XMM-Newton} data. Our goal is to constrain the turbulent properties in the ICM. Given the quality of the data that we analyze, several specific goals arise: constraining the amplitude of fluctuations, inferring turbulent Mach numbers, and deriving a hydrostatic mass bias from the Mach numbers. To pilot this methodology we have studied two massive ($M_{500} \approx 1 \times 10^{15}$ M$_{\odot}$) clusters at redshifts $0.2 < z < 0.3$. A shock is known to exist in one system \citep[SPT-CLJ0638-5358;][]{botteon2018}, but the direct impact of that shock on the fluctuations is subdominant relative to other fluctuations in the corresponding region. For our baseline results, we adopt a circular surface brightness model (spherical ICM model) and subsequently we explore the impact of using elliptical surface brightness models.
    
    The amplitude of fluctuations we recover in these two clusters is larger than those seen in several other studies \citep[e.g.,][]{churazov2012,Gaspari2013_PS,Zhuravleva2015,hofmann2016,arevalo2016,eckert2017,hernandez2023}, but in line with other studies which probe similar physical scales \citep[e.g.,][]{khatri2016,dupourque2023}. We estimate Mach numbers for each ring based on three methods (scaling relations): (1) relating the peak of the amplitude spectrum to a Mach number, (2) relating the integrated power spectrum to a Mach number, and (3) combining the two previous methods by taking the average of their respectively derived Mach numbers. The Mach numbers inferred from the two separate scaling relations are generally in agreement, except for Ring 2 of SPT-CLJ0638-5358 which has a high peak in its amplitude spectrum of density fluctuations and a comparatively low integrated power spectrum of density fluctuations.
    
    %To estimate the hydrostatic mass bias we opt to derive amplitude spectra of surface brightness fluctuations in two regions (rings), where we calculate Mach numbers in each region and associated logarithmic slopes (with respect to radius) of the Mach numbers. 

    We calculate hydrostatic biases for each set of Mach numbers (i.e. for each method of deriving a Mach number) and find similar bias values across the three methods (as calculated for Ring 2) in each cluster, which we interpret as the hydrostatic bias for $M_{500}$ of each cluster. While the two clusters studied here, SPT-CLJ0232-4421 and SPT-CLJ0638-5358, are ostensibly dynamically relaxed and disturbed, respectively, they both have fairly low hydrostatic biases. Moreover, the disturbed system has the lower hydrostatic bias. The disturbed cluster has greater non-thermal pressure than the relaxed cluster, as inferred from the larger Mach numbers. However, the steep logarithmic slope of the Mach numbers in the disturbed cluster yields a lower hydrostatic bias. If indeed the hydrostatic mass bias is low, especially for SPT-CLJ0638-5358, we interpret this as an endorsement of the notion that hydrostatic mass estimates can themselves be (transiently) boosted from a corresponding (transient) boost in the underlying SZ or X-ray signal. 

    The inferred hydrostatic mass biases were relatively insensitive to the extent of Ring 1 relative to Ring 2, with the exception of SPT-CLJ0638-5358, where toggling the ring extents shows that prominent fluctuations appear beyond $R_{500} / \sqrt{2}$, notably in the northwestern quadrant. When fitting elliptical models, our inferred fluctuations decrease by $5-40$\%. However, our hydrostatic mass bias estimates did not change in a corresponding fashion. This is due to the fact that the logarithmic slope of the Mach numbers did not change dramatically and it plays a critical role in modulating the derived value of the hydrostatic mass bias. This may suggest that the hydrostatic mass bias can be robustly calculated across different elliptical geometries assumed, though this needs to be assessed across a broader sample, as will be done in our study of the full sample of SPT-selected clusters with archival \textit{XMM-Newton} data.
    %Hydrostatic mass biases for $M_{500}$ of two respective clusters are determined to be $0.14 \pm 0.07$ and $0.03 \pm 0.41$ \MG{I really don't understand why such a low value for the disturbed system!}. That is, our dynamically disturbed cluster has a bias near to zero (though with a large uncertainty). This small hydrostatic mass bias might reflect the briefly-lived and disproportionate increase in Compton $Y$, relative to $M$, during a merger \citep[e.g.,][]{wik2008,krause2012}.

    %\MG{To be expanded with more results}

\begin{acknowledgments}

  Charles Romero is supported by NASA ADAP grant 80NSSC19K0574 and Chandra grant G08-19117X. E. Bulbul acknowledges financial support from the European Research Council (ERC) Consolidator Grant under the European Union’s Horizon 2020 research and innovation program (grant agreement CoG DarkQuest No 101002585). MG acknowledges partial support by HST GO-15890.020/023-A, the \textit{BlackHoleWeather} program, and NASA HEC Pleiades (SMD-1726). RK acknowledges support from the Smithsonian Institution, the Chandra High Resolution Camera Project through NASA contract NAS8-03060, and NASA Grants 80NSSC19K0116, GO1-22132X, and GO9-20109X. PN was supported by NASA contract NAS8-03060. The authors thank the anonymous referee for comments.

\end{acknowledgments}

\vspace{5mm}
\facilities{SPT, \textit{XMM-Newton}}

\software{astropy \citep{astropy2013,astropy2018,astropy2022},  emcee \citep{foreman2013}, pyproffit\citep{eckert2017}, ESAS \citep{snowden2008}
          }
\bibliography{references}{}
\bibliographystyle{aasjournal}

\appendix

\section{The importance of modelling a cool core}
\label{sec:appendix_cc}

    \begin{figure}[!h]
        \begin{center}
            \includegraphics[width=0.47\textwidth]{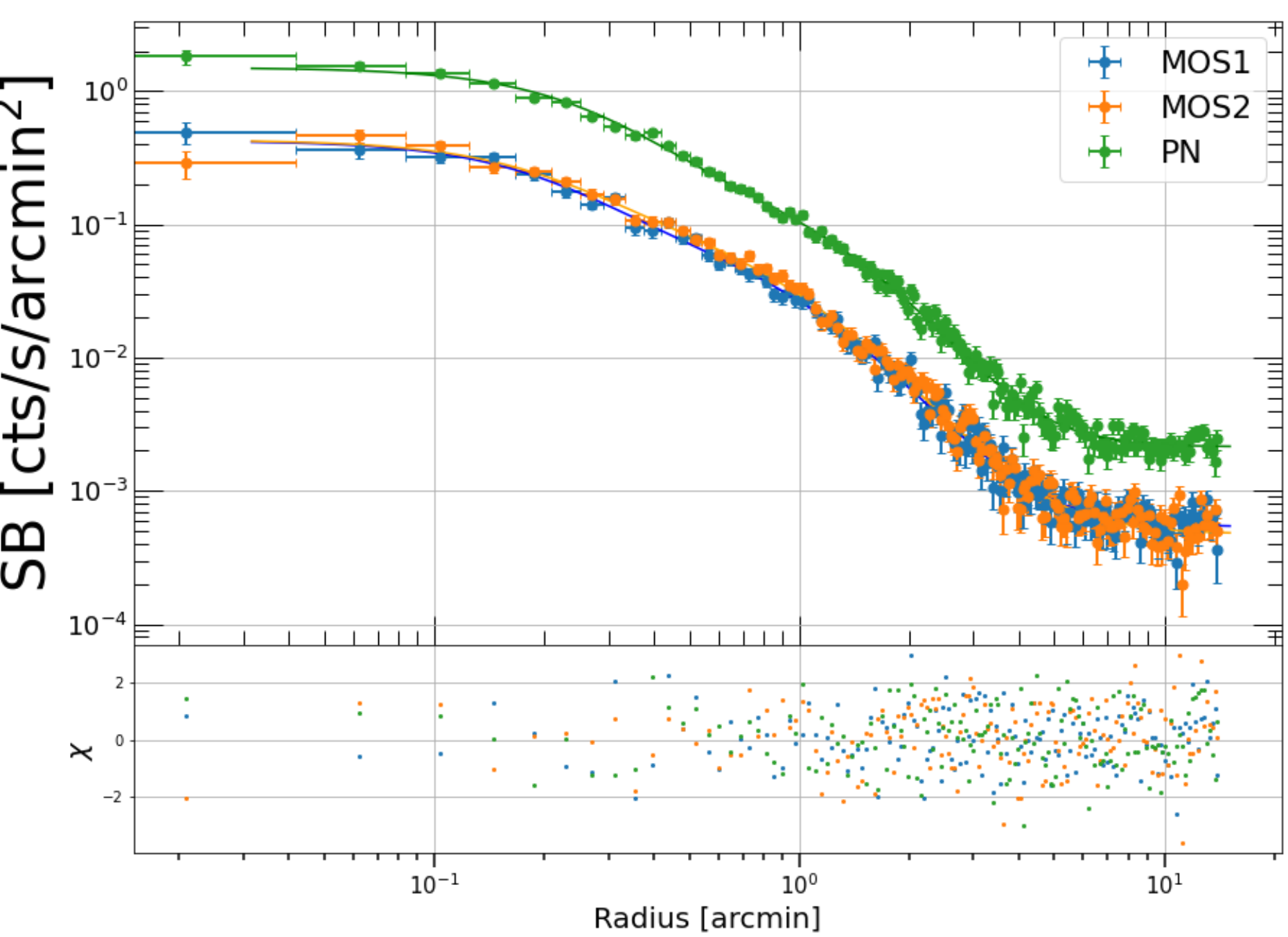}
            \includegraphics[width=0.47\textwidth]{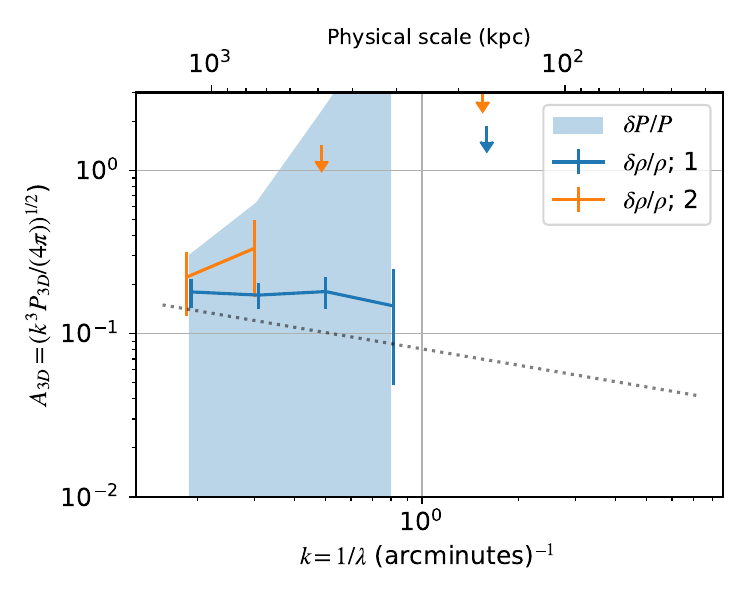}
        \end{center}
        \caption{\textbf{Left:} Surface brightness profiles for SPT-CLJ0232-4421 by CCD for ObsID 0042340301 in the 0.4-1.25 keV band. \textbf{Right:} Resultant amplitude spectra when adopting a double $\beta$-model; cf. Figure~\ref{fig:SZvsXraySpectra}.}
        \label{fig:DoubleBetaResults}
    \end{figure}

    In Section~\ref{sec:Xray_analysis}, we note that the $\beta$-model leaves residuals in the radial profile of SPT-CLJ0232-4421 owing to its cool-core. Here, we investigate the impact on the resultant spectra when adopting a model with more free parameters. In particular, we adopt a double $\beta$-model as parameterized in \lstinline{pyproffit}:
    \begin{equation}
        S(r) = S_0 [(1+(r/r_{c,1})^2)^{-3 \beta + 0.5} +  R(1+(r/r_{c,2})^2)^{-3 \beta + 0.5}] + B,
        \label{eqn:double_beta}
    \end{equation}
    where $S_0$, $\beta$, and $B$ are as in Equation~\ref{eqn:XR_beta}, and if we make a restriction that $r_{c,2} < r_{c,1}$, then we may consider $r_{c,2}$ to be the characteristic radius of the cool-core (and $r_{c,1}$ could be considered the extended core radius). The $R$ parameter allows for a ratio between the normalizations of the two $\beta$-models. Note, the model makes no restriction of which $r_{c}$ is smaller; the choice is arbitrary.

       We perform the profile fits as in Section~\ref{sec:Xray_analysis} and find that the residuals, as seen in residual radial profiles (Figure~\ref{fig:DoubleBetaResults}), are notably reduced. While Figure~\ref{fig:DoubleBetaResults} shows just profiles for ObsID 0042340301 in the 0.4-1.25 keV band, the results (especially the residuals) are very similar for both ObsIDs and both energy bands. We note that $\beta \sim 0.8$ across the CCDs, ObsIDs, and two energy bands with the double $\beta$-model. In comparison, the $\beta$ values were $\sim 0.56$ when fitting the single $\beta$ model. Although we do not wish to scrutinize the fitted parameter values, we highlight the larger $\beta$ value as it indicates steeper slopes, which is highly relevant. The lower panel in Figure~\ref{fig:DoubleBetaResults} shows the resultant amplitude spectra (comparable to Figure~\ref{fig:SZvsXraySpectra}) when using the double $\beta$-model for SPT-CLJ0232-4421 (the SZ-derived spectrum remains unchanged). Table~\ref{tbl:dbl-beta-ps_products} clearly quantifies the relevant spectral parameters; when comparing with Table~\ref{tbl:ps_products}, it is clear that both rings show an increase in fluctuations. How can this be?

    \begin{table}[!h] 
        \centering  
        \begin{tabular}{c | c c c c c} 
            & A$_{\text{3D}}(k_{\rm peak})$ & $\sigma_{\text{3D}}$ & $\sigma_{\ln}$ & $k_{\text{peak}}$ & $\lambda_{\text{peak}}$ (kpc) \\ 
        \hline 
        Ring 1 & $0.18 \pm 0.04$ & $0.21 \pm 0.03$ & $0.20 \pm 0.03$ & 0.49 & 516 \\ 
        Ring 2 & $0.33 \pm 0.17$ & $0.19 \pm 0.09$ & $0.19 \pm 0.10$ & 0.30 & 837
        \end{tabular} 
        \caption{Key properties of the amplitude spectra in SPT-CLJ0232-4421 when surface brightness profile model is a double $\beta$-model. The $k_{\text{peak}}$ column is in units of inverse arcminutes.}  
        \label{tbl:dbl-beta-ps_products}
    \end{table} 

\begin{figure}[!h]
        \begin{center}
            \includegraphics[width=0.47\textwidth]{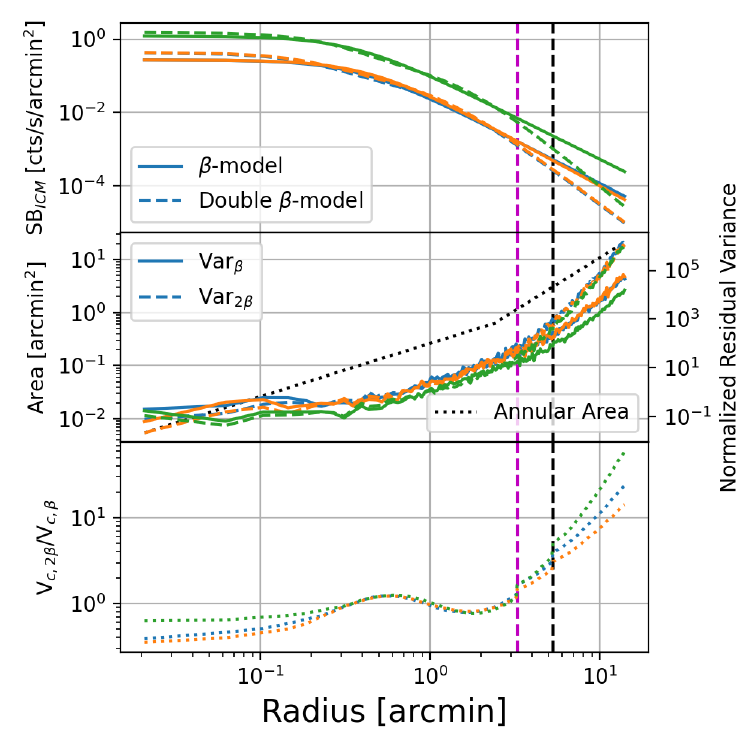}
        \end{center}
        \caption{Profiles from SPT-CLJ0232-4421. In all panels, the blue, orange, green coloring corresponds to MOS1, MOS2, and PN CCDs (as in Figure~\ref{fig:DoubleBetaResults}). \textbf{Top:} A comparison of single $\beta$-model vs. double $\beta$-model surface brightness profiles without the contribution from the background. \textbf{Middle:} Normalized variance within annuli, i.e. $Var(\delta S/\bar{S})$ within the same annuli used to derive the observed surface brightness profiles. Var$_{2\beta} = Var(\delta S/\bar{S})$ for the case of $\bar{S}$ being defined by the double $\beta$-model, and Var$_{\beta}$ is the corresponding normalized variance for the single $\beta$-model. The black curve denotes the area per annulus. \textbf{Bottom:} The comparison of the cumulative, area-weighted, variance of of the normalized residuals from the double $\beta$-model relative to that of the single $\beta$-model. The magenta dashed line denotes the division between Ring1 and Ring2; the black dashed line denotes $R_{500}$ (and the outer edge of Ring2).}
    \label{fig:WaitButWhy}
    \end{figure}

    The larger $\beta$ value noted earlier is the fundamental reason for this increase in fluctuations. Of course, it is not the sole culprit; the large region plays a role as well. Figure~\ref{fig:WaitButWhy} shows several profiles which help to elucidate the mechanisms at play. The top panel shows the profiles of the surface brightness profiles with the background subtracted. This is the profile by which the residual image is divided to obtain the normalized fluctuation image. Consequentially, for equal fluctuations ($\delta S$) at large radii, the normalized fluctuations ($\delta S/\bar{S}$) will be larger for the profile with the steeper slope. This is seen in the middle panel, where the dashed lines (corresponding to the normalized fluctuations for the double $\beta$-model. The bottom panel compares the cumulative sum of the product of the colored curves and the black curve in the middle panel. That is, the cumulative area-weighted, normalized variance serves as a proxy for the power in the fluctuations. This proxy is more reflective of the power at smaller scales. Although this proxy does not fully reveal how the power changes at difference scales, it nicely demonstrates how an excess in power can occur despite the (unnormalized) residual profile (falsely) suggesting smaller fluctuations.

    \begin{table}[!h] 
        \centering  
        \begin{tabular}{c | c c c} 
         & $-b_{\mathcal{M},\text{peak}}$ & $-b_{\mathcal{M},\text{int}}$ & $-b_{\mathcal{M},\text{comb}}$ \\ 
        \hline 
        Ring 1 & $-0.41 \pm 1.30$ & $0.12 \pm 0.19$ & $-0.06 \pm 0.70$ \\ 
        Ring 2 & $0.30 \pm 0.28$ & $0.09 \pm 0.09$ & $0.20 \pm 0.29$ \\ 
        \end{tabular} 
        \caption{Derived hydrostatic biases when using a double $\beta$-model for SPT-CLJ0232-4421.}  
        \label{tbl:dbl-beta-bms}
    \end{table} 

    If we continue our analysis to the derivation of hydrostatic biases, we find that the adoption of a double $\beta$-model makes a more substantial difference than the elliptical considerations if looking at the recovered values and ignore the uncertainties. However, the uncertainties are still large and thus statistically we should not be consider the differences significant.

\section{Deprojection Details}
\label{sec:appendix_deproj}

    We note two concerns related to the deprojection of our 2D spectra to 3D: (1) how do we deal with the dependence of the Window function on projected cluster-centric radius, $\theta$ and (2) how valid is the approximation (given in Equation~\ref{eqn:deproj_approx}) to Equation~\ref{eqn:deproj}.

    \subsection{Averaging within a region}

\begin{figure}[!h]
            \begin{center}
                \includegraphics[width=0.47\textwidth]{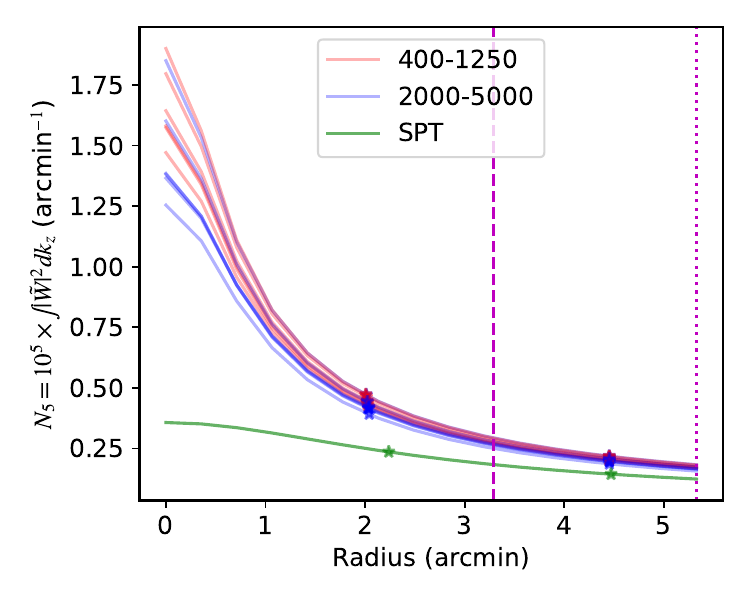}
                \includegraphics[width=0.47\textwidth]{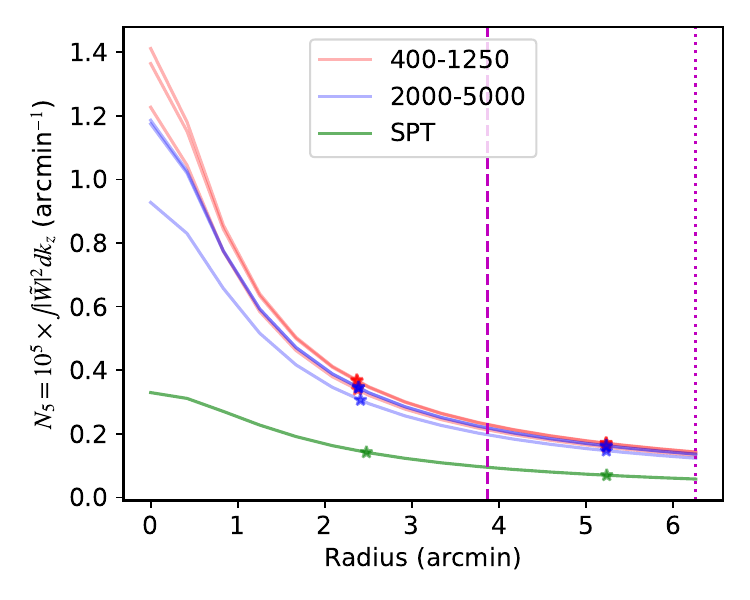}
            \end{center}
            \caption{\textbf{Left:} Integrated window functions, $N(\theta)$ and weighted averages within rings for SPT-CLJ0232-4421. \textbf{Right:} Same as left, but for SPT-CLJ0638-5358. For the X-ray data $N(\theta)$ is plotted for each CCD and each ObsID by energy band indicated as by their range in eV.}
            \label{fig:WeightedN}
        \end{figure}

        If we, for the moment, take the approximation given by Equation~\ref{eqn:deproj_approx} to be valid, we still have the issue that $N(\theta)$ is a function of $\theta$ (see Figure~\ref{fig:WeightedN}) and we have a range of values of $\theta$ within our annuli. We posit that the appropriate way to average $N(\theta)$ is to weight the values by area subtended by an annulus of radius $\theta$ and width $d\theta$. In practice, we sample 16 equally spaced lines of sight in the range $[0,R_{500}]$ and weight by area between these bins, given the line of sight at $\theta=0$ zero weight. For each ring we calculate an effective $N$ (shown as stars in Figure~\ref{fig:WeightedN}; hereafter $N_{\text{eff}}$) as the weighted average of $N(\theta)$ as above for those $\theta$ which lie within the ring.

        It is clear that $N$ can vary substantially from the inner edge of an annulus (or $R=0$ for the inner circle) to the outer edge of a respective annulus. As noted in \citet{Zhuravleva2015}, the maximal differences can be used to set a maximal uncertainty in $N_{\text{eff}}$. 
    
   %     Let us consider the extreme values of $N(\theta)$ within a ring relative to $N_{\text{eff,i}}$. 
   %     
   % \begin{table}[!h] 
   %     \centering  
   %     \begin{tabular}{c c | c c c c} 
   %         Dataset & $\alpha$ & \multicolumn{4}{c}{B$_\text{A}(k_{500})$} \\
   %          & & \multicolumn{2}{c}{SPT-CLJ 0232-4432} & \multicolumn{2}{c}{SPT-CLJ 0638-5531} \\
   %          & & $\sigma_{-}$ & $\sigma_{+}$ & $\sigma_{+}$ & $\sigma_{-}$ \\
   %         %  & & B$_\text{A,SZ}(k_{500}) & B$_\text{A,XR}(k_{500}) \\ 
   %         \hline 
   %         \multirow{2}{*}{x-ray}  & Ring 1 & 25 & 260 & 25 & 250 \\
   %                                 & Ring 2 & 16 & 35  & 16 & 24 \\
   %         \multirow{2}{*}{SZ}     & Ring 1 & 21 & 51  & 13 & 31 \\
   %                                 & Ring 2 & 14 & 20  & 17 & 26 \\
   %         \hline
   %     \end{tabular} 
   %     \caption{Bias values at $k_{500}$ }  
   %     \label{tbl:NsigmaUL}
   % \end{table} 

   % [[2.57513129 0.24548282]  [0.35469053 0.163336  ]]
   %   <> <> <> <>  SZ N's  <> <> <> <> 
   %   [0.51417016 0.20733784] [0.20237281 0.14163787]

    \subsection{Validity of approximation}
    \label{sec:deproj_validity}

\begin{figure}[!h]
        \begin{center}
            \includegraphics[width=0.47\textwidth]{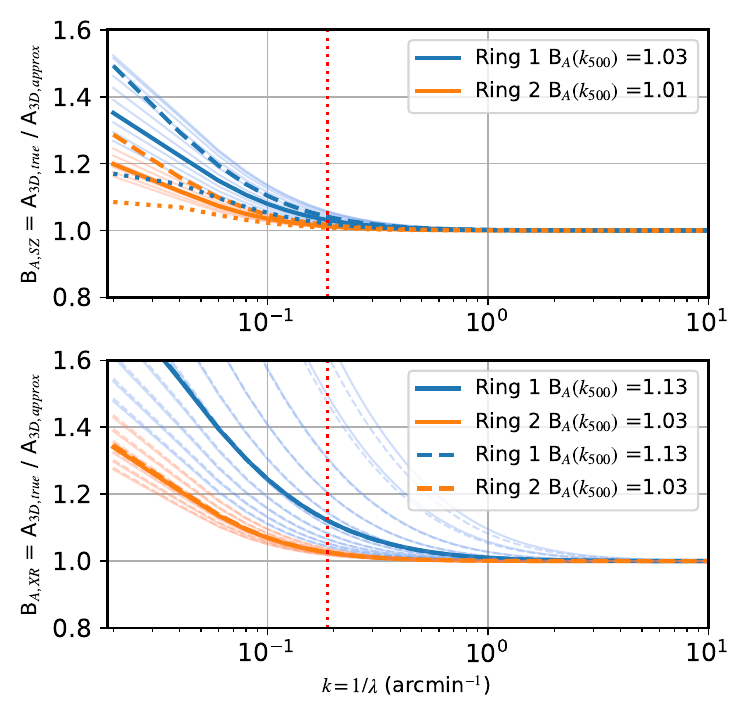}
            \includegraphics[width=0.47\textwidth]{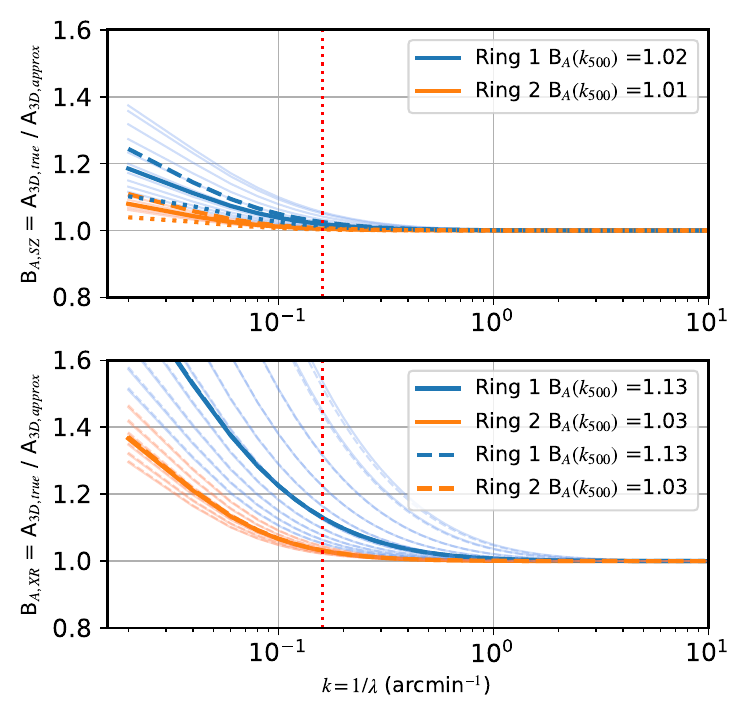}
        \end{center}
        \caption{\textbf{Left:} Bias in magnitude spectra for SPT-CLJ0232-4421; top panel is with respect to the SZ (SPT) window functions and bottom panel is with respect to X-ray (\textit{XMM-Newton}) window functions. \textbf{Right:} Same as left, but for SPT-CLJ0638-5358. In all panels the thin and faint lines are for a specific line-of-sight and assuming an underlying 3D spectral index of $\alpha = 3$. The thick and darker lines are the weighted averages within their respective rings. For the top (SZ) plots, the solid thick lines are the weighted averages with an underlying 3D spectral index of $\alpha = 3$; the dotted line corresponds to $\alpha = 2$ and the dashed corresponds to $\alpha = 4$. In the bottom panels the solid and dashed lines correspond to our LE (0.4-1.25 keV) and HE (2.0 - 5.0 keV) bands respectively and are nearly identical. The legend indicates the bias values at scales of $R_{500}$, which is marked in all plots as the red dotted vertical line.}
    \label{fig:WindowApprox}
    \end{figure}

        To test the validity of Equation~\ref{eqn:deproj_approx} we can perform the full integration of Equation~\ref{eqn:deproj} with an assumed 3D power spectrum, which we take to be a power law with a cutoff at $k_c$:
        \begin{equation}
            P_{\text{3D}} = P_0 e^{k_c/k} k^{-\alpha},
            \label{eqn:PL_Wcutoff}
        \end{equation}
        where $P_0$ is the normalization (for our purposes, this is arbitrary), and $\alpha$ is the spectral index. We take $k_c = 1/(5 R_{500})$, which is well away from the largest scales that we sample. As discussed in Section~\ref{sec:PS_analysis}, the slope $\alpha = 3$ is of notable concern to deriving the peak in the magnitude spectra, but we should also be concerned with slopes shallower and steeper than that. Figure~\ref{fig:WindowApprox} shows the bias curves for SZ and X-ray when assuming $\alpha = 3$.

    %\textcolor{red}{These corrections assuming $\alpha=3$ are not applied. At most they matter at the 20\% level in Ring 1. }

    \begin{table}[!h] 
        \centering  
        \begin{tabular}{c c | c c c c} 
            Dataset & $\alpha$ & \multicolumn{4}{c}{B$_{\text{A}}(k_{500})$} \\
            \hline
             & & \multicolumn{2}{c}{SPT-CLJ0232-4432} & \multicolumn{2}{c}{SPT-CLJ0638-5531} \\
             & & Ring 1 & Ring 2 & Ring 1 & Ring 2 \\
            %  & & B$_\text{A,SZ}(k_{500}) & B$_\text{A,XR}(k_{500}) \\ 
            \hline 
            \multirow{3}{*}{SZ}    & 2 & 1.02 & 1.01 & 1.01 & 1.00 \\
                                    & 3 & 1.03 & 1.01 & 1.02 & 1.01 \\
                                    & 4 & 1.04 & 1.02 & 1.03 & 1.01 \\
            \hline
            \multirow{3}{*}{X-ray} & 2 & 1.09 & 1.02 & 1.09 & 1.02 \\
                                  & 3 & 1.13 & 1.03 & 1.13 & 1.03 \\
                                  & 4 & 1.16 & 1.04 & 1.16 & 1.04 \\                       
            \hline
            %\multirow{3}{*}{SZ}    & 2 & 1.01 & 1.02 & 1.00 & 1.01 \\
            %                        & 3 & 1.01 & 1.03 & 1.01 & 1.02 \\
            %                        & 4 & 1.02 & 1.04 & 1.01 & 1.03 \\
            %\hline
            %\multirow{3}{*}{X-ray} & 2 & 1.02 & 1.09 & 1.02 & 1.09 \\
            %                      & 3 & 1.03 & 1.13 & 1.03 & 1.13 \\
            %                      & 4 & 1.04 & 1.16 & 1.04 & 1.16 \\                       
            %\hline
            \end{tabular} 
        \caption{Bias values at $k_{500}$ }  
        \label{tbl:BA_k500}
    \end{table}

\section{Masking effects}
\label{sec:appendix_masking}

    \begin{figure}[!h]
        \begin{center}
            \includegraphics[width=0.47\textwidth]{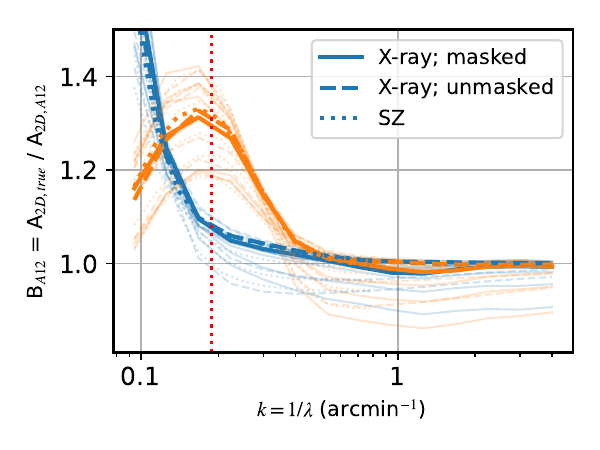}
            \includegraphics[width=0.47\textwidth]{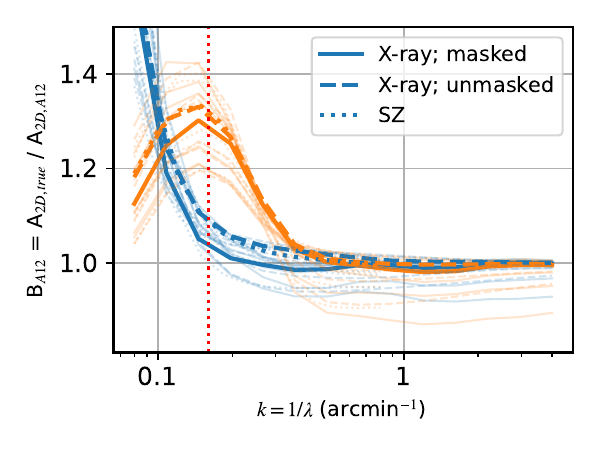}
        \end{center}
        \caption{Colors indicate Rings as in Figure ~\ref{fig:WindowApprox}. \textbf{Left:} Bias in magnitude spectra for SPT-CLJ0232-4421 due to masking of annuli in all cases. For the X-ray lines, the legend indicates if point sources and chip gaps are also masked, "masked", or not: "unmasked". \textbf{Right:} Same as left, but for SPT-CLJ0638-5358. In all panels the thin and faint lines are for spectra indices $2 <= \alpha <= 4$; the bold lines indicate $\alpha = 3$. The scale (inverse) of $R_{500}$ is indicated by the red dotted vertical line.}
    \label{fig:AnnularBiasIndices}
    \end{figure}
    
    The $\Delta$-variance method presented in \citet{arevalo2012}, like other $\Delta$-variance methods, is intended to deal with gaps in an image and thus is a major motivation for using this method on the \textit{XMM-Newton} data which has chip gaps and requires masking point sources. For simpler masks (e.g., a circular mask for our inner ring) an alternative is to apodize the circular mask \citep[e.g][]{koch2019}, though there remains the choice of apodization and whatever choice is made, there is still an effect on the recovered power spectrum, principally in normalization\footnote{See, for example \url{https://turbustat.readthedocs.io/en/latest/tutorials/applying_apodizing_functions.html}}. Apodization can similarly be applied to annular masks and in all cases the effect should be controlled for the specific mask and apodization choice for an assumed underlying power spectrum or power spectra. Apodization is not a panacea and ultimately the inclusion of chip gaps and point sources would become too much for this approach.

    \begin{figure}[!h]
        \begin{center}
            \includegraphics[width=0.47\textwidth]{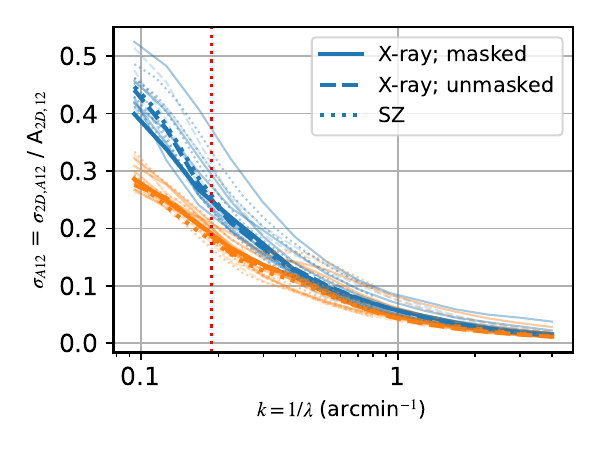}
            \includegraphics[width=0.47\textwidth]{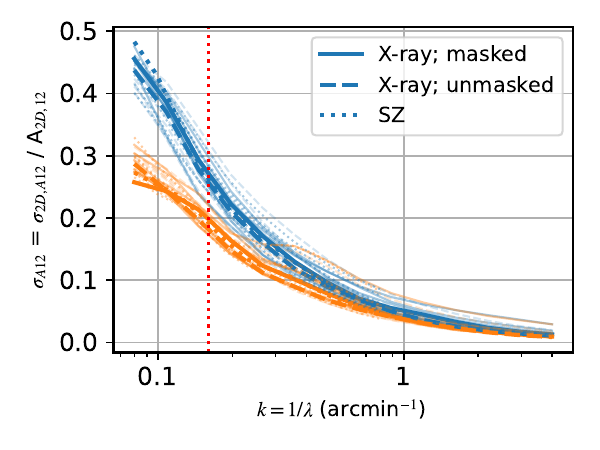}
        \end{center}
        \caption{Colors indicate Rings as in Figure ~\ref{fig:WindowApprox}. \textbf{Left:} Sample variance, or rather standard deviation $\sigma$, in magnitude spectra for SPT-CLJ0232-4421 due to masking of annuli in all cases. For the X-ray lines, the legend indicates if point sources and chip gaps are also masked, "masked", or not: "unmasked". \textbf{Right:} Same as left, but for SPT-CLJ0638-5358. In all panels the thin and faint lines are for spectra indices $2 <= \alpha <= 4$; the bold lines indicate $\alpha = 3$. The scale (inverse) of $R_{500}$ is indicated by the red dotted vertical line.}
    \label{fig:AnnularDispIndices}
    \end{figure}
    
    Thus, we return again to the $\Delta$-variance method presented in \citet{arevalo2012} and opt to use this for both our \textit{XMM-Newton} and SPT datasets. Again, while it is designed to work with arbitrary masks, and was shown in \citet{arevalo2012} to recover the power spectra shape very well for arbitrary masks for a range of spectral indices, we attempt to account for any biases in the recovered spectra due to the masking. For instance, it should not be surprising that if a region (mask) does not cover a large spatial scale, then the recovered power spectrum will be underestimated at that scale (this can be seen in Appendix B of \citet{romero2023}). Here, we check our ability to recover power spectra of varying spectral indices, maintaining $1/k_c = 5 R_{500}$, as in the previous section, given our choice of annuli by generating image realizations described by the various power spectra. We further test the recovery of these power spectra when we include chip gaps and point source masking. Though we might expect the scale recovery to be independent of pixel size, we decided to test with pixelizations corresponding to those in the X-ray and SZ datasets.

    \begin{figure}[!h]
        \begin{center}
            \includegraphics[width=0.47\textwidth]{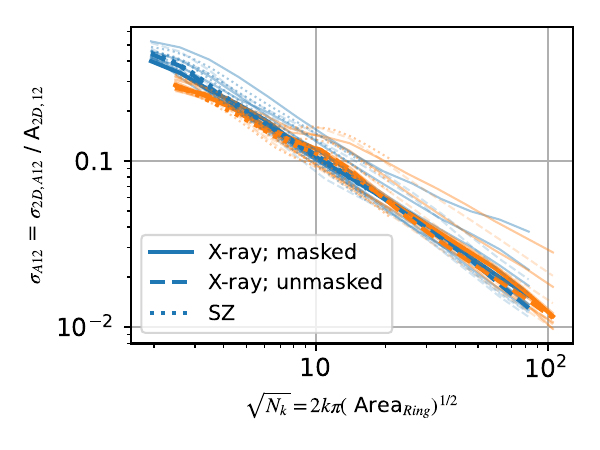}
            \includegraphics[width=0.47\textwidth]{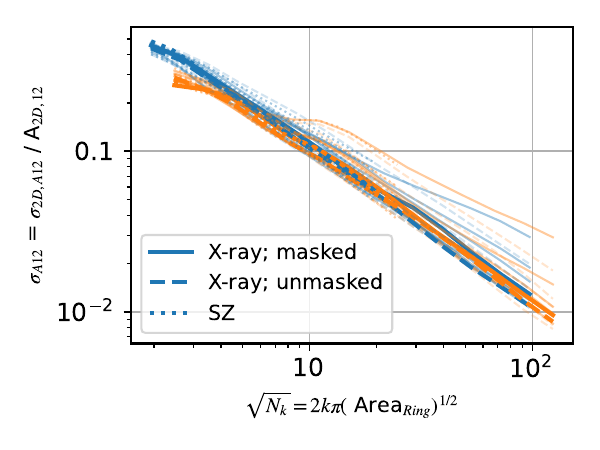}
        \end{center}
        \caption{Rather than scaling the dispersion against scale, we scale against $\sqrt{N_k}$, where $N_k$ is the number of sampling elements (if taken as a square with length $1/(2 \pi k)$). This definition of $N_k$ provides the relation: $\sigma_{\text{A12}} = 1 / \sqrt{N_k}$.}
    \label{fig:AnnularDispNsamp}
    \end{figure}
    
    It should be noted that for a given realization for a given scale $l = 1/k$ which may be ``fully" sampled, as in there is at least one circle of diameter $l$, sample variance will be notable. Indeed, even when a scale is sampled by several circles of corresponding diameter there will be some variance. For this reason, we iterate over 200 realizations, sampling to larger scales than reported for our data in Section~\ref{sec:PS_analysis}. The bias is taken as the average recovered power spectra divided by the input power spectra and accounting for the known spectral index normalization bias \citep[][]{arevalo2012}. Prior to spectral measurements we do not smooth our images by the instrument PSF; thus we do not need to be concerned with beam corrections \citep{romero2023}. For a given input spectrum the sample dispersion ($\sigma$) is taken as the standard deviation of the spectral measurements across realizations and normalized by the input power spectrum at its normalization bias. We report biases (B$_{\text{A12}}$; see Figure~\ref{fig:AnnularBiasIndices}) and dispersions ($\sigma_{\text{A12}}$\footnote{Note that true sample variance is independent of the method of power spectrum estimation used. However, the reported dispersion is itself measured with the A12 method, so we give it this subscript.}; see Figure~\ref{fig:AnnularDispIndices} relative to the (average) recovered magnitude spectra.

    %We note that the dispersions shown in Figure~\ref{fig:AnnularDispIndices} are normalized to the recovered spectra and thus account for recovery bias.

    \begin{figure}[!h]
        \begin{center}
            \includegraphics[width=0.47\textwidth]{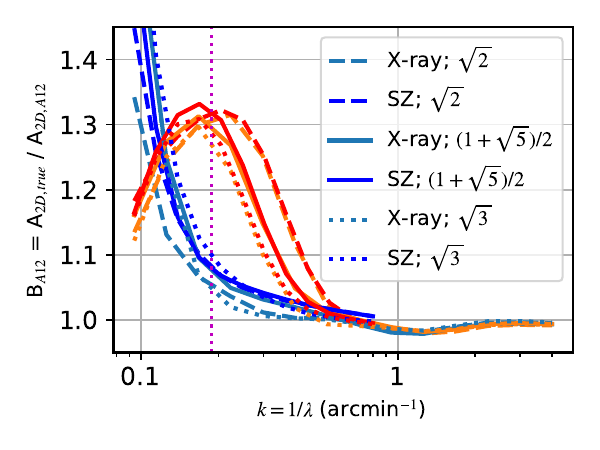}
            \includegraphics[width=0.47\textwidth]{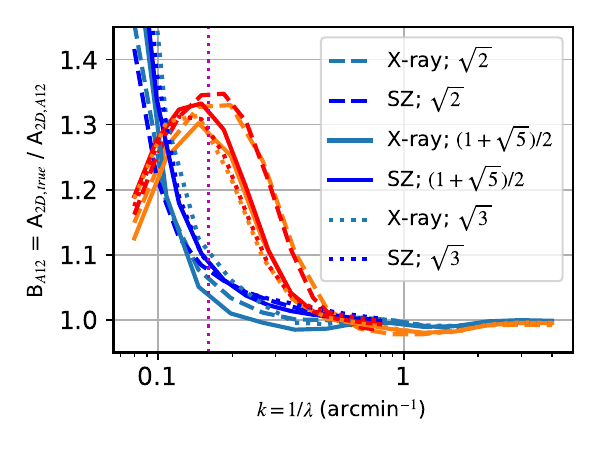}
            \includegraphics[width=0.47\textwidth]{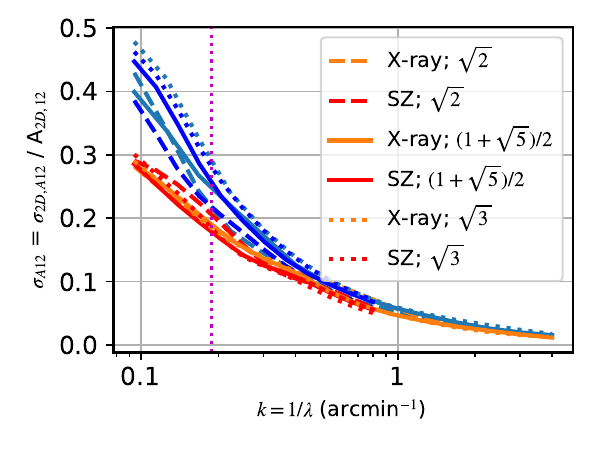}
            \includegraphics[width=0.47\textwidth]{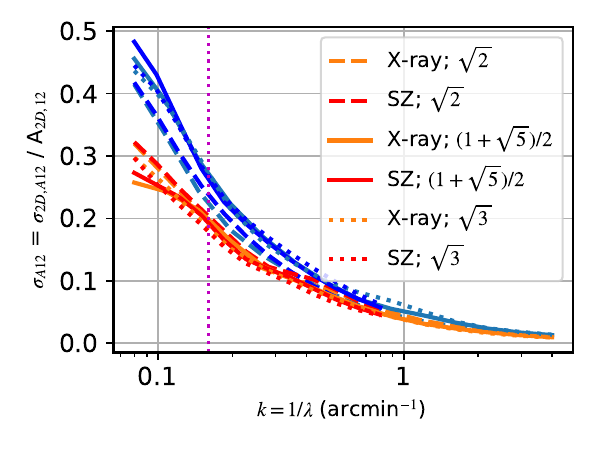}
        \end{center}
        \caption{\textbf{Left:} Bias (upper plot) and sample variance, or rather standard deviation $\sigma$, (lower plot) in magnitude spectra for SPT-CLJ0232-4421 due to masking of annuli in all cases and masking chip gaps and point sources for the X-ray cases. \textbf{Right:} Same as left, but for SPT-CLJ0638-5358. The scale (inverse) of $R_{500}$ is indicated by the magenta dotted vertical line.}
    \label{fig:AnnularVaryRadii}
    \end{figure}

    Figures~\ref{fig:AnnularBiasIndices} and \ref{fig:AnnularDispIndices} indicate that both the bias and dispersion depend on the underlying spectral index. Interestingly, there appears to be a dependence on the pixel size. This is likely due to the coarser mask with the coarser pixelization; the radii of concern are between 12 and 25 pixels in radius for the SPT (SZ) pixelization (0.25$^{\prime}$). We also see that accounting for the masking of chip gaps and point sources, while not dominant relative to the annular masking, is important. Though Figure~\ref{fig:AnnularBiasIndices} presents the sample variance relative to the scales, we can recast this to account for the area in each ring and find (see Figure~\ref{fig:AnnularDispNsamp} that the sample variance scales with the number of sampling elements as it should.

    While the sampling bias is dependent on the underlying spectral index and the sample variance appears independent of the underlying spectral index, both the sampling bias and sample variance should depend on the choice of rings. Though not discussed in Section~\ref{sec:Choices}, these biases should also be considered when choosing regions. We present sample bias and variance for both clusters using different rings in Figure~\ref{fig:AnnularVaryRadii}. The numbers present in the legend refer to the factor, $\phi$ which is used to define the separation between the inner ring (circle) and outer annulus, $R_1 = R_{500}/\phi$. As suggested in Section~\ref{sec:Choices}, within some basic constraints, the exact choice of regions (annuli) does not seem to be critical when considering the recovery of a range of scales on noiseless (synthetic) data.

    In this section, we have analyzed the recover of mock spectra which are effectively simple power-laws so as to remain relatively agnostic about the spectral shape. When exploring other other spectral shapes, especially with values $k_c$ consistent with our inferred injection scales, we find small differences relative to our current approach. In the case of a higher signal-to-noise spectrum, correcting for both spectral shape and masking may be warranted. 
    %However, with better constraints on spectral shape, one can revisit such mock calculations with more appropriate power spectral models and iteratively correct for both masking and spectral shape biases.

\section{SPT-CLJ0638-5358 By Quadrant}
\label{sec:appendix_slices}

    The shock to the southwest of the cluster center \citep{botteon2018} in SPT-CLJ0638-5358 would in itself suggest there should be elongation of the cluster along a northeast-to-southwest axis and indeed we find this (Figure~\ref{fig:EllipticityParams}). In Section~\ref{sec:Choices} we discussed the impact of fitting and subtracting elliptical surface brightness models. Here we investigate keeping a circular model but further subdividing regions to allow for more localized interrogations of fluctuations. 

    \begin{table}[!h] 
   \centering  
   \begin{tabular}{c c | c c | c c} 
             &        & \multicolumn{2}{c}{Ring 1} & \multicolumn{2}{c}{Ring 2} \\
    Quadrant & $\phi$ & $\mathcal{M}_{\text{3D,peak}}$ & $-b_{\mathcal{M}}$ & $\mathcal{M}$ & $-b_{\mathcal{M}}$ \\ 
    \hline
    \multirow{3}{*}{Q1} & $\sqrt{3}$ & $ 0.57 \pm 0.08 $ & $-0.02 \pm 0.22$ & $0.88 \pm 0.40$ & $0.20 \pm 0.19$ \\
                & $(1 + \sqrt{5})/2$ & $0.68 \pm 0.10$ & -- & -- & -- \\
                        & $\sqrt{2}$ & $0.74 \pm 0.17$ & -- & -- & -- \\
    \hline
    \multirow{3}{*}{Q2} & $\sqrt{3}$ & $0.77 \pm 0.20$ & $0.06 \pm 0.36$ & $1.07 \pm 0.45$ & $0.30 \pm 0.23$ \\
                & $(1 + \sqrt{5})/2$ & $0.82 \pm 0.25$ & -- & -- & -- \\
                        & $\sqrt{2}$ & $0.97 \pm 0.12$ & -- & -- & -- \\              
    \hline
    \multirow{3}{*}{Q3} & $\sqrt{3}$ & $0.41 \pm 0.12$ & $-0.10 \pm 0.18$ & $0.92 \pm 0.46$ &  $0.10 \pm 0.23$ \\
                & $(1 + \sqrt{5})/2$ & $0.50 \pm 0.12$ & -- & -- & -- \\
                        & $\sqrt{2}$ & $0.60 \pm 0.21$ & -- & -- & -- \\
    \hline
    \multirow{3}{*}{Q4} & $\sqrt{3}$ & $0.95 \pm 0.19$ & $-1.38 \pm 3.25$ & $2.26 \pm 0.83$ &  $0.33 \pm 0.53$ \\
                & $(1 + \sqrt{5})/2$ & $0.93 \pm 0.27$ & $-1.36 \pm 4.17$ & $2.19 \pm 0.99$ & $0.25 \pm 0.85$ \\
                        & $\sqrt{2}$ & $1.07 \pm 0.49$ & $0.21 \pm 0.72$ & $1.30 \pm 0.53$ & $0.42 \pm 0.31$ \\
    \hline
    \end{tabular} 
   \caption{Mach numbers and corresponding hydrostatic mass biases; empty fields arise due to a lack of points in our amplitude spectra above $2\sigma$.}  
   \label{tbl:bms_quads_0638}
\end{table} 

    \begin{figure}[!h]
        \begin{center}
            \includegraphics[width=0.47\textwidth]{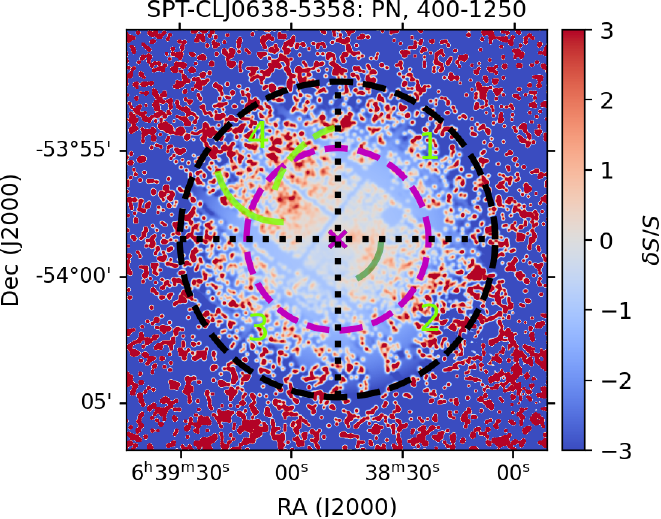}
        \end{center}
        \caption{The same as Figure~\ref{fig:FractionalResiduals} for SPT-CLJ0638 now annotated to show the quadrants.}
        \label{fig:Quadrants_SPT0638}
    \end{figure}

    In particular, we opt to subdivide our rings into quadrants and also revisit the choice of radius between Ring 1 and Ring 2. As is evidenced by some empty entries in Table~\ref{tbl:bms_quads_0638}, the data is not deep enough to produce significant (here taken as just a $2\sigma$ threshold) for all cases and quadrants explored. Even so, this analysis finds larger Mach numbers in the southwest and northeast quadrants. While this is true for Ring 1 it is more strikingly so in Ring 2, but the uncertainties are also larger in Ring 2. We find it notable that (1) the northeast quadrant, i.e. quadrant 4, recovers larger Mach numbers than quadrant 2 (wherein lies the known shock \citealt{botteon2018}) and (2) that the Mach number drops sharply when adopting $\phi = \sqrt{2}$ relative to the other ring separations. This indicates that there are prominent fluctuations between $3.^{\prime}1 < r < 3.^{\prime}3$ (potentially spanning a larger range than that). This could indicate some structure more aligned with the dashed chartreuse curve in Figure~\ref{fig:FractionalResiduals} than the solid chartreuse curve.

    We note that the residuals responsible for the dominant fluctuations in quadrant 4 are thus at larger radii than the SE shock found in \citep{botteon2018}; thus their search via Gaussian gradient magnitude (GGM) may not have yielded a detection in part due to the decrease in surface brightness at larger radii. Admittedly the fluctuations do not appear as a canonical brightness jump, which would also reduce the prominence in a GGM image. Indeed, we might take the ambiguity from the SPT residuals as being suggestive that the gas dynamics is more complicated than a classical merger shock, which even in favorable geometries such as in Abell 2146 \citep[e.g.,][]{russell2012,russell2022} the heating of shocks can be quite complicated \citep{chadayammuri2022}.

\section{Intermediate data products}
\label{sec:appendix_Supplemental}

    Although we do make inferences from the logarithmic pressure and Mach slopes, we present them in Table~\ref{tbl:log_slopes} for transparency in the values use to derive hydrostatic mass biases that are reported in Table~\ref{tbl:bms}. In particular, these values are used in Equations~\ref{eqn:mach_bias} and \ref{eqn:Pnt_mach}.

    \begin{table}[!h] 
   \centering  
   \begin{tabular}{c c | c c c c} 
    & & $d \ln P / d \ln r$ & $d\ln \mathcal{M}_{\text{peak}} / d \ln r$ & $d\ln \mathcal{M}_{\text{int}} / d \ln r$ & $d\ln \mathcal{M}_{\text{comb}} / d \ln r$ \\ 
   \hline 
   \multirow{2}{*}{SPT-CLJ0232-4421} & Ring 1 & $-0.75 \pm 0.01$ & \multirow{2}{*}{$-0.511 \pm 0.70$} & \multirow{2}{*}{$-0.91 \pm 1.07$} & \multirow{2}{*}{$-0.66 \pm 1.07$} \\ 
    & Ring 2 & $-3.42 \pm 0.09$ &  &  &  \\ 
    \hline
    \multirow{2}{*}{SPT-CLJ0638-5358} &  Ring 1 & $-1.21 \pm 0.01$ & \multirow{2}{*}{$1.50 \pm 0.83$} & \multirow{2}{*}{$0.73 \pm 0.72$} & \multirow{2}{*}{$1.26 \pm 1.23$} \\ 
    & Ring 2 & $-3.04 \pm 0.11$ &  &  &  \\ 
   \end{tabular} 
   \caption{The logarithmic slopes as inferred for the circular cluster model and presented in Section~\ref{sec:PS_results}. In particular these values are used in Equations~\ref{eqn:mach_bias} and \ref{eqn:Pnt_mach}. } 
   \label{tbl:log_slopes}
\end{table}

\section{Mass estimates for our pilot clusters}
\label{appendix:mass_estimates}

The clusters chosen for our pilot survey are quite massive ($M_{500} \sim 10^{15}$ M$_{\odot}$) and consequently have been observed by many facilities and have many mass estimations. We do not intend to review the details of each mass estimation method used here, but broadly characterize the mass estimation as either a total mass estimate or a hydrostatic mass (whether explicitly calculating mass via hydrostatic equilibrium or building of some relation which was established using the assumption of hydrostatic equilibrium). Our goal is then to contextualize our hydrostatic mass bias values obtained through our fluctuation analysis with hydrostatic mass bias values obtained by other methods. In addition to masses available in the literature, we perform another version of our pressure profile fitting (Section~\ref{sec:betaFittingSPT}) where we fit a universal pressure profile \citep[UPP;][]{arnaud2010} such that it is a function of $M_{500}$ and redshift, $z$, where the former is allowed to vary and the latter is fixed. We report our obtained values of $M_{500}$ in Table~\ref{tbl:lit_masses} and note that by assuming a fixed profile shape, the associated uncertainties are artificially reduced.

\begin{table}[!h] 
   \centering  
   \begin{tabular}{c c c c c | c c} 
Reference & Facility(ies) & Cosmology & Quantity & Type & SPT-CLJ0232-4421 & SPT-CLJ0638-5358 \\ 
\hline 
Bleem15       & SPT$^{A}$ & Fiducial$^{a,b}$ & $M_{500}$  & \textcolor{teal}{Total} & $ 12.01_{-1.80}^{+1.80}$ & $ 12.01_{-1.81}^{+1.81}$ \\ 
Hilton21      & ACT & Fiducial$^{a}$ &  $M_{500,uncorr}$  & \textcolor{teal}{HE} & $ 10.12_{-1.75}^{+2.12}$  & $ 10.47_{-1.91}^{+2.33}$ \\ 
Hilton21      & ACT & Fiducial$^{a}$ & $M_{500,c}$  & \textcolor{teal}{\underline{HE}}& $ 8.73_{-1.43}^{+1.71}$  & $ 8.91_{-1.53}^{+1.85}$ \\ 
Hilton21      & ACT$^{D}$ & Fiducial$^{a}$ & $M_{500,cal}$  & \textcolor{teal}{\underline{Total}} & $ 12.30_{-1.43}^{+2.70}$  & $ 12.54_{-1.53}^{+2.88}$ \\ 
Bocquet19     & SPT$^{B}$  & Fiducial$^{a,b}$ & $M_{500}$ & \textcolor{teal}{Total} & $ 11.30_{-1.36}^{+1.11}$ & $ 11.29_{-1.36}^{+1.10}$ \\ 
Bulbul19      & SPT$^{C}$ & Fiducial$^{c}$ & $M_{500}$  & \textcolor{teal}{\underline{Total}} & $ 9.45_{-1.10}^{+1.16}$ & $ 9.42_{-1.09}^{+1.18}$ \\ 
Salvati22     &\textit{Planck}+SPT & Fiducial$^{d}$ & $M_{500,fixed}$ & \textcolor{teal}{Total} & $ 11.16_{-1.14}^{+1.52}$  & $ 9.51_{-1.11}^{+1.81}$ \\ 
Bleem15       & SPT$^{A}$ & Planck2014$^{e}$ & $M_{500}$ & \textcolor{green}{Total} & $14.67 \pm 2.29$ & $ 14.67 \pm 2.30$ \\ 
Planck16      & \textit{Planck} & Planck2016$^{f}$ & $M_{500}$ & \textcolor{green}{HE} & 7.54 & 6.83 \\
Melin21       & \textit{Planck}+SPT & Planck2020$^{g}$ & $M_{500}$ & \textcolor{green}{HE} & $8.29_{-0.23}^{+0.23}$ & $7.72_{-0.23}^{+0.23}$ \\
Salvati22     &\textit{Planck}+SPT & $\nu \Lambda$CDM$^{h}$ & $M_{500,free}$ & \textcolor{red}{Total} & $10.35_{-2.63}^{+2.72}$  & $ 9.24_{-2.40}^{+2.44}$ \\ 
Tarr{\'\i}o19 & \textit{Planck}+\textit{ROSAT} & Fiducial$^{a}$ & $M_{500}$ & \textcolor{red}{HE} & $ 7.82_{-0.96}^{+0.90}$ & $ 8.61_{-0.67}^{+0.64}$ \\ 
Piffaretti11  & \textit{ROSAT} & Fiducial$^{a}$ & $M_{500}$ & \textcolor{red}{HE} & $ 6.13$ & $ 6.88$ \\ 
Mantz10       & \textit{ROSAT}+\textit{Chandra} & Fiducial$^{a}$ & $M_{500}$ & \textcolor{red}{\underline{HE}} & $12.7 \pm 2.5$ & $10.3 \pm 1.4$ \\
Zhang08       & \textit{XMM} & Fiducial$^{a}$ & $M_{500}^{\clubsuit}$ & \textcolor{red}{\underline{HE}} & $8.43 \pm 2.48$ & -- \\
Zhang08       & \textit{XMM} & Fiducial$^{a}$ & $M_{500}^{\diamondsuit}$ & \textcolor{red}{HE} & $7.66 \pm 2.20$ & -- \\
Fox22         & \textit{HST} & Fiducial$^{a}$ & $M_{500,\text{sl}}^{\dagger}$ & \textcolor{red}{\underline{Total}} &  $7.54_{-0.32}^{+0.33}$  & -- \\
%Hernandez23   & DES+\textit{Planck} & Fiducial$^{a}$ & $M_{500,\lambda}^{\bowtie}$ & \textcolor{red}{Total} & -- & $8.41$ \\
CoMaLitV      & \textit{ROSAT} & Fiducial$^{a}$ & $M_{500,\text{X,wl}}^{\ddagger}$ & \textcolor{red}{\underline{Total}} & $7.41 \pm 0.59$ & $7.96 \pm 0.67$ \\
Klein19       & WFI$^{E}$ & Fiducial$^{a}$ & $M_{500}^{\heartsuit}$ & \textcolor{red}{Total} & $5.13_{-1.69}^{+1.94}$ & -- \\
Klein19       & WFI$^{E}$ & Fiducial$^{a}$ & $M_{500}^{\spadesuit}$ & \textcolor{red}{Total} & $4.08_{-1.31}^{+1.67}$ & -- \\
This work     & SPT & Fiducial$^{a}$ & $M_{500,{\text{UPP}}}^{\bowtie}$ & \textcolor{red}{\underline{HE}} & $8.36 \pm 0.05 $ & $7.70 \pm 0.05$ \\
\hline
Bulbul19      & \textit{XMM} & Fiducial$^{a}$ & $M_{\text{gas},500}$ & \textcolor{red}{\underline{Gas}} & $1.67_{-0.08}^{+0.08} $ & $0.97_{-0.21}^{+0.21}$ \\ 
Mantz10       & \textit{ROSAT}+\textit{Chandra} & Fiducial$^{a}$ & $M_{\text{gas},500}$ & \textcolor{red}{\underline{Gas}} & $1.45 \pm 0.25$ & $1.18 \pm 0.13$ \\
Zhang08       & \textit{XMM} & Fiducial$^{a}$ & $M_{\text{gas,}500}$ & \textcolor{red}{\underline{Gas}} & $0.89\pm 0.09$  & -- \\
   \end{tabular} 
   \caption{\textbf{Cosmologies:} $^{a}$$\Lambda$CDM with $\Omega_m = 0.3$, $\Omega_{\Lambda} = 0.7$, and $H_0 = 70$ km/s/Mpc. $^{b}$ Added stipulation that $\sigma_8 = 0.8$. $^{c}$$\Lambda$CDM with $\Omega_m = 0.3$, $\Omega_{\Lambda} = 0.7$, and $H_0 = 67.74$ km/s/Mpc and $\sigma_8 = 0.82$. $^d$Flat $\nu\Lambda$CDM with $\Omega_m = 0.3$, $\sigma_8 = 0.8$, $(1-b)_{\text{SZ}} = 0.58$. $^{e}$Planck2014 cosmology \citep{planck2014_XVI}: $H_0 = 67.3$ km/s/Mpc; $\Omega_m = 0.315$; $\Omega_{\Lambda} = 1-\Omega_m$, $\sigma_8 = 0.84$; $^{f}$Planck2016 cosmology \citep{planck2016_XIII}: $H_0 = 67.5$ km/s/Mpc; $\Omega_m = 0.312$; $\Omega_{\Lambda} = 0.688$, $\sigma_8 = 0.815$; $^{g}$Planck2020 cosmology \citep{planck2020_VI}: $H_0 = 67.4$ km/s/Mpc; $\Omega_m = 0.315$; $\Omega_{\Lambda} = 0.685$, $\sigma_8 = 0.811$. $^{h}$$\Omega_m = 0.29$, $\Omega_{\Lambda} = 0.71$, $H_0 = 61.3$ km/s/Mpc, and $\sigma_8 = 0.76$. \textbf{Facilities:} $^{A}$Additional facilities are used for cluster confirmation in the catalog; X-ray data from \textit{ROSAT} has a small weight in cluster masses via the fitted $Y_X - M$ relation. $^{B}$As with $^{A}$, but \textit{Chandra} is used rather than \textit{ROSAT} data; data from Magellan and \textit{HST} are also used for weak lensing measurements. $^{C}$External facilities are as in \citet{deHaan2016}, except that \textit{XMM} data is used instead of \textit{Chandra}. $^{D}$Richness-based weak-lensing calibration from DES \citep{mcclintock2019}.
   $^{E}$The Wide-Field Imager at the 2.2m MPG/ESO telescope.  \textbf{Citation keys:} Bleem15: \citet{bleem2015}, Bocquet19: \citet{bocquet2019}, Bulbul19: \citet{bulbul2019}, Hilton21: \citet{hilton2021}, Melin21: \citet{melin2021} Salvati22: \citet{salvati2022}, Planck16: \citet{planck2016_XXVII}, Tarr{\'\i}o19: \citet{tarrio2019}, Piffaretti11: \citet{piffaretti2011}, Mantz10: \citet{mantz2010b}, Zhang08: \citet{zhang2008}, Fox22: \citet{fox2022},  CoMaLitV:\citet{sereno2017}, Klein19: \citet{klein2019}
   \textbf{Quantities:} $^{\dagger}$sl = strong lensing. $^{\ddagger}$Masses come from the MCXC estimates \citep{piffaretti2011} and have been scaled using a weak-lensing calibration. $^{\bowtie}$Assumes the UPP shape is fixed to that in \citet{arnaud2010}. $^{\clubsuit}$Hydrostatic equilibrium mass profile from $n_e$,$T_e$ profiles. $^{\diamondsuit}$Assumes an $L_X$-$M$ relation derived from the work's sample. $^{\heartsuit}$Background galaxies selected by detection-optimization with distance and purity cuts. $^{\spadesuit}$Taking the conservative background selection without a prior on the concentration. \textbf{Types:} Masses are deemed either hydrostatic (HE) or total (Total) and are color-coded to signify masses which appear to be comparable. If the type is underlined, then it is used in the calculation of a hydrostatic mass bias (Figure~\ref{fig:HydMassBiases}).} 
   \label{tbl:lit_masses}
\end{table} 

We have tried to identify groups of mass estimates within which we consider the comparison of hydrostatic masses (denoted with HE) to total masses to be appropriate and have color coded these green, teal, or red; in the red sample we also include the gas masses which can be used to estimate the hydrostatic bias with an accompanying HE mass. Furthermore, in the red sample, our desire is to have a total mass estimate from lensing data. Unfortunately SPT-CLJ0638-5358 appears to be without a lensing mass estimate in the literature. In lieu of this, we have used the mass as determined by a weak-lensing calibrated $L_X$-$M$ relation \citep{sereno2017}. The distinction between the green and teal groups is the assumed cosmology, where we deem the different Planck cosmologies to be sufficiently close to each other for comparison. For the fiducial cosmologies we exclude the uncorrected (uncorr) mass estimates from \citet{hilton2021}, which do not account for up-scatter in mass estimates with respect to the underlying cluster mass function, though it should be evident that a hydrostatic mass bias derived with that value will be lower than the bias value obtained with the corrected mass estimate. Additionally, we apply the small correction based on $h$, $(0.6774/0.7)^{-1}$, to the masses from \citet{bulbul2019} for comparison with similar mass estimates.
% Hernandez23: \citet{hernandez2023}, Apparently they just rescaled Planck masses by 1.0/0.8 = 1.25. Bah. Not particularly insightful for probing individual clusters.
% The same could be said for CoMaLit.

\end{document}